\documentclass[lettersize,journal]{IEEEtran}
\usepackage{amsmath,amsfonts}
\usepackage{algorithmic}
\usepackage{algorithm}
\usepackage{array}
\usepackage[caption=false,font=normalsize,labelfont=sf,textfont=sf]{subfig}
\usepackage{textcomp}
\usepackage{stfloats}
\usepackage{url}
\usepackage{verbatim}
\usepackage{graphicx}
\usepackage{cite}
\usepackage{booktabs}
\usepackage{tabularx}
\usepackage{adjustbox}
\usepackage{tabulary}
\usepackage{tabu}
\usepackage{array}
\usepackage{multirow}
\usepackage{booktabs}
\usepackage{tabularx}
\usepackage{threeparttable}
\begin{document}

\title{A BTR-Based Approach for  Detection\\ of Infrared Small Targets}

\author{Ke-Xin Li
      
\thanks{This paper was produced by the IEEE Publication Technology Group. They are in Piscataway, NJ.}
\thanks{Manuscript received April 19, 2021; revised August 16, 2021.}}

\markboth{Journal of \LaTeX\ Class Files,~Vol.~14, No.~8, August~2021}%
{Shell \MakeLowercase{\textit{et al.}}: A Sample Article Using IEEEtran.cls for IEEE Journals}

\maketitle

\begin{abstract}
Infrared small target detection plays a crucial role in military reconnaissance and air defense systems. However, existing low-rank sparse based methods still face high computational complexity when dealing with low-contrast small targets and complex dynamic backgrounds mixed with target-like interference. To address this limitation, we reconstruct the data into a fourth-order tensor and propose a new infrared small target detection model based on bilateral tensor ring decomposition, called BTR-ISTD. The approach begins by constructing a four-dimensional infrared tensor from an image sequence, then utilizes BTR decomposition to effectively distinguish weak spatial correlations from strong temporal-patch correlations while simultaneously capturing interactions between these two components. This model is efficiently solved under the proximal alternating minimization (PAM) framework. Experimental results demonstrate that the proposed approach outperforms several state-of-the-art methods in terms of detection accuracy, background suppression capability, and computational speed.
\end{abstract}


\section{Introduction}
\IEEEPARstart{I}{nfrared} small target detection, as a pivotal technology in the domains of national security \cite{ref1} and public security \cite{ref2}, has been extensively applied over the past decades in numerous critical fields including landmine detection, early warning systems, nighttime navigation, and precision-guided weaponry\cite{ref3,ref4,ref5,ref6}. Typically occupying only a few pixels and embedded within complex backgrounds with strong interference, effectively distinguishing infrared small targets from the background necessitates the simultaneous utilization of both spatial local correlation and temporal continuity. Existing detection methods can be broadly categorized into two classes based on the "information domain" employed: single-frame infrared image detection and multi-frame infrared image detection.
\subsection{Related Work}
Single-frame methods rely solely on the current frame's data to determine the presence and location of a target. Their fundamental principle lies in exploiting the distinctive characteristics of targets and backgrounds in the spatial domain, such as shape, texture, and intensity contrast. Pioneering work by
Goldfarb and Qin et al.\cite{ref7} investigated the infrared patch-tensor (IPT) model and proposed the tensor robust principal component analysis (TRPCA) model. Gao et al.\cite{ref8} introduced an infrared small target detection method based on the infrared patch-image (IPI) model and low-rank sparse matrix decomposition, which achieves more stable and superior detection performance under varying backgrounds and target sizes through stable principal component optimization and adaptive segmentation. Dai et al.\cite{ref9} further addressed the insufficient observation of strong edges in the IPI model by employing more accurate low-rank background estimation and nonnegative constraints, thereby significantly enhancing target saliency and clutter suppression in extremely complex backgrounds. 
\begin{figure}[!h]
\centering
% 第一行
\begin{minipage}{0.35\linewidth}
\centering
\subfloat[]{\includegraphics[width=1.3\linewidth]{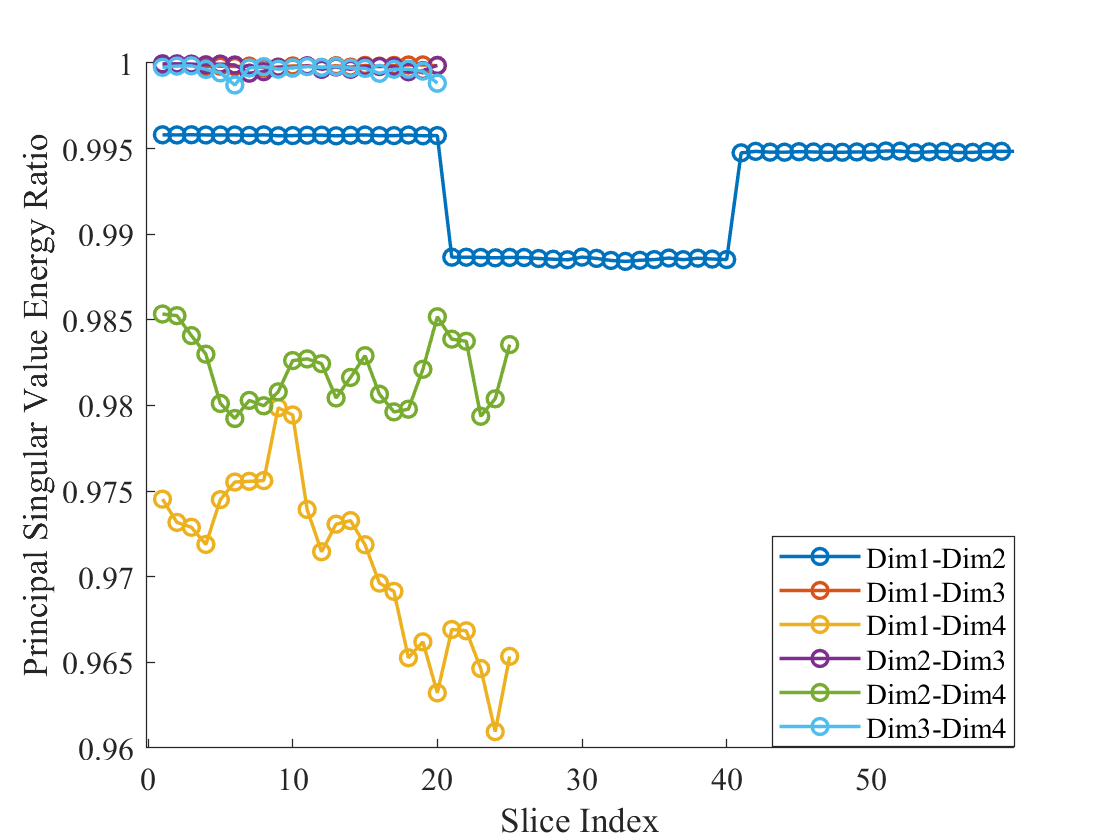}%
\label{fig_first_case}}
\end{minipage}
\hspace{0.16\linewidth}
%\hfill
\begin{minipage}{0.35\linewidth}
\centering
\subfloat[]{\includegraphics[width=1.3\linewidth]{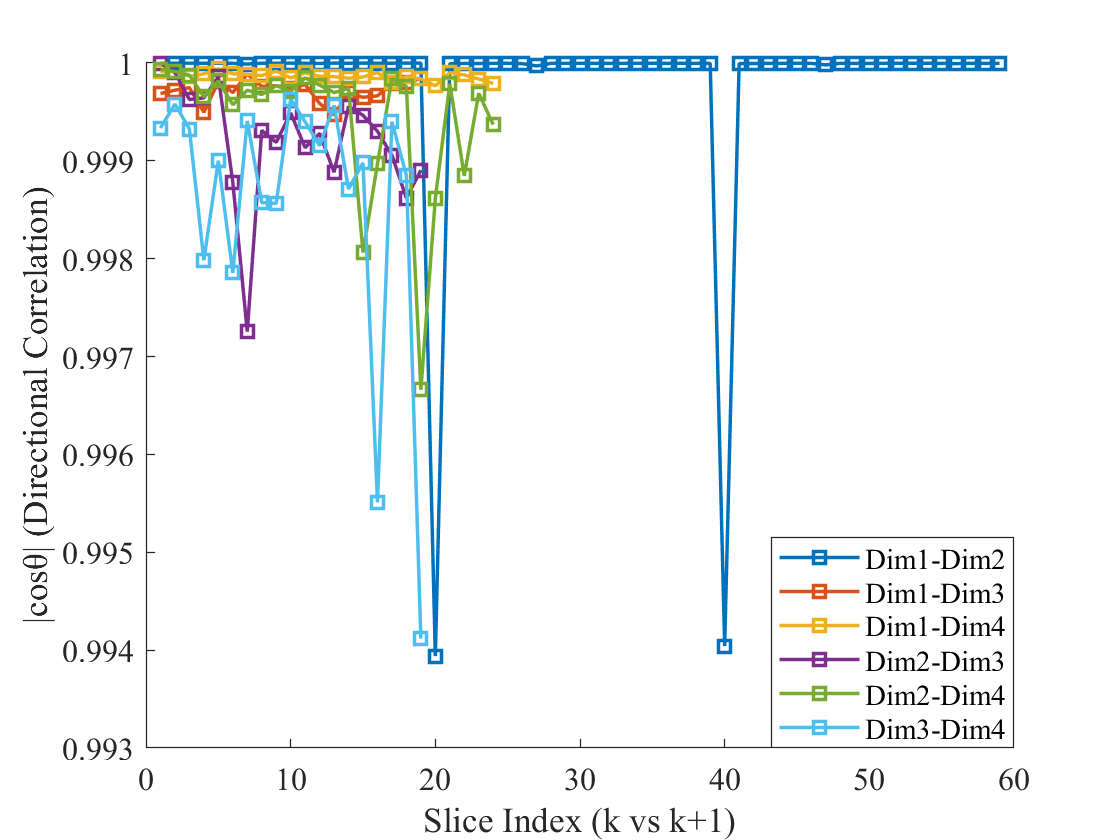}%
\label{fig_second_case}}
\end{minipage}
\caption{Correlation Analysis Between Tensor Dimensions of Small Infrared Targets.(a)SVD Energy Ratio Comparison. (b)SVD Directional Correlation}
\label{fig_sim}
\end{figure}
Subsequently, the tensor, as a higher-dimensional extension of the matrix, was introduced to address the poor real-time performance issues associated with matrix-based methods, which often rely heavily on singular value decomposition.
For instance, Dai and Wu et al.\cite{ref10} incorporated element-wise local structure adaptive weights into the infrared patch tensor (IPT) model, called reweighted IR patch-tensor (RIPT) model. Zhang et al.\cite{ref11} proposed a method that combines the partial sum of the tensor nuclear norm (PSTNN) with directional residual-weighted dual-neighborhood local contrast, while Zhu et al.\cite{ref12} developed tensor completion with top-hat regularization (TCTHR) by exploiting target structural priors and background autocorrelation. These methods improved background suppression and target–background separation under complex scenes. Nevertheless, single-frame approaches remain constrained by the absence of temporal information, making it difficult to distinguish targets from clutter in sea–sky, cloud, and wave environments.

Multi-frame detection methods exploit temporal characteristics of targets from consecutive image sequences while integrating spatial information, making it easier to detect extremely dim and small targets compared with single-frame approaches. Liu et al.\cite{ref13} proposed an infrared small target detection method based on spatio-temporal tensor (STT) decomposition, where a sliding window is used to construct a spatio-temporal cube and formulate a low-rank–sparse tensor model, effectively separating targets from backgrounds in complex scenes. Building upon STT, Zhang et al.\cite{ref14} introduced corner-aware spatio-temporal tensor (ECA-STT) decomposition by incorporating corner importance measures and nonlocal tensor total variation, which mitigates the interference of strong edges and corners in the background. Liu et al.\cite{ref15} further developed a nonconvex tensor low-rank approximation (NTLA) method, which adaptively reweights different singular values for more accurate background estimation, and integrated asymmetric spatio-temporal total variation (ASTTV) regularization to enhance robustness against complex scenarios. Luo et al.\cite{ref16} presented an improved method, IMNN-LWEC, which constructs tensors from non-overlapping spatio-temporal blocks and introduces an improved multi-modal weighted tensor nuclear norm (IMWTNN) for more precise background separation, while combining local weighted entropy contrast (LWEC) and tubal sparsity regularization to further enhance target saliency and suppress background interference. Yin et al.\cite{ref17} proposed a three-dimensional low-rank and sparse tensor decomposition framework, which distinguishes targets from pseudo-target structures using 3DST, models background low-rankness with 3DLogTNN, and enforces background smoothness via 3DTV, thereby achieving superior detection and background suppression in complex scenes.
\subsection{Motivation}
However, three-dimensional methods often struggle to simultaneously capture multi-level correlations across local-global and spatiotemporal dimensions. To address this issue, Wu et al. [18]proposed a fourth-order tensor-based infrared small target detection method. This approach utilizes tensor ring decomposition to balance spatio-temporal information and employs weighted kernel norm constraints to separate low-rank backgrounds from sparse targets. Although this TR method enhances background suppression while preserving spatio-temporal correlations, it exhibits several limitations: overly simplified cross-dimensional correlation handling, high computational complexity, and insufficient adaptability to complex dynamic backgrounds and long sequences. 

To overcome these shortcomings, we begin by exploring correlations between fourth-order tensor dimensions. Based on singular value decomposition (SVD), we construct two key metrics to characterize structural dependencies among different dimension combinations. For any given pair of dimensions, we unfold the tensor along that dimension and perform SVD decomposition on each slice of the resulting matrix sequence. This yields the proportion of principal singular value energy for each slice and the directional consistency between principal singular vectors of adjacent slices. Fig. 1(a) displays the principal singular value energy distribution curves for different dimensionality combinations. The horizontal axis represents the slice index, corresponding to the sequential matrix slice number after unfolding along that dimensionality combination. The vertical axis shows the principal singular value energy distribution, reflecting the proportion of energy accounted for by the first singular value in the current slice. Higher values indicate more prominent and stable principal structures. Fig. 1(b) presents the directional consistency curve of principal singular vectors. The horizontal axis again denotes slice indices, but here it compares adjacent slices (k and k+1). The vertical axis shows the cosine value of the angle between the principal singular vectors, $|\cos\theta|$, which describes the magnitude of the directional variation between adjacent slices. Values closer to 1 indicate more stable structural orientation and stronger inter-dimensional correlation.

The results reveal that dimensions 1–2 and 3–4 exhibit the highest proportion of principal singular value energy among all dimension pairs, with the smoothest variation. This indicates these two dimension pairs maintain more consistent and prominent dominant structural features across different slices. Concurrently, the direction consistency curves in Fig. 1 reveal that the $|\cos\theta|$ values for these two dimension pairs remain consistently elevated with minimal fluctuations. This indicates their principal singular directions exhibit the least variation across consecutive slices, reflecting a more stable directional structure. The consistency of principal singular value energy distribution and the stability of principal direction consistency both support the same conclusion: for infrared small target tensors, dimensions 1–2 and 3–4 exhibit the strongest structural correlation, demonstrating significant energy concentration and directional consistency. Therefore, based on these findings, this study proposes an infrared small target detection model based on bidirectional TR.

\begin{figure*}[!t]
\centering
\includegraphics[width=0.95\textwidth]{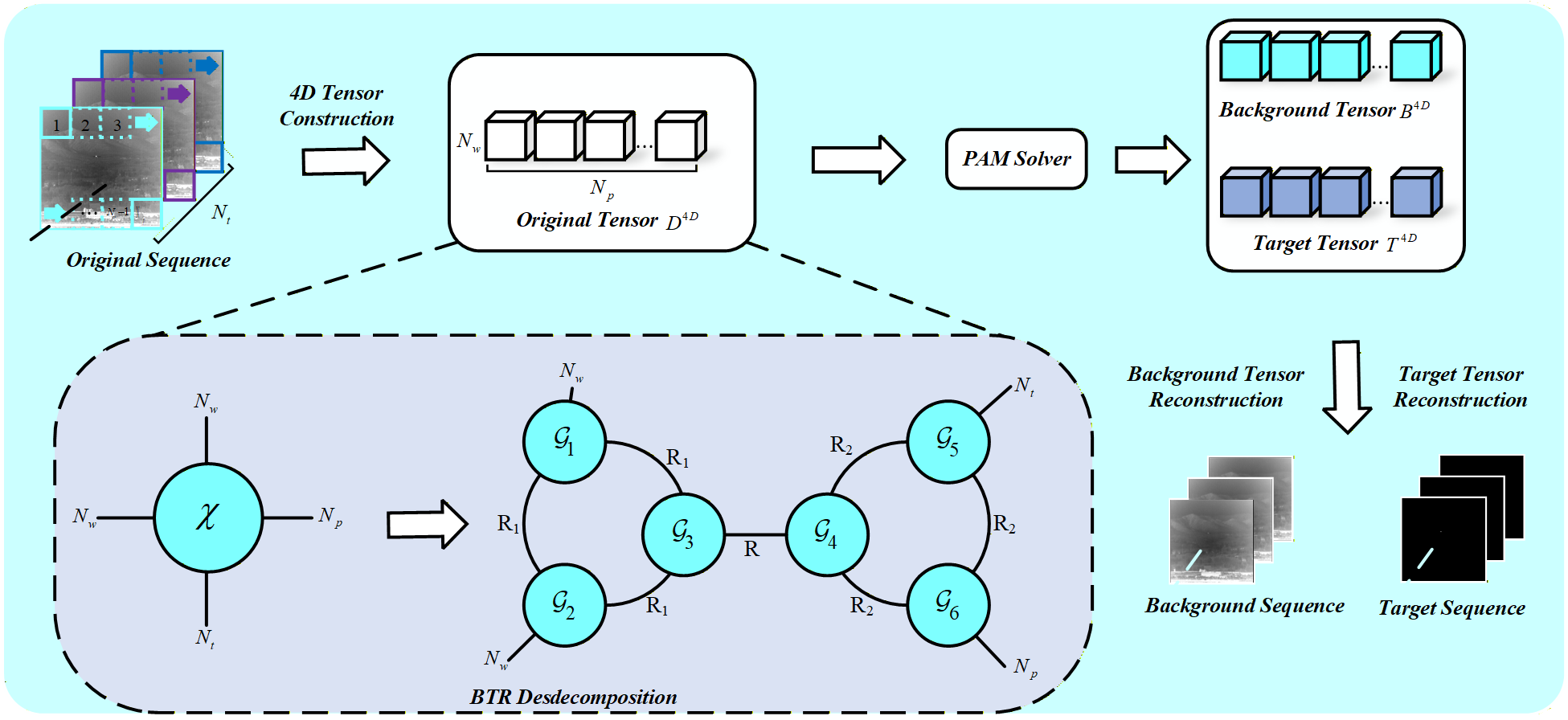}
\caption{The overall procedure of proposed methods in BTR-ISTD.}
\label{fig_sim}
\end{figure*}

In summary, our work makes the following two key contributions:
\begin{itemize}
  \item Using an SVD-based correlation analysis approach,  we systematically examines the dimensional coupling characteristics within infrared small-target tensors from the perspectives of energy dominance and directional consistency. The experimental results demonstrate that the infrared small-target data exhibit the strongest structural correlation along dimensions 1–2 and 3–4, significantly higher than that observed in other dimension pairs. This finding reveals intrinsic structural patterns within the infrared small-target tensor and provides a reliable basis for subsequent tensor-model design, dimension-selection strategies, and feature-constraint formulation.
  \item We develop a novel infrared small target detection model based on Bidirectional Tensor Ring (BTR) decomposition, building upon previous research through a transfer learning approach. Our model effectively captures multi-dimensional correlations across local-to-global and spatial-to-temporal dimensions in 4D infrared data using efficient BTR decomposition, significantly reducing computational and storage complexity. Additionally, by incorporating a specifically designed Parallel Alternating Minimization (PAM) optimization algorithm, we establish theoretical convergence guarantees for the proposed method.
  \item Extensive experimental results demonstrate that our approach achieves superior performance in infrared small target detection and improves computational efficiency by orders of magnitude compared to state-of-the-art tensor-based methods (see Table II for details).
\end{itemize}
\subsection{Outlines}
The remainder of this article is organized as follows. Section II introduces the notations and provides preliminary knowledge. Section III constructs an infrared small target detection model based on BTR decomposition and develops a PAM-based solving algorithm with theoretical convergence guarantees. Section IV conducts extensive experiments to evaluate the performance of the proposed method. Finally, Section V concludes the article.
\section{Notation And Priliminaries}
We use bold lowercase letters, e.g., $\mathbf{x}$, bold uppercase letters, e.g., $\mathrm{X}$, and calligraphic uppercase letters, e.g., $\mathcal{X}$, to denote vectors, matrices, and tensors respectively. As an example, a tensor of order $N = 4$ with size $I_1 \times I_2 \times I_3 \times I_4$ is denoted by $\mathcal{X} \in \mathbb{R}^{I_1 \times I_2 \times I_3 \times I_4}$. Meanwhile, $\|\mathcal{X}\|_1$ indicates the $\ell_1$ norm of tensor $\mathcal{X}$, and $\|\mathcal{X}\|_F$ is adopted to describe Frobenius norm of tensor $\mathcal{X}$. The Frobenius norm of a tensor $\mathcal{X} \in \mathbb{R}^{I_1 \times I_2 \times \cdots \times I_N}$ is 
defined as $\|\mathcal{X}\|_F=\sqrt{\sum_{i_1,i_2,\cdots,i_N}|\mathcal{X}(i_1,i_2,\cdots,i_N)|^2}$.

Tensor unfolding, also referred to as matricization, reorganizes the data from the original tensor into a matrix structure by mapping a designated subset of its dimensions to the rows and projecting the remaining dimensions to the columns.For instance,  
an $I_1 \times I_2 \times I_3 \times I_4$ tensor can be restructured into an $I_1 \times I_2 I_3 I_4$ matrix, represented as $X_{[1|2,3,4]} = \text{Unfold}(\mathcal{X}, [1|2,3,4])$.  
It can also be restructured into an  
$I_1 I_2 \times I_3 I_4$ matrix, denoted by $X_{[1,2|3,4]} = \text{Unfold}(\mathcal{X}, [1,2|3,4])$.

The tensor contraction operation is performed between two given tensors to generate a new tensor by pairing specific dimensions, multiplying corresponding elements, and summing over the paired indices. For instance, given a third-order tensor \(\mathcal{X} \in \mathbb{R}^{I \times J \times K}\) and fourth-order tensor \(\mathcal{Y} \in \mathbb{R}^{M \times I \times N \times K}\), the tensor contraction between the 1st and 3rd dimensions of \(\mathcal{X}\) and the 2nd and 4th dimensions of \(\mathcal{Y}\) yields a tensor \(Z = \mathcal{X} \times_{1,3}^{2,4} \mathcal{Y} \in \mathbb{R}^{J \times M \times N}\), whose elements are calculated as
\begin{equation*}
\mathcal{Z}(j,m,n)=\sum_{i=1}^I\sum_{k=1}^K\mathcal{X}(i,j,k)\mathcal{Y}(m,i,n,k).
\end{equation*}
\textbf{\textit{Theorem 1:}}  \cite{ref19} Let tensors \(\mathcal{X} \in \mathbb{R}^{I_1 \times I_2 \times \cdots \times I_N}\) and \(\mathcal{Y} \in \mathbb{R}^{J_1 \times J_2 \times \cdots \times J_M}\), then we have
\begin{align*}
    &Z = \mathcal{X} \times^m_n \mathcal{Y}\Leftrightarrow\\
    & Z_{[1: N-d|N-d+1:N+M-2d]} = \mathcal{X}_{[n_1:n_2|n_3:n_4]}\times\mathcal{Y}_{[m_1:m_2|m_3:m_4]}
\end{align*}
where $n$ and $m$ are arbitrary rearrangements of $(1, 2, \dots, N)$ and $(1, 2, \dots, M)$, respectively.

Theorem 1 implies that tensor contraction is computationally equivalent to matrix multiplication. For example, given two tensors $\mathcal{X} \in \mathbb{R}^{I \times J \times K}$ and $\mathcal{Y} \in \mathbb{R}^{M \times I \times N \times K}$, the contraction $Z = \mathcal{X} \times_{1,3}^{2,4} \mathcal{Y} \in \mathbb{R}^{J \times M \times N}$ is equivalent to the matrix multiplication $Z_{[1|2,3]} = X_{[2|1,3]} Y_{[2,4|1,3]} \in \mathbb{R}^{J \times MN}$.
\section{Proposed Algorithm And Solving Algorithm}
As illustrated in Fig. 2, the construction of the 4D tensor initiates with extracting $N_{p}$ spatial local patches of size ${N_{w}\times N_{w}}$ from each single-frame infrared image using a sliding window technique. These patches are first stacked along the third dimension (frame index $N_{t}$) to form a three-dimensional sub-tensor. Subsequently, multiple such sub-tensors from consecutive frames are concatenated along the fourth dimension (patch index $N_{p}$), ultimately yielding a four-dimensional infrared tensor $\mathcal{D}^{4D}$ with dimensions ${N_{w}\times N_{w}\times N_{t}\times N_{p}}$. Correspondingly, the novel 4D infrared tensor $\mathcal{D}^{4D}$ can be formulated as: 
\begin{equation}
\label{deqn_ex1a}
{\mathcal{D}}^{4D}={\mathcal{B}}^{4D}+{\mathcal{T}}^{4D}
\end{equation}
Where $\mathcal{B}^{4D}$ and $T^{4D}\in\mathbb{R}^{N_{w}\times N_{w}\times N_{t}\times N_{p}}$, respectively, represent the background tensor, the target tensor.

Given the properties of a low-rank background and sparse targets, the following model can be formulated:
\begin{equation}
\begin{split}
&\min_{\mathcal{B}^{4D}, \mathcal{T}^{4D}}\  \mathrm{rank}(\mathcal{B}^{4D}) + \lambda_1 \| \mathcal{T}^{4D} \|_0 \\
&\mathrm{s.t.}\ \mathcal{D}^{4} = \mathcal{B}^{4D} + \mathcal{T}^{4D}.
\end{split}
\end{equation}

In the formulated model, the first term constitutes a low-rank component that characterizes the continuously varying background of infrared images, the second term represents a sparse component that captures the sparse nature of dim and small targets within the high-dimensional tensor, and $\lambda_1$ serves as a positive penalty parameter that regulates the noise component.
\subsection{BTR Decomposition for Infrared Small Target Detection}
First, we redefine the BTR decomposition to adapt it for 4D infrared small target tensors.\\
\textit{\textbf{Definition 1}(BTR Decomposition):}Let   $\mathcal{X}\in\mathbb{R}^{N_{w}\times N_{w}\times N_{t}\times N_{p}}$ be an fourth-order tensor such that
\begin{equation}
\label{deqn_ex1a}
\begin{split}
\mathcal{X} &= [\mathrm{TR}(\mathcal{G}_{1}, \mathcal{G}_{2}, \mathcal{G}_{3})] \times_{3}^{1} [\mathrm{TR}(\mathcal{G}_{4}, \mathcal{G}_{5}, \mathcal{G}_{6})] \\
&= [\mathcal{G}_{1} \times_{3}^{1} \mathcal{G}_{2} \times_{4,1}^{1,3} \mathcal{G}_{3}] \times_{3}^{1} [\mathcal{G}_{4} \times_{3}^{1} \mathcal{G}_{5} \times_{4,1}^{1,3} \mathcal{G}_{6}].
\end{split}
\end{equation}

Where $\mathcal{G}_{1}\in\mathbb{R}^{R_{1}\times N_{w}\times R_{1}}$, $\mathcal{G}_{2}\in\mathbb{R}^{R_{1}\times N_{w}\times R_{1}}$, $\mathcal{G}_{3}\in\mathbb{R}^{R_{1}\times R\times R_{1}}$ , $\mathcal{G}_{4}\in\mathbb{R}^{R_{2}\times R\times R_{2}}$, $\mathcal{G}_{5}\in\mathbb{R}^{R_{2}\times N_{t}\times R_{2}}$ and $\mathcal{G}_{6}\in\mathbb{R}^{R_{2}\times N_{p}\times R_{2}}$ are called BTR factors. Then we call (3) a BTR decomposition of and simply denote it by ${\mathcal{X}}=\mathrm{BTR}(\{{\mathcal{G}}_{k}\}_{k=1}^{6})$.

In our previous study \cite{ref20}, we first proposed the "bilateral tensor ring" structure, which not only reduces computational and storage costs but also preserves global dependencies across dimensions while maintaining local correlations according to the strength of correlations between dimensions. Considering that infrared small targets exhibit spatial sparsity (typically occupying less than 0.1\% of pixels) and temporal continuity (maintaining motion trajectories across consecutive frames), and based on previous correlation analysis experiments, it can be observed that the spatial dimension combinations and the time-patch dimension combinations of infrared small targets exhibit weak and strong correlations, respectively.In Remark 1, we explain the fundamental mechanism of BTR decomposition.\\
\textit{\textbf{Remark 1} (Correlation Representation):}
Given a fourth-order infrared small target tensor $\mathcal{X}\in\mathbb{R}^{N_{w}\times N_{w}\times N_{t}\times N_{p}}$, the tensor is divided into two parts, the correlation strength on both sides controlled by $R_1$ and $R_2$, respectively: the TR decomposition on the left effectively captures the spatial continuity of the background, while the TR decomposition on the right simultaneously models the separation characteristics of the target and background in both the temporal domain and the image block dimension. The interactive rank $R$ serves as a key bridge, enabling the adaptive separation of sparse target features from redundant background information.(please see Fig. 2)

In summary, BTR decomposition not only distinguishes between the weak spatial correlation and the strong spectraltemporal correlations but also concurrently captures the interdependencies between these two components.Leveraging this advantage, we adopt the BTR framework to construct an infrared small target detection model in our subsequent work.

As the rank function constitutes a non-convex objective function, the $l_0$ norm is employed as a convex alternative to the  $l_1$ norm.Based on (3), the BTR-ISTD model is constructed by introducing BTR as follows:
\begin{equation}
\label{deqn_ex1a}
\begin{split}
&\min_{\mathcal{G}_k, \mathcal{B}^{4D}, \mathcal{T}^{4D}} \frac{\alpha}{2} \left\| \mathcal{B}^{4D} - \mathrm{BTR}(\{\mathcal{G}_k\}_{k=1}^{6}) \right\|_F^2\\
&\quad\quad\quad\quad + \lambda_1 \left\| \mathcal{T}^{4D} \right\|_1 \\
&\mathrm{s.t.} \mathcal{D}^{4D} = \mathcal{B}^{4D} + \mathcal{T}^{4D},
\end{split}
\end{equation}

where $\alpha > 0 $ is a parameter.
\subsection{Model Optimization}
Based on definition1, two auxiliary factors are introduced:$\mathcal{A}\in\mathbb{R}^{N_{w}\times N_{w}\times R}$ and $\mathcal{B}\in\mathbb{R}^{R\times N_{t}\times N_{p}}$.
\begin{equation}
\label{deqn_ex1a}
\begin{split}
&\min_{\mathcal{A}, \mathcal{B}, \mathcal{G}_k, \mathcal{B}^{4D}, \mathcal{T}^{4D}} \frac{\alpha}{2} \left\| \mathcal{B}^{4D} - \mathcal{A} \times_3^1 \mathcal{B} \right\|_F^2 + \lambda_1 \left\| \mathcal{T}^{4D} \right\|_1, \\
&\quad \mathrm{s.t.}\mathcal{D}^{4D} = \mathcal{B}^{4D} + \mathcal{T}^{4D},\mathcal{A} = \mathrm{TR}(\{\mathcal{G}_k\}_{k=1}^3), \\
& \ \quad\quad\mathcal{B} = \mathrm{TR}(\{\mathcal{G}_k\}_{k=4}^6).
\end{split}
\end{equation}

By introducing a penalty term, problem (5) can be rewritten as the following unconstrained problem:
\begin{equation}
\label{deqn_ex1a}
\begin{split}
\min_{\mathcal{A}, \mathcal{B}, \mathcal{G}_k, \mathcal{B}^{4D}, \mathcal{T}^{4D}} 
&\frac{\alpha}{2} \left\| \mathcal{B}^{4D} - \mathcal{A} \times_3^1 \mathcal{B} \right\|_F^2 
+ \lambda_1 \left\| \mathcal{T}^{4D} \right\|_1 \\
&+ \frac{\beta_1}{2} \left\| \mathcal{A} - \mathrm{TR}(\{\mathcal{G}_k\}_{k=1}^3) \right\|_F^2 \\
&+ \frac{\beta_2}{2} \left\| \mathcal{B} - \mathrm{TR}(\{\mathcal{G}_k\}_{k=4}^6) \right\|_F^2 \\
&+ \frac{\beta_3}{2} \left\| \mathcal{D}^{4D} - (\mathcal{B}^{4D} + \mathcal{T}^{4D}) \right\|_F^2.
\end{split}
\end{equation}

\begin{figure*}[!t]
\centering

\begin{minipage}{0.155\linewidth}
\centering
\includegraphics[width=1\linewidth]{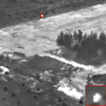}
\label{fig_first_case}
\end{minipage}
\hfill
\begin{minipage}{0.155\linewidth}
\centering
\includegraphics[width=1\linewidth]{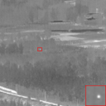}
\label{fig_second_case}
\end{minipage}
\hfill
\begin{minipage}{0.155\linewidth}
\centering
\includegraphics[width=1\linewidth]{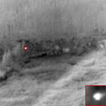}
\label{fig_third_case}
\end{minipage}
\hfill
\begin{minipage}{0.155\linewidth}
\centering
\includegraphics[width=1\linewidth]{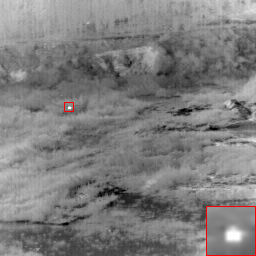}
\label{fig_fourth_case}
\end{minipage}
\hfill
\begin{minipage}{0.155\linewidth}
\centering
\includegraphics[width=1\linewidth]{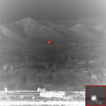}
\label{fig_fifth_case}
\end{minipage}
\hfill
\begin{minipage}{0.155\linewidth}
\centering
\includegraphics[width=1\linewidth]{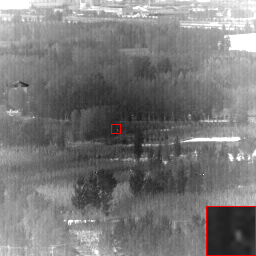}
\label{fig_sixth_case}
\end{minipage}

\begin{minipage}{0.155\linewidth}
\centering
\includegraphics[width=1\linewidth]{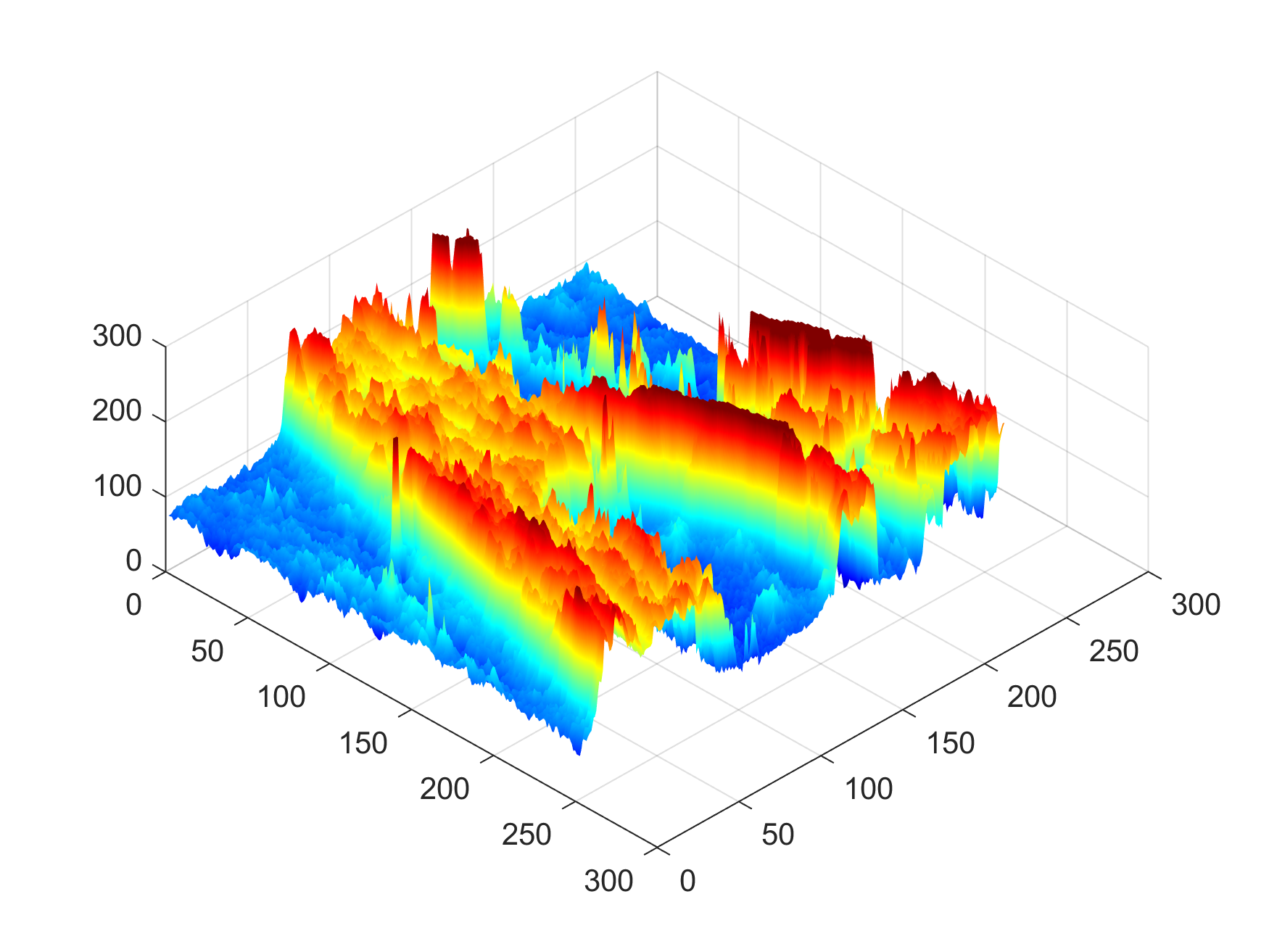}
\label{fig_lrsr_result}
\end{minipage}
\hfill
\begin{minipage}{0.155\linewidth}
\centering
\includegraphics[width=1\linewidth]{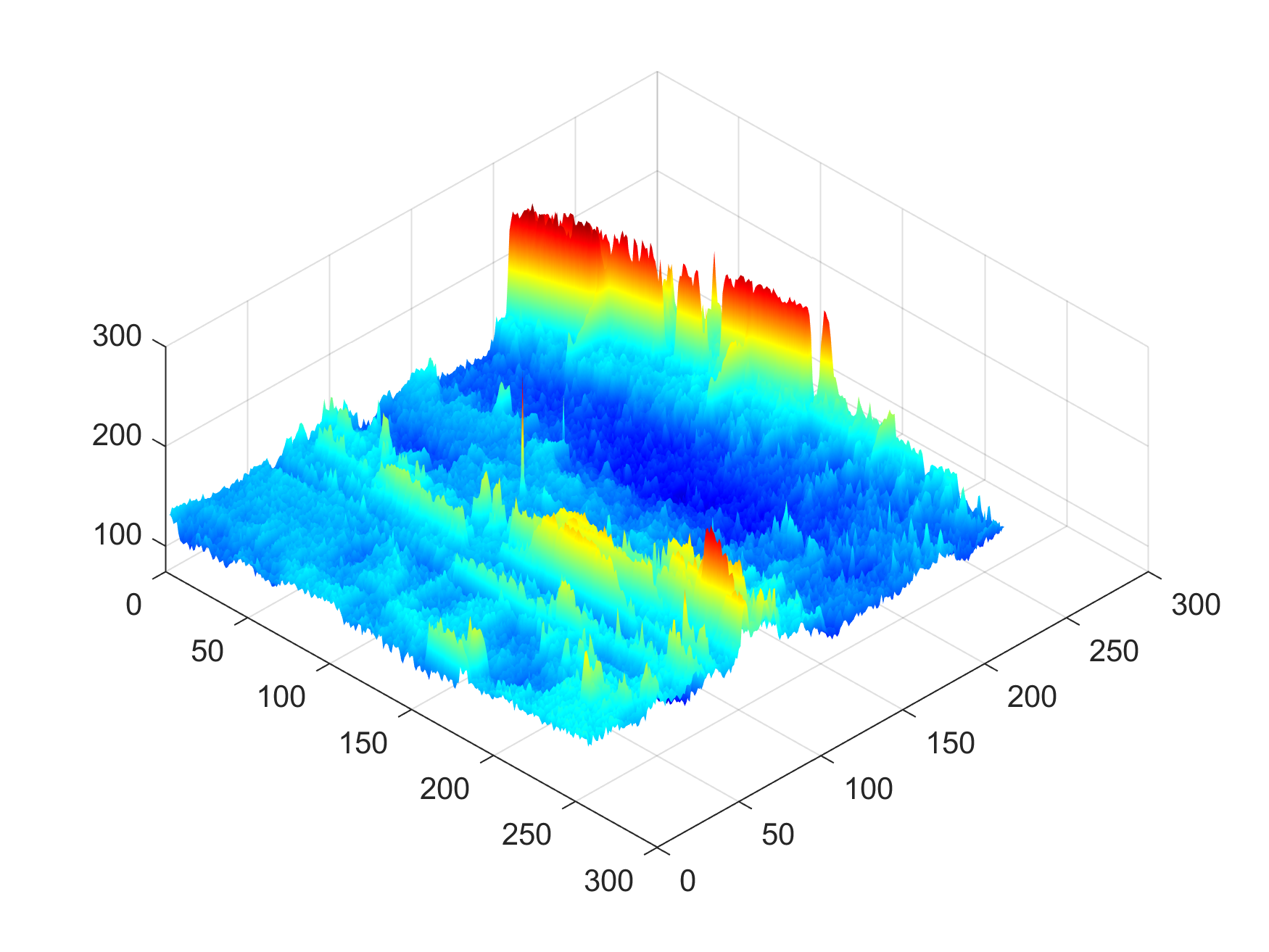}
\label{fig_srws_result}
\end{minipage}
\hfill
\begin{minipage}{0.155\linewidth}
\centering
\includegraphics[width=1\linewidth]{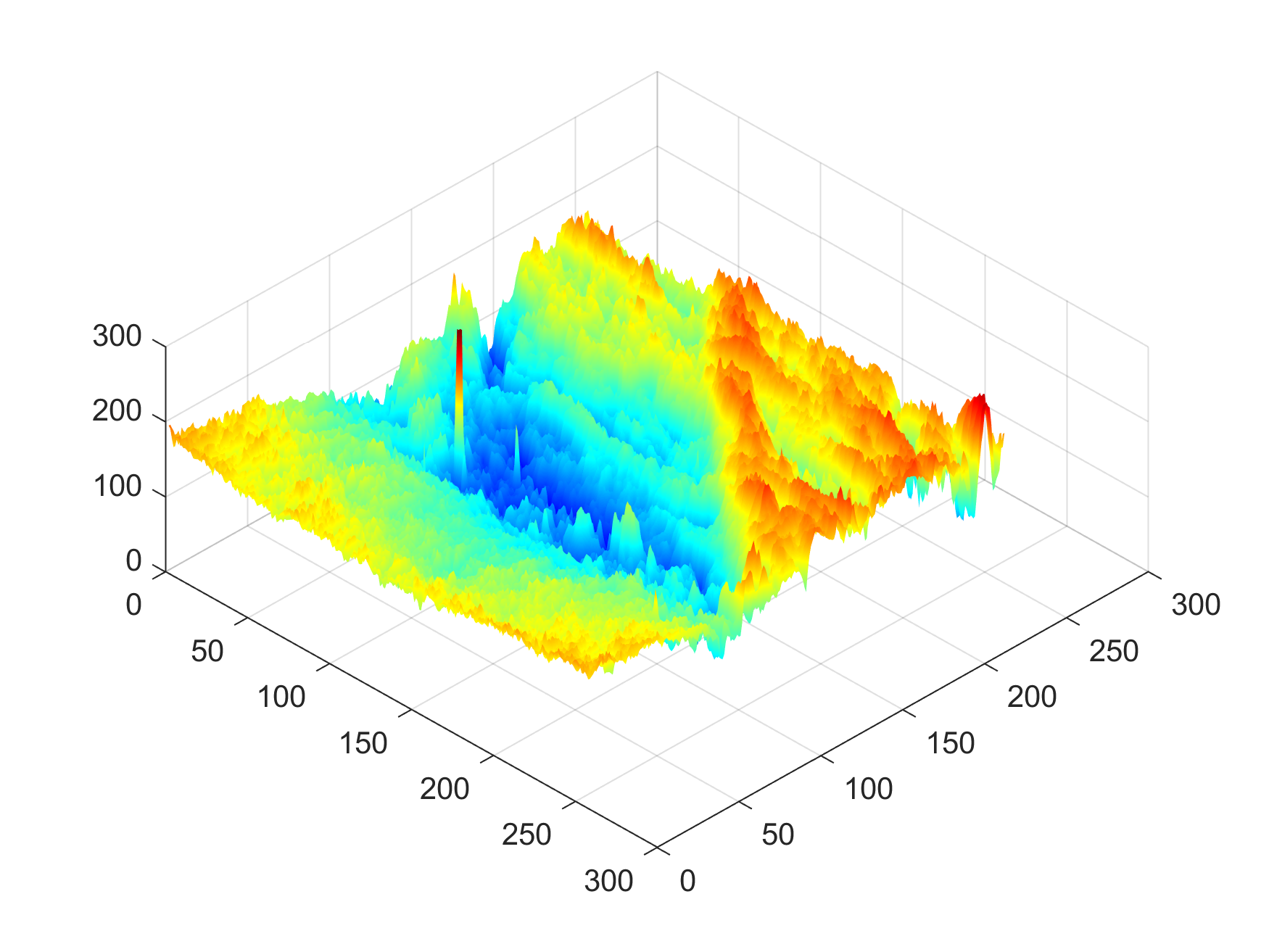}
\label{fig_our_result}
\end{minipage}
\hfill
\begin{minipage}{0.155\linewidth}
\centering
\includegraphics[width=1\linewidth]{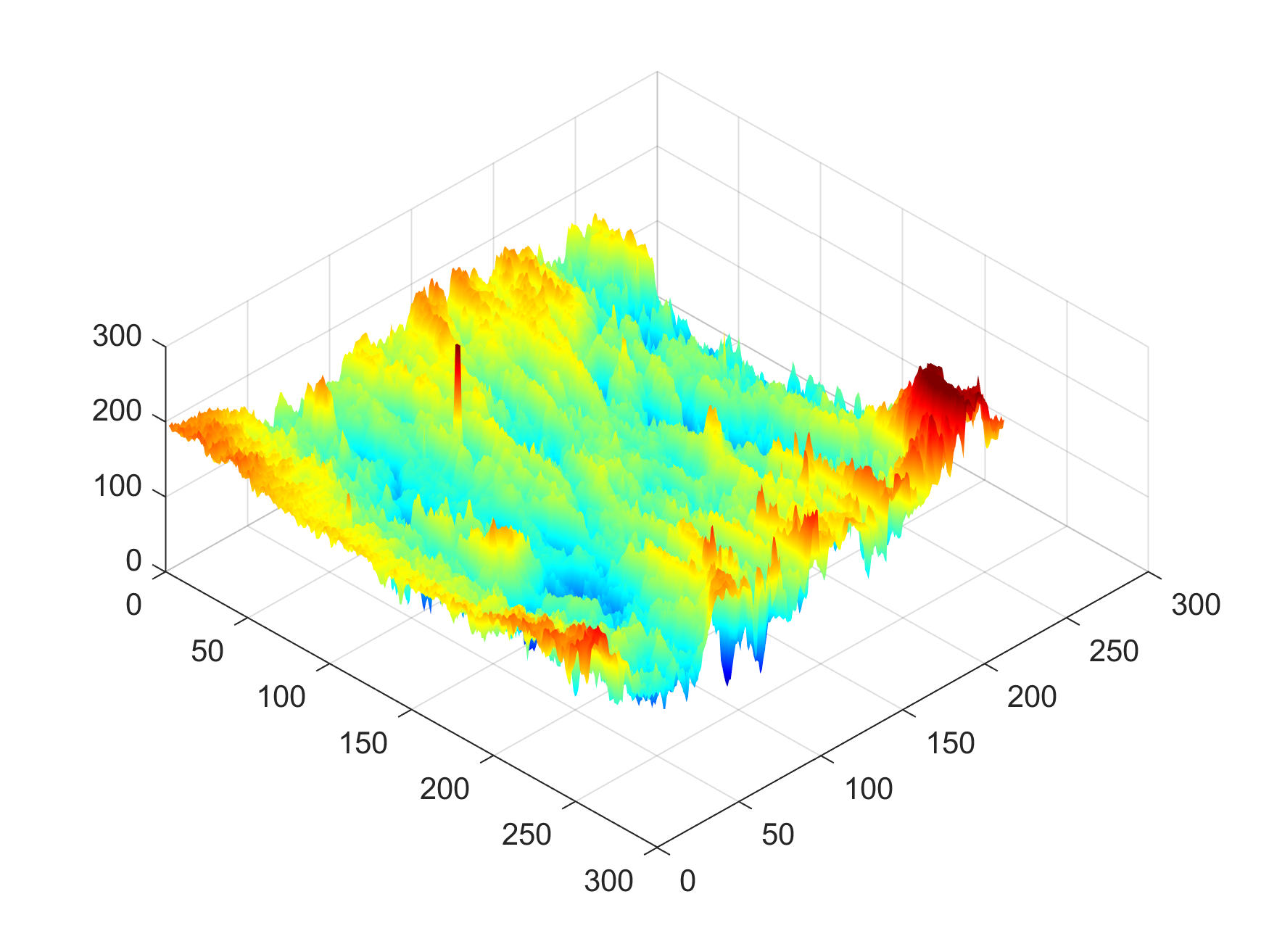} 
\label{fig_lrsr_seq4}
\end{minipage}
\hfill
\begin{minipage}{0.155\linewidth}
\centering
\includegraphics[width=1\linewidth]{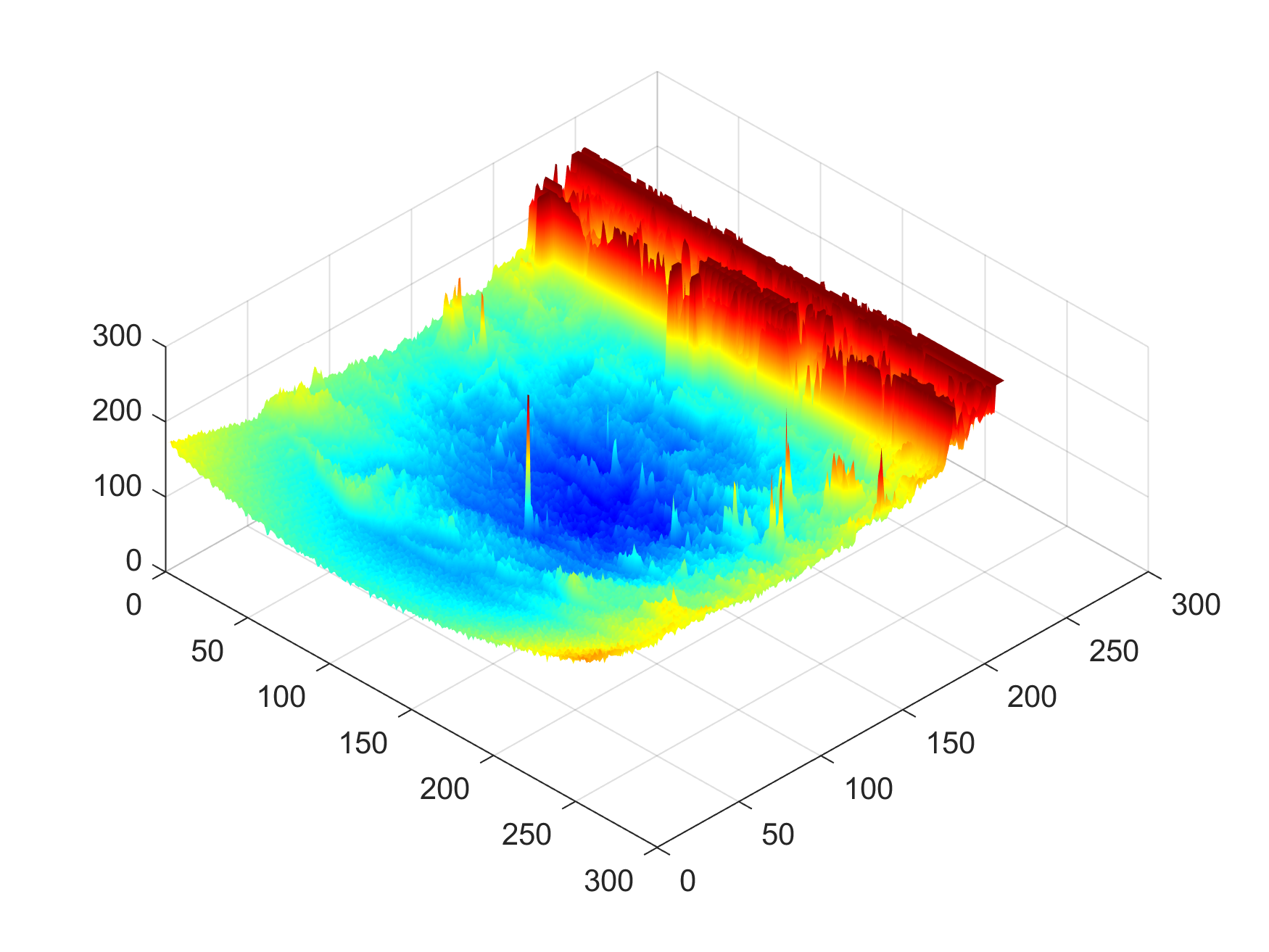} 
\label{fig_srws_seq4}
\end{minipage}
\hfill
\begin{minipage}{0.155\linewidth}
\centering
\includegraphics[width=1\linewidth]{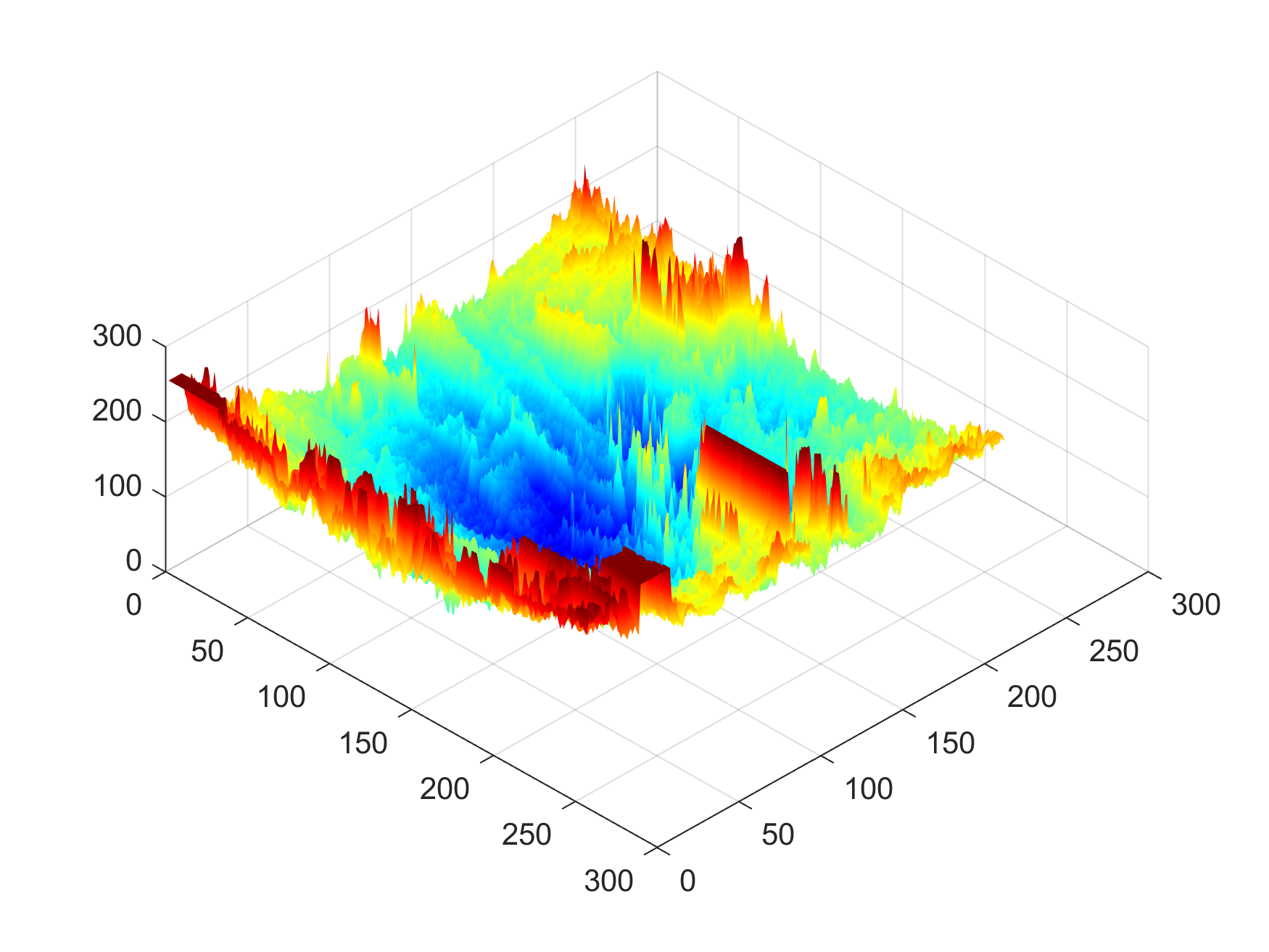} 
\label{fig_our_seq4}
\end{minipage}

\caption{Representative images of sequences 1–6 and their corresponding 3-D surface plot.}
\label{fig_sim}
\end{figure*}
The proximal alternating minimization (PAM) algorithm is employed to solve problem (6), and its solution can be obtained by alternately updating the following equations:
\begin{equation}
\begin{cases}
\mathcal{A}=\underset{\mathcal{A}}{\operatorname*{\mathrm{arg}\operatorname*{min}}}f(\mathcal{A},\mathcal{B},\mathcal{G}_k,\mathcal{B}^{4D},\mathcal{T}^{4D})+\frac{\rho}{2}\|\mathcal{A}-\hat{\mathcal{A}}\|_F^2
\\
\mathcal{B}=\underset{\mathcal{B}}{\operatorname*{\mathrm{arg}\operatorname*{min}}}f(\mathcal{A},\mathcal{B},\mathcal{G}_k,\mathcal{B}^{4D},\mathcal{T}^{4D})+\frac{\rho}{2}\|\mathcal{B}-\hat{\mathcal{B}}\|_{F}^{2}  \\
\mathcal{G}_k = \underset{\mathcal{G}_k} {\operatorname*{\mathrm{arg}\operatorname*{min}}}f(\mathcal{A},\mathcal{B},\mathcal{G}_k,\mathcal{B}^{4D},\mathcal{T}^{4D}) + \frac{\rho}{2} \|\mathcal{G}_k - \hat{\mathcal{G}}_k\|_F^2 \\
\mathcal{B}^{4D}=\underset{\mathcal{B}^{4D}}{\operatorname*{\mathrm{arg}\operatorname*{min}}}f(\mathcal{A},\mathcal{B},\mathcal{G}_k,\mathcal{B}^{4D},\mathcal{T}^{4D}) + \frac{\rho}{2} \|\mathcal{B}^{4D} - \hat{\mathcal{B}}^{4D}\|_F^2 \\
\mathcal{T}^{4D}=\underset{\mathcal{T}^{4D}}{\operatorname*{\mathrm{arg}\operatorname*{min}}}f(\mathcal{A},\mathcal{B},\mathcal{G}_k,\mathcal{B}^{4D},\mathcal{T}^{4D}) + \frac{\rho}{2} \|\mathcal{T}^{4D} - \hat{\mathcal{T}}^{4D}\|_F^2
\end{cases}
\end{equation}
\subsubsection{Update $\mathcal{A}$}
The $\mathcal{A}$-subproblem can be written as
\begin{equation}  
\label{deqn_ex1a}
\begin{aligned}
\underset{\mathcal{A}}{\operatorname*{\mathrm{arg}\operatorname*{min}}}&\frac{\alpha}{2}{\left\|\mathcal{B}^{4D}-\mathcal{A}\times_3^1\mathcal{B}\right\|}_F^2+\frac{\beta_1}{2}{\left\|\mathcal{A}-TR(\left\{\mathcal{G}_k\right\}_{k=1}^3)\right\|}_F^2\\
&+\frac{\rho}{2}{\left\|\mathcal{A}-\hat{\mathcal{A}}\right\|}_F^2
\end{aligned}
\end{equation}

According to Theorem 1 , (8) can be rewritten in the following matrix form:
\begin{equation}
\underset {\mathbf{A}}{\operatorname*{\mathrm{arg}\operatorname*{min}}}\frac{\alpha}{2}\|\mathbf{X-AB}\|_F^2+\frac{\beta_1}{2}\|\mathbf{A-C}\|_F^2+\frac{\rho}{2}\|\mathbf{A-\hat{A}}\|_F^2
\end{equation}
Where $\mathbf{X}=\mathrm{Unfold}(\mathcal{X},[1,2\mid3,4])\in\mathbb{R}^{HW\times BT}$, $\mathbf{A}=\mathrm{Unfold}(\mathcal{A},[1,2\mid3])\in\mathbb{R}^{HW\times R}$, $\mathbf{B}=\mathrm{Unfold}(\mathcal{B},[1\mid2,3])\in\mathbb{R}^{R\times BT}$ , and $\mathbf{C}=\mathrm{Unfold}(\mathrm{TR}(\{\mathcal{G}_k\}_{k=1}^3),[1,2\mid3])\in\mathbb{R}^{HW\times R}$. 

Since the objective function of (9) is differentiable, we can obtain the solution as
\begin{equation}
    \mathbf{A}=(\alpha\mathbf{XB}^\mathrm{T}+\beta_1\mathbf{C}+\rho\hat{\mathbf{A}})(\alpha\mathbf{BB}^\mathrm{T}+(\beta_1+\rho)\mathbf{I})^{-1}.
\end{equation}
\subsubsection{Update $\mathcal{B}$}
The $\mathcal{B}$-subproblem can be written as
\begin{equation}  
\label{deqn_ex1a}
\begin{aligned}
\underset{\mathcal{B}}{\operatorname*{\mathrm{arg}\operatorname*{min}}}&\frac{\alpha}{2}{\left\|\mathcal{B}^{4D}-\mathcal{A}\times_3^1\mathcal{B}\right\|}_F^2+\frac{\beta_2}{2}{\left\|\mathcal{B}-TR(\left\{\mathcal{G}_k\right\}_{k=4}^6)\right\|}_F^2\\
&+\frac{\rho}{2}{\left\|\mathcal{B}-\hat{\mathcal{B}}\right\|}_F^2.
\end{aligned}
\end{equation}

According to Theorem 1 , (11) can be rewritten in the following matrix form:
\begin{equation}
\underset {\mathbf{B}}{\operatorname*{\mathrm{arg}\operatorname*{min}}}\frac{\alpha}{2}\|\mathbf{X-AB}\|_F^2+\frac{\beta_2}{2}\|\mathbf{B-D}\|_F^2+\frac{\rho}{2}\|\mathbf{B-\hat{B}}\|_F^2,
\end{equation}
Where $\mathbf{D}=\mathrm{Unfold}(\mathrm{TR}(\{\mathcal{G}_k\}_{k=4}^6),[1\mid2,3])\in\mathbb{R}^{R\times BT}$, the objective function of (12) is also differentiable, and thus we can obtain the solution as
\begin{equation}
   \mathbf{B}=(\alpha\mathbf{A}^\mathrm{T}\mathbf{A}+(\beta_2+\rho)\mathbf{I})^{-1}(\alpha\mathbf{A}^\mathrm{T}\mathbf{X}+\beta_2\mathbf{D}+\rho\mathbf{\hat{B}}).
\end{equation}
\subsubsection{Update $\mathcal{G}_k$}
For $k=1,2,3$ ,the $\mathcal{G}_k$-subproblem can be written as
\begin{equation}
\underset{\mathcal{G}_k}{\operatorname*{\mathrm{arg}\operatorname*{min}}}\frac{\beta_1}{2}{\left\|\mathcal{A}-TR(\left\{\mathcal{G}_k\right\}_{k=1}^3)\right\|}_F^2+\frac{\rho}{2}{\left\|\mathcal{G}_k-\hat{\mathcal{G}}_k\right\|}_F^2.
\end{equation}

According to Theorem 1, (14) can be rewritten in the following matrix form:
\begin{equation}
\underset{\mathbf{G}_k}{\operatorname*{\operatorname*{argmin}}}\frac{\beta_1}{2}\|\mathbf{A}_k-\mathbf{G}_k\mathbf{M}_k\|_F^2+\frac{\rho}{2}\|\mathbf{G}_k-\mathbf{\hat{G}}_k\|_F^2.
\end{equation}

The objective function of (15) is also differentiable, and thus we can obtain the solution as
\begin{equation}
    \mathbf{G}_k=(\beta_1\mathbf{A}_k\mathbf{M}_k^\mathrm{T}+\rho\mathbf{\hat{G}}_k)(\beta_1\mathbf{M}_k\mathbf{M}_k^\mathrm{T}+\rho\mathbf{I})^{-1}.
\end{equation}

For $k=4,5,6$, the $\mathcal{G}_k$-subproblem can be written as
\begin{equation}
\underset{\mathcal{G}_k}{\operatorname*{\mathrm{arg}\operatorname*{min}}}\frac{\beta_2}{2}{\left\|\mathcal{B}-TR(\left\{\mathcal{G}_k\right\}_{k=4}^6)\right\|}_F^2+\frac{\rho}{2}{\left\|\mathcal{G}_k-\hat{\mathcal{G}}_k\right\|}_F^2.
\end{equation}

According to Theorem 1 , (17) can be rewritten in the following matrix form:
\begin{equation}
\underset{\mathbf{G}_k}{\operatorname*{\operatorname*{argmin}}}\frac{\beta_2}{2}\|\mathbf{B}_k-\mathbf{G}_k\mathbf{M}_k\|_F^2+\frac{\rho}{2}\|\mathbf{G}_k-\mathbf{\hat{G}}_k\|_F^2.
\end{equation}

The objective function of (18) is also differentiable, and thus we can obtain the solution as
\begin{equation}
    \mathbf{G}_k=(\beta_2\mathbf{B}_k\mathbf{M}_k^\mathrm{T}+\rho\mathbf{\hat{G}}_k)(\beta_2\mathbf{M}_k\mathbf{M}_k^\mathrm{T}+\rho\mathbf{I})^{-1}.
\end{equation}
\subsubsection{Update $\mathcal{B}^{4D}$}
The $\mathcal{B}^{4D}$-subproblem can be written as
\begin{equation}
\label{deqn_ex1a}
\begin{aligned}
\underset{\mathcal{B}^{4D}}{\operatorname*{\operatorname*{\arg\min}}}&\frac{\alpha}{2}{\left\|\mathcal{B}^{4D}-\mathcal{A}\times_{3}^{1}\mathcal{B}\right\|}_{F}^{2}+\frac{\beta_{3}}{2}{\left\|\mathcal{D}^{4D}-(\mathcal{B}^{4D}+\mathcal{T}^{4D})\right\|}_{F}^{2}\\
&+\frac{\rho}{2}{\left\|\mathcal{B}^{4D}-\hat{\mathcal{B}}^{4D}\right\|}_{F}^{2}.
\end{aligned} 
\end{equation}

Problem (20) has the exact solution by taking the derivative of $\mathcal{B}^{4D}$:
\begin{equation}
\label{deqn_ex1a}
\begin{aligned}
\hat{\mathcal{B}}^{4D}=\frac{\alpha(\mathcal{A}\times_3^1\mathcal{B})+\beta_3(\mathcal{D}^{4D}-\mathcal{T}^{4D})+\rho\mathcal{B}^{4D}}{\alpha+\beta_3+\rho}
\end{aligned} 
\end{equation}
\subsubsection{Update $\mathcal{T}^{4D}$}
The $\mathcal{T}^{4D}$-subproblem can be written as
\begin{equation}
\begin{aligned}
\underset{\mathcal{T}^{4D}}{\operatorname*{\operatorname*{\arg\min}}}&\lambda_1\left\|T^{4D}\right\|_1+\frac{\beta_3}{2}\left\|\mathcal{D}^{4D}-(\mathcal{B}^{4D}+\mathcal{T}^{4D})\right\|_F^2\\
&+\frac{\rho}{2}\left\|\mathcal{T}^{4D}-\hat{\mathcal{T}}^{4D}\right\|_F^2,
\end{aligned}
\end{equation}
which has the following solution:
\begin{equation}
    \hat{\mathcal{T}}^{4D}=\mathrm{shrink}_1\left(\mathcal{T}_*^{4D},\frac{\lambda_1}{\beta_3+\rho}\right),
\end{equation}
where
\begin{equation}
T_*^{4D}=\frac{\beta_3(\mathcal{D}^{4D}-\mathcal{B}^{4D})+\rho\hat{\mathcal{T}}^{4D}}{\beta_3+\rho},
\end{equation}
and
\begin{equation}
    [\mathrm{shrink}_1(\mathcal{X},\xi)]_{i,j,m}=\mathrm{sign}(x_{i,j,m})\max(\mid x_{i,j,m}\mid-\xi,0).
\end{equation}

Finally, the complete solution process of the proposed model is described in Algorithm 1.
\begin{algorithm}[H]
\caption{PAM-Based Algorithm for Proposed Model}\label{alg:pam_algorithm}
\begin{algorithmic}[1]
\REQUIRE 4D tensor $\mathcal{D}^{4D}$, the ranks $R$, $R_1$ and $R_2$, and parameters 
$\alpha = 1$, $\lambda_1 = 0.1$, $\beta_1 = 1$, $\beta_2 = 1$, $\beta_3 = 2$ and $\rho = 0.01$

\ENSURE: Let $\mathcal{B}^{4D} = \mathcal{D}^{4D}$, 
$\mathcal{T}^{4D} = 0$, and initialize $\mathcal{A}$, $\mathcal{B}$, and 
$\mathcal{G}_k$ for $k = 1, 2, \cdots, 6$ with all elements equal to 1.

\FOR{$i = 1$ to $20$}
\STATE Update A via (10) and
 reshape it to 
$\mathcal{A} \in \mathbb{R}^{N_{w}\times N_{w}\times R}$

\STATE Update B via (13) and reshape it to 
$\mathcal{B} \in \mathbb{R}^{R\times N_{t}\times N_{p}}$

\STATE Update $\mathcal{G}_1$, $\mathcal{G}_2$, and $\mathcal{G}_3$ via (16), and reshape them to 
$\mathcal{G}_1 \in \mathbb{R}^{R_1 \times N_{w} \times R_1}$, 
$\mathcal{G}_2 \in \mathbb{R}^{R_2 \times N_{w} \times R_2}$, and 
$\mathcal{G}_3 \in \mathbb{R}^{R_3 \times R \times R_3}$

\STATE Update $\mathcal{G}_4$, $\mathcal{G}_5$, and $\mathcal{G}_6$ via (19), and reshape them to 
$\mathcal{G}_4 \in \mathbb{R}^{R_2 \times R \times R_2}$, 
$\mathcal{G}_5 \in \mathbb{R}^{R_2 \times N_{t} \times R_2}$, and 
$\mathcal{G}_6 \in \mathbb{R}^{R_2 \times N_{p} \times R_2}$

\STATE Update $\mathcal{B}^{4D}$ via (21)

\STATE Update $\mathcal{T}^{4D}$ via (23)

\IF{convergence condition satisfied}
\STATE Break the loop
\ENDIF
\ENDFOR

\RETURN $\mathcal{B}^{4D}$, $\mathcal{T}^{4D}$
\end{algorithmic}
\end{algorithm}

\section{Experiment and Results}
In this part, to evaluate the effectiveness of the proposed BTR-ISTD model, six real infrared small target sequences were employed, and its performance was compared with six representative detection methods. All experiments were conducted using MATLAB R2023b.

\subsection{Experimental Setting}
To evaluate the performance of the proposed method in complex real-world scenarios, we performed experiments on six representative infrared sequence sets and compared the results with six existing detection methods. These sequences were obtained from the publicly available IASTD dataset \cite{ref21}, which covers diverse backgrounds such as sky, ground, mountains, forests, and buildings. Targets exhibited varying sizes, shapes, and motion characteristics. The detailed information of each sequence set is summarized in Table I, and representative frames are illustrated in Fig. 2.

\begin{table*}[!t]
\caption{Detailed Characteristics of all Sequences}
\label{tab:sequences}
\centering
\resizebox{\textwidth}{!}{
\begin{tabu} to \linewidth {X[0.5,c] X[0.8,c] X[1,c] X[2,c] X[2,c]}
\toprule
Seq & Frames & Image Size & Target Information & Background Information \\
\midrule
1 & 100 & 256×256 & small, slow-moving & highlighted interference \\
2 & 100 & 256×256 & low nonlocal contrast & forest, noise, artificial structures \\
3 & 100 & 256×256 & move fast & forest, target-like clutter \\
4 & 100 & 256×256 & small, dim, regular & thick bushes, noise, heavy clutters \\
5 & 100 & 256×256 & tiny, slow-moving & artificial structures, bright road \\
6 & 100 & 256×256 & tiny,slow-moving & artificial structures, target-like clutter \\
\bottomrule
\end{tabu}
}
\end{table*}
To effectively evaluate the performance of the proposed BTR-ISTD model, the three-dimensional receiver operating characteristic (3-D ROC) curve is adopted as the evaluation metric. In this metric, the threshold is treated as the independent variable , while $\mathcal{\mathit{P}}_D$ and $\mathcal{\mathit{P}}_F$  serve as the dependent variables. $\mathcal{\mathit{P}}_D$ and $\mathcal{\mathit{P}}_F$  are defined as follows:
\begin{equation}
\label{deqn_ex1a}
\begin{split}
\mathit{P}_D = \frac{\text{number of detected targets}}{\text{number of actual targets}}
\end{split}
\end{equation}
\begin{equation}
\label{deqn_ex1a}
\begin{split}
\mathit{P}_F = \frac{\text{number of false pixels detected}}{\text{total pixels in whole image}}
\end{split}
\end{equation}

Simultaneously, projecting the 3D ROC curve onto different planes yields three two-dimensional curves: the ($\mathcal{\mathit{P}}_D$ and $\mathcal{\mathit{P}}_F$) ROC curve, the ($\mathcal{\mathit{P}}_D$ and $\tau$) ROC curve, and the ($\mathcal{\mathit{P}}_F$ and $\tau$) ROC curve, where $\tau$ denotes the background suppression rate. These curves provide three additional quantitative metrics: the Area Under the Curve of Signal-to-Noise Probability Ratio ($\mathrm{AUC}_{\mathrm{SNPR}}$), the Area Under the Curve of Target Detection–Background Suppression ($\mathrm{AUC}_{\mathrm{TD}\text{-}\mathrm{BS}}$), and the Area Under the Curve of Overall Detection Probability ($\mathrm{AUC}_{\mathrm{ODP}}$). Together, they offer a more comprehensive assessment of detection performance.
\begin{equation}
\label{deqn_ex1a}
\begin{split}
\mathrm{AUC}_{\mathrm{SNPR}}=\frac{\mathrm{AUC}_{(D,\tau)}}{\mathrm{AUC}_{(P,\tau)}}\in\left[0,+\infty\right]
\end{split}
\end{equation}
\begin{equation}
\label{deqn_ex1a}
\begin{split}
\mathrm{AUC}_{\mathrm{TD}\text{-}\mathrm{BS}}=\mathrm{AUC}_{(D,\tau)}-\mathrm{AUC}_{(F,\tau)}\in[-1,1]
\end{split}
\end{equation}
\begin{equation}
\label{deqn_ex1a}
\begin{split}
\mathrm{AUC}_{\mathrm{ODP}}=\mathrm{AUC}_{(D,\tau)}+\left(1\mathrm{-AUC}_{(P,\tau)}\right)\in[0,2].
\end{split}
\end{equation}

The baseline methods selected for comparison consist of three single-frame approaches, namely IPI, PSTNN, and RIPT, and three multi-frame approaches, namely ECA-STT, STWRT, and 4D-TR.

\begin{figure*}[!h]
\centering
\setlength{\tabcolsep}{2pt}

\begin{tabular}{@{}*{8}{c}@{}}
% 第1行
\includegraphics[width=0.115\linewidth]{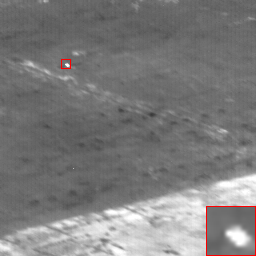} &
\includegraphics[width=0.115\linewidth]{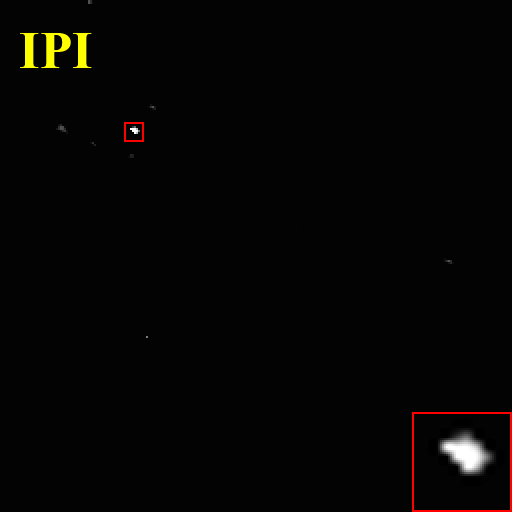} &
\includegraphics[width=0.115\linewidth]{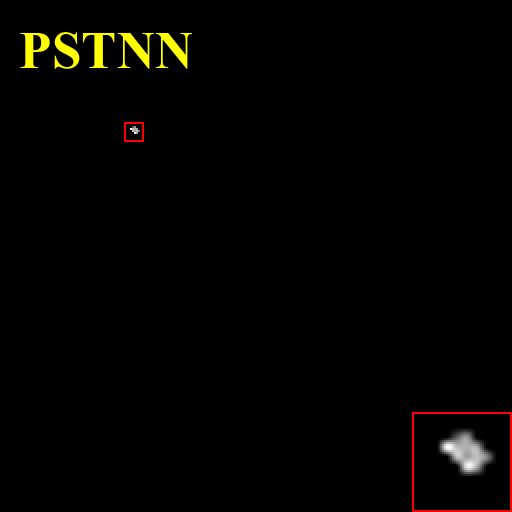} &
\includegraphics[width=0.115\linewidth]{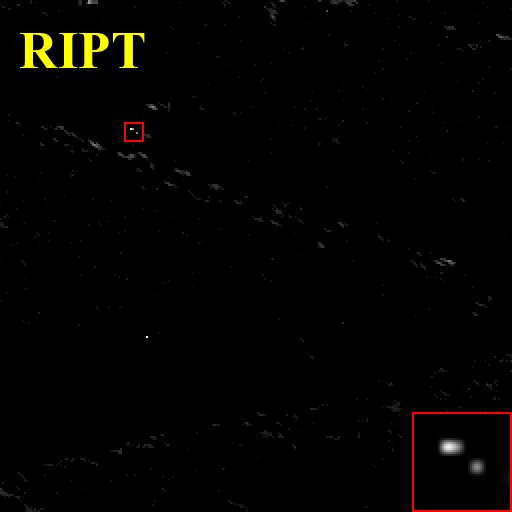} &
\includegraphics[width=0.115\linewidth]{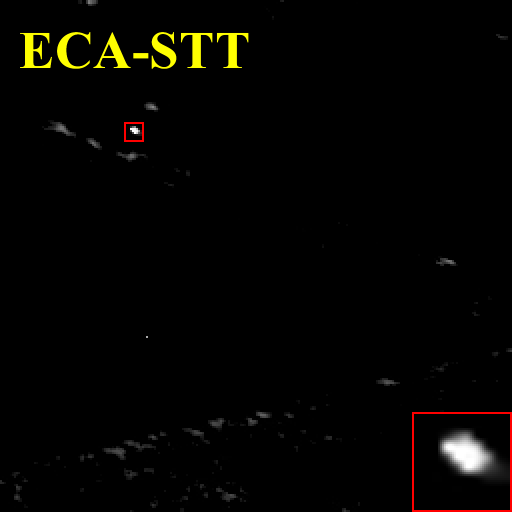} &
\includegraphics[width=0.115\linewidth]{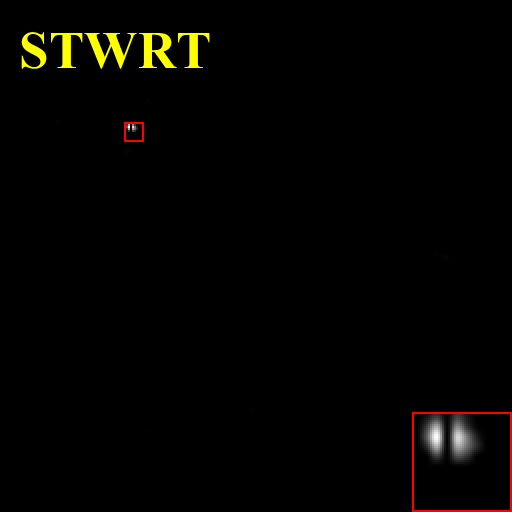} &
\includegraphics[width=0.115\linewidth]{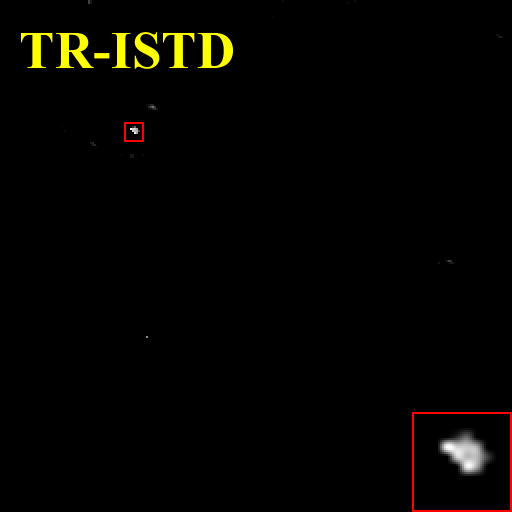} &
\includegraphics[width=0.115\linewidth]{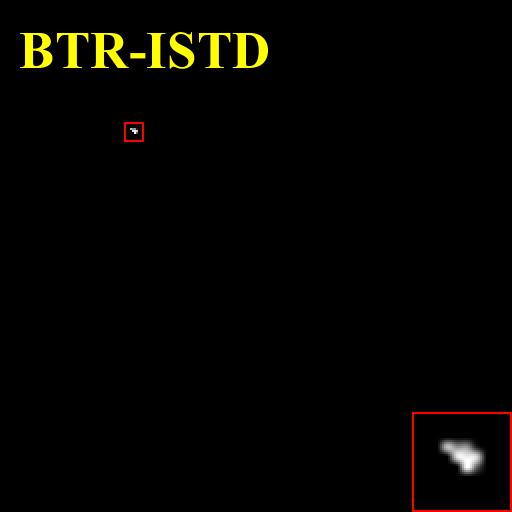} \\[-4pt]

\includegraphics[width=0.115\linewidth]{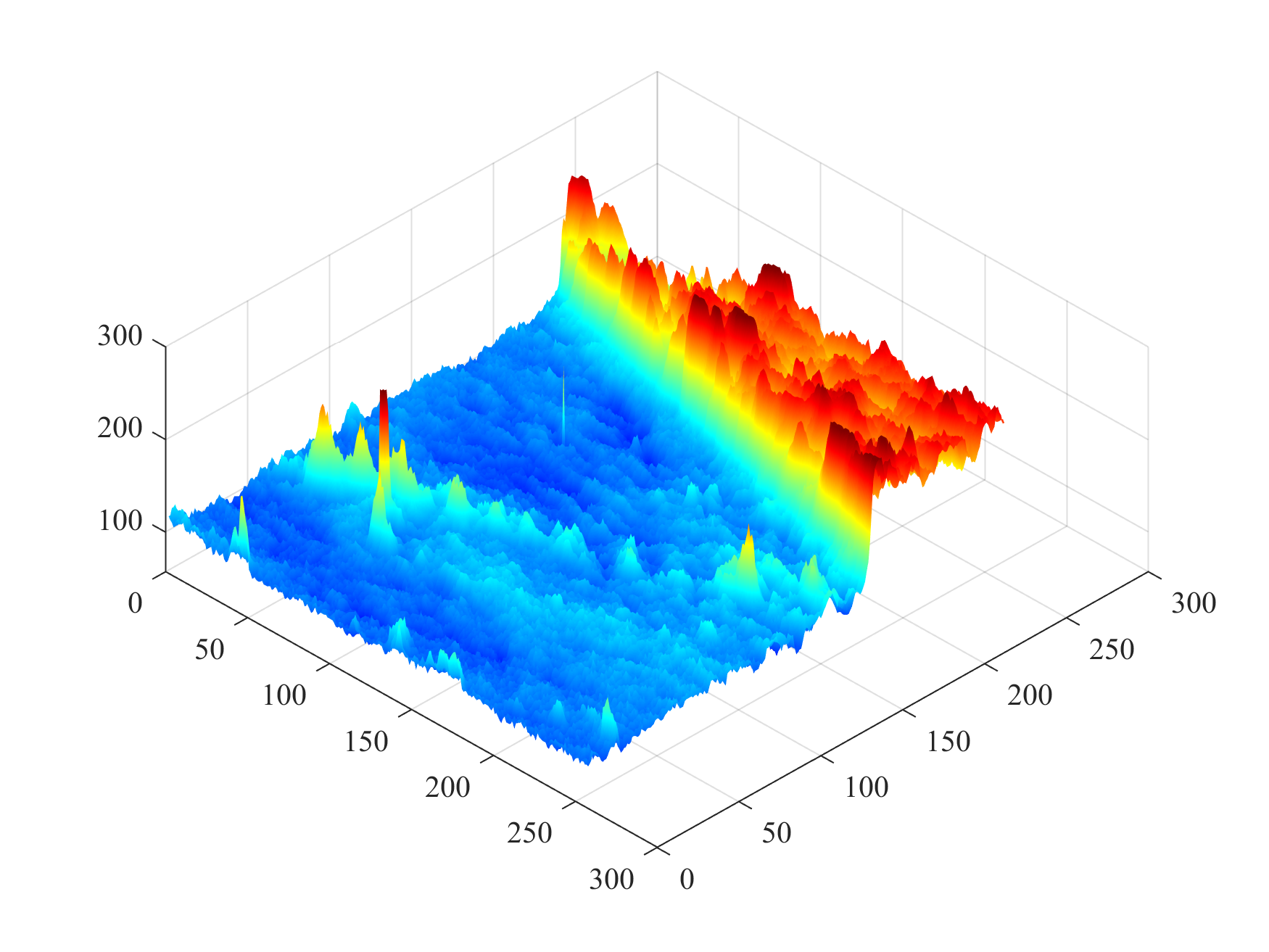} &
\includegraphics[width=0.115\linewidth]{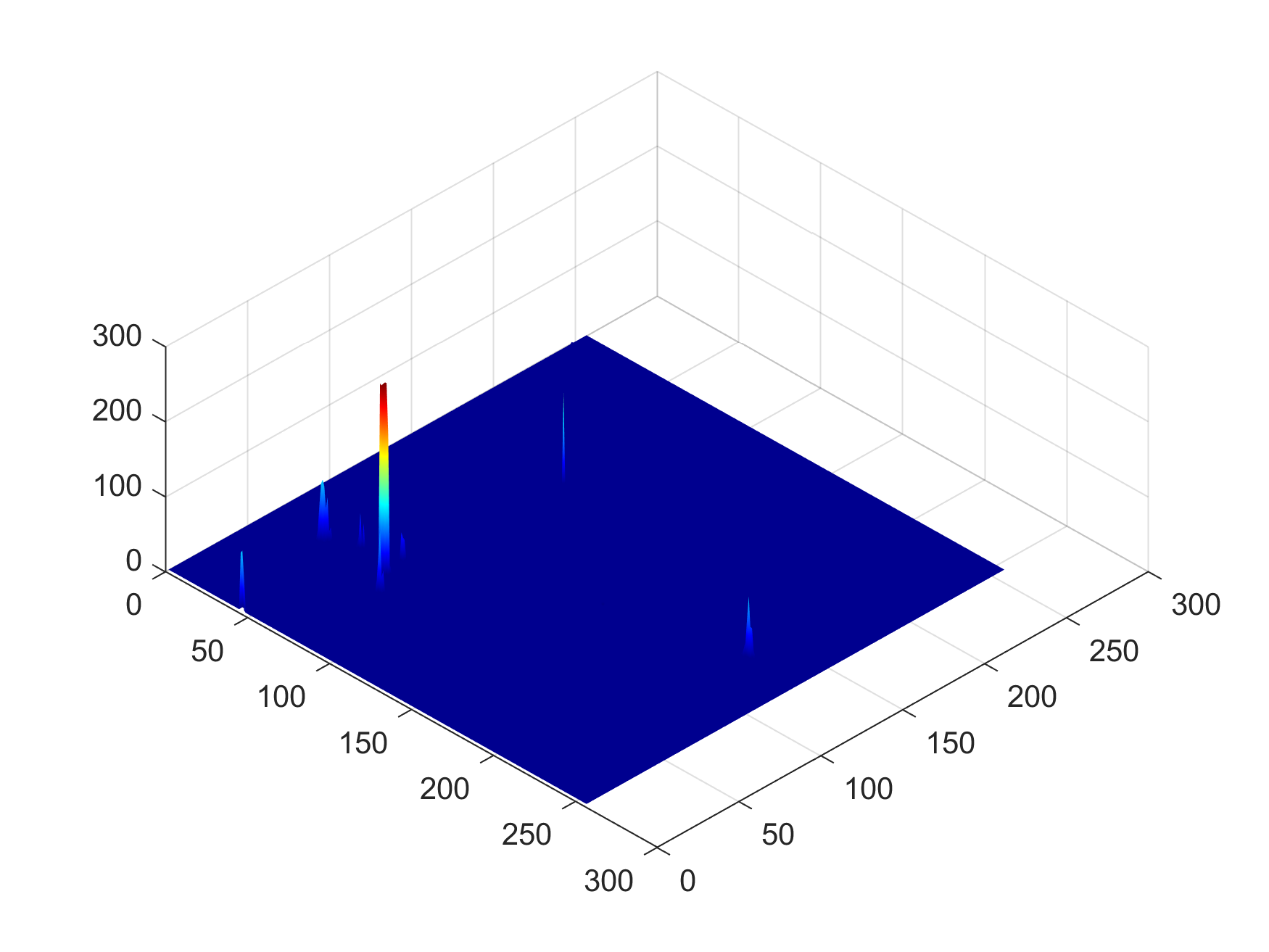} &
\includegraphics[width=0.115\linewidth]{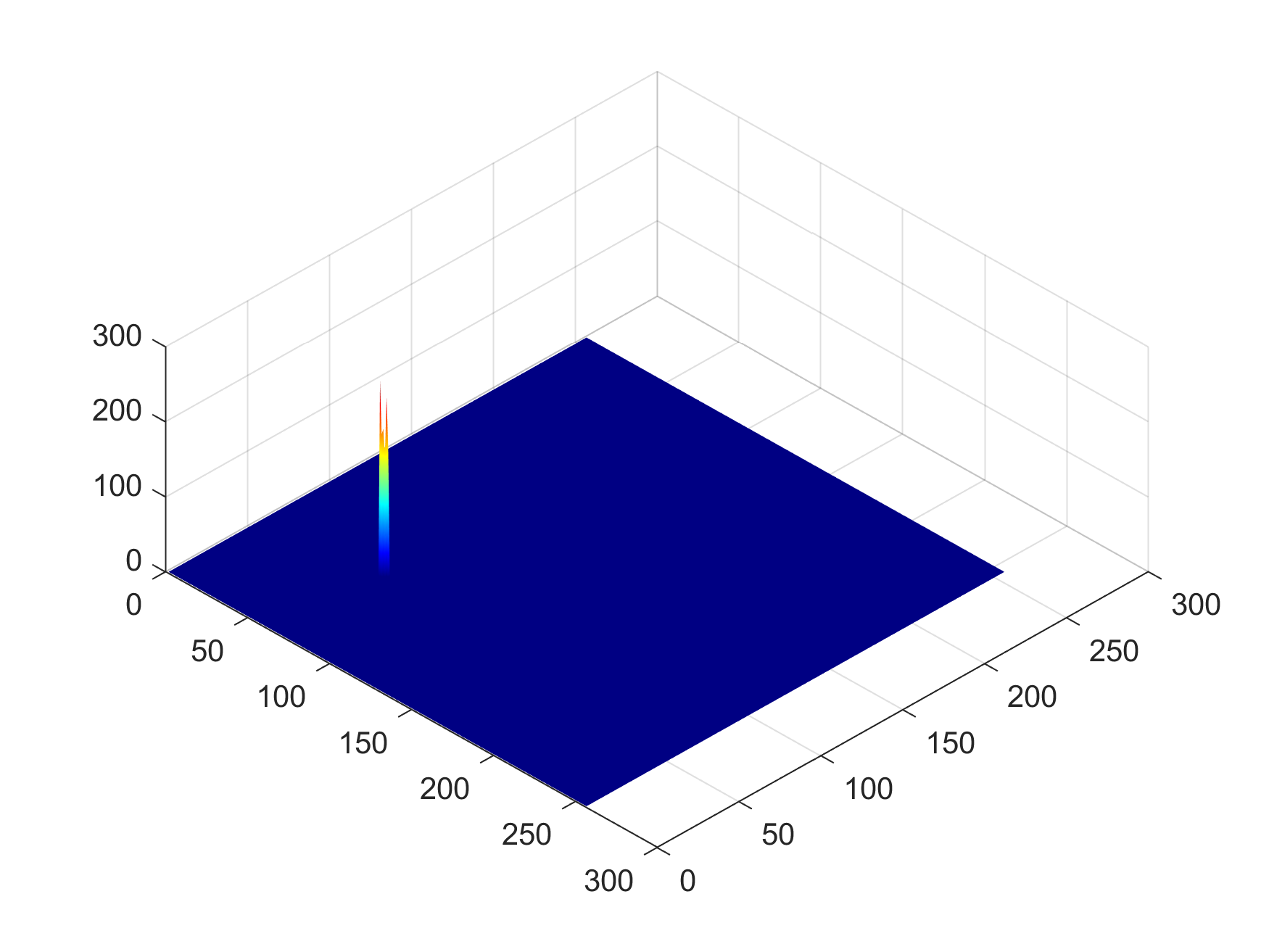} &
\includegraphics[width=0.115\linewidth]{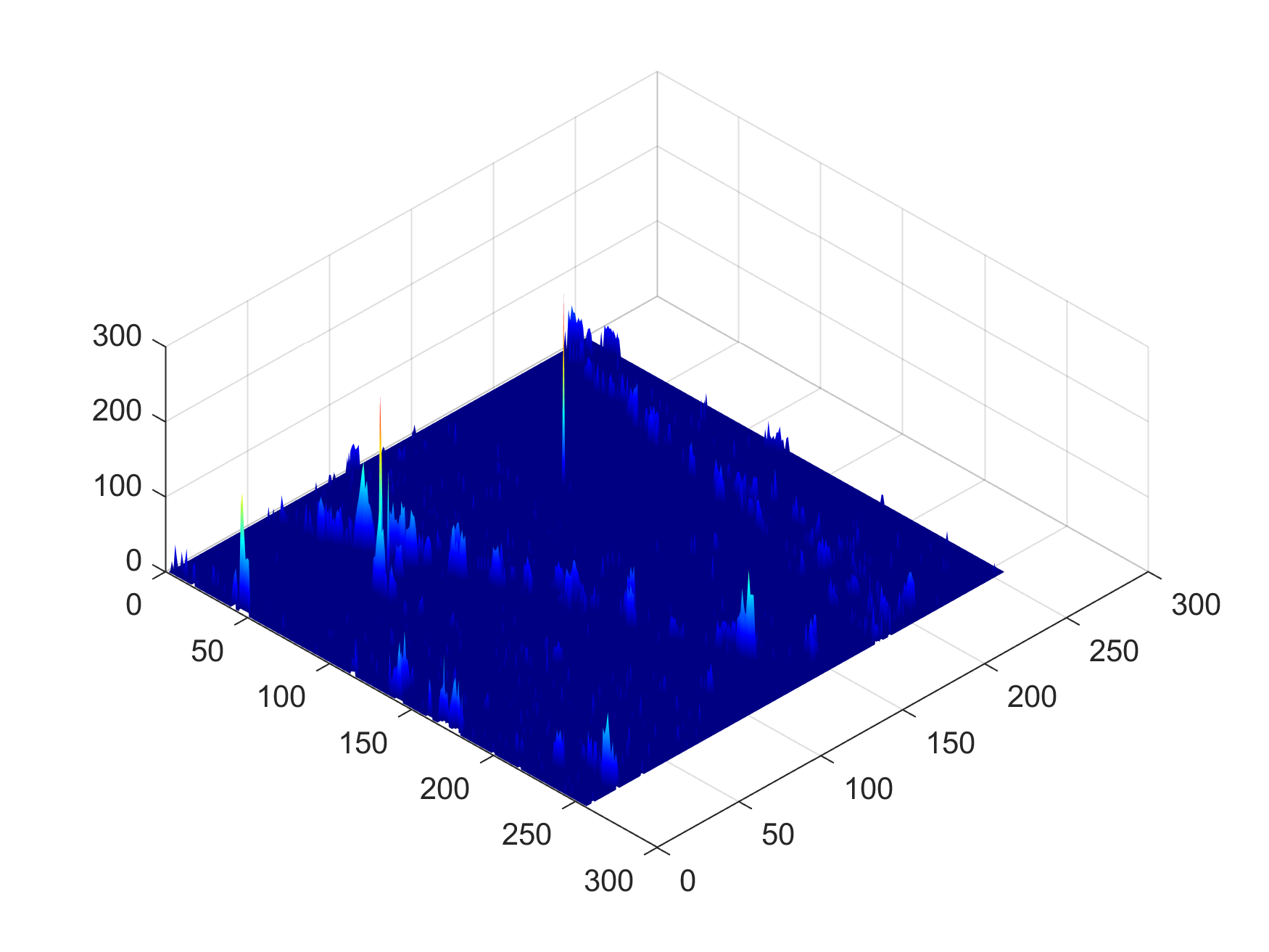} &
\includegraphics[width=0.115\linewidth]{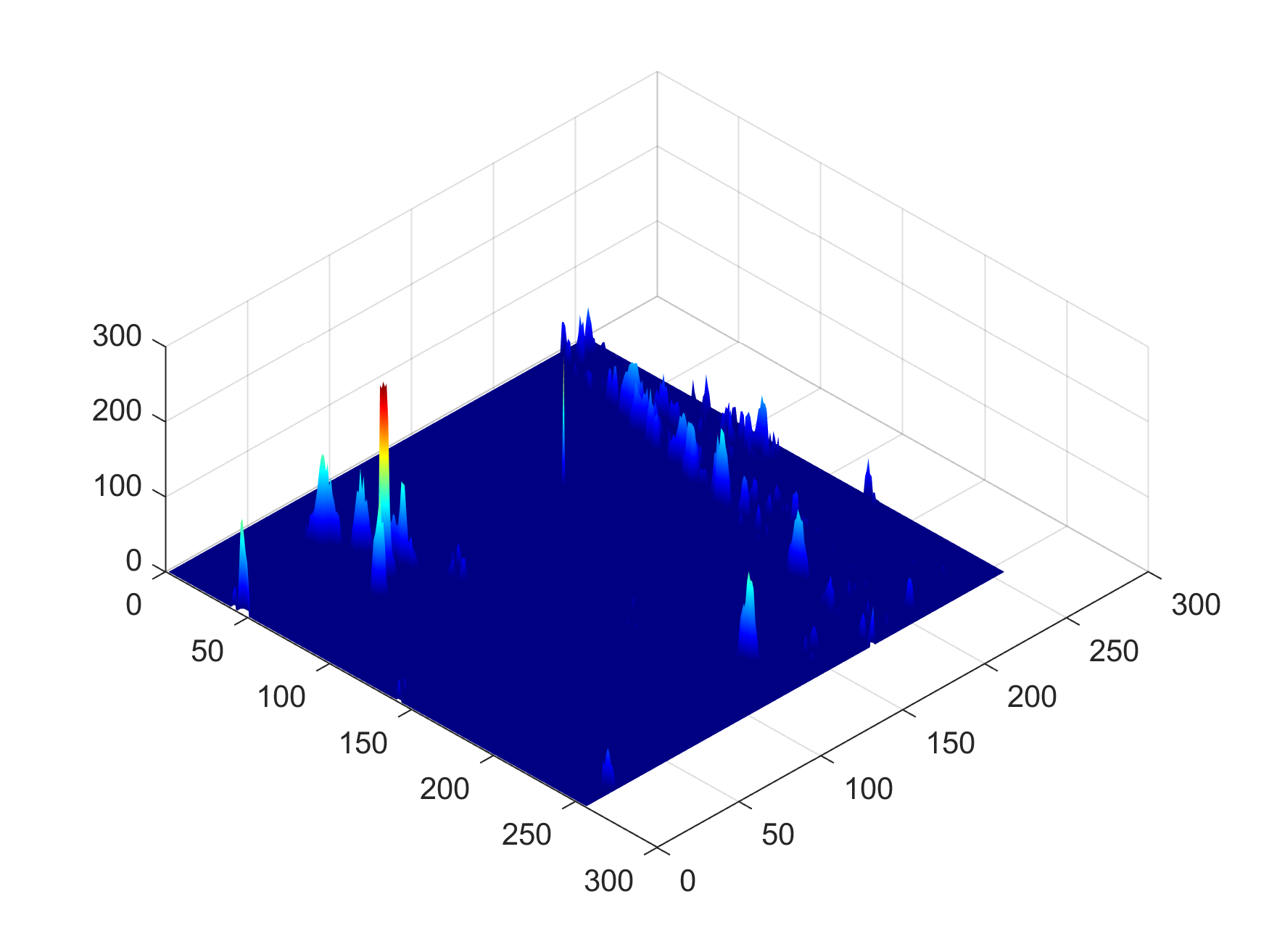} &
\includegraphics[width=0.115\linewidth]{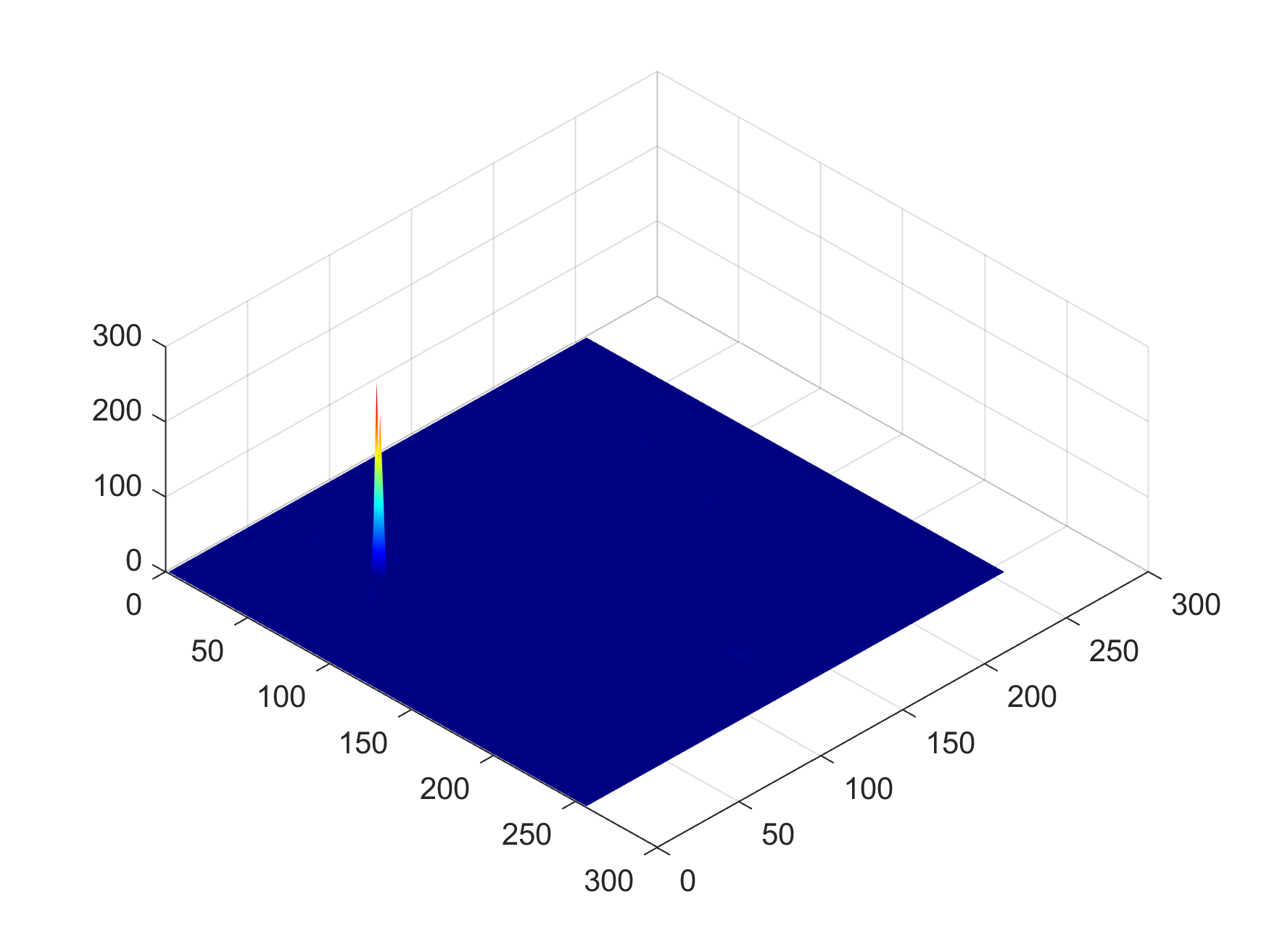} &
\includegraphics[width=0.115\linewidth]{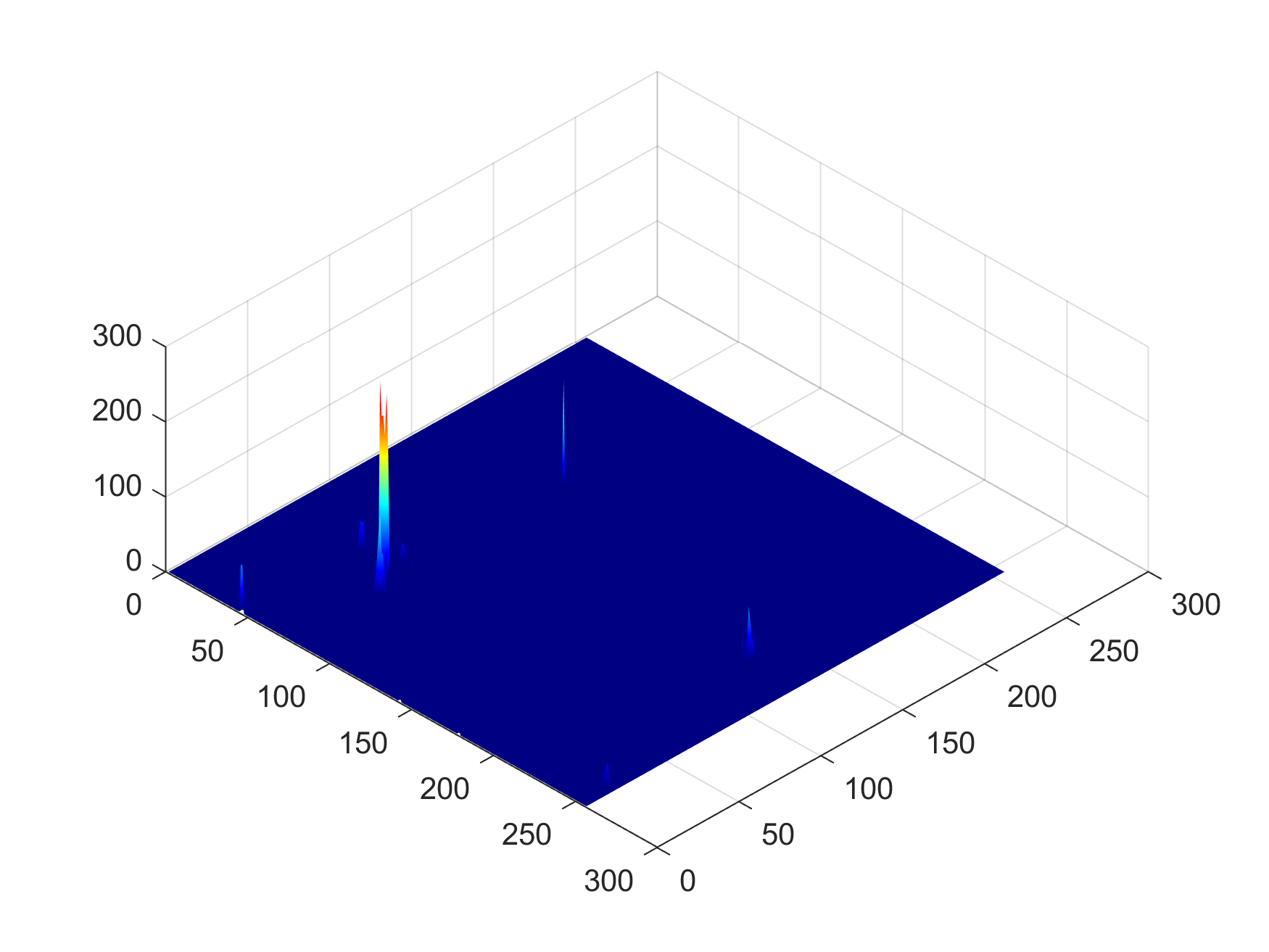} &
\includegraphics[width=0.115\linewidth]{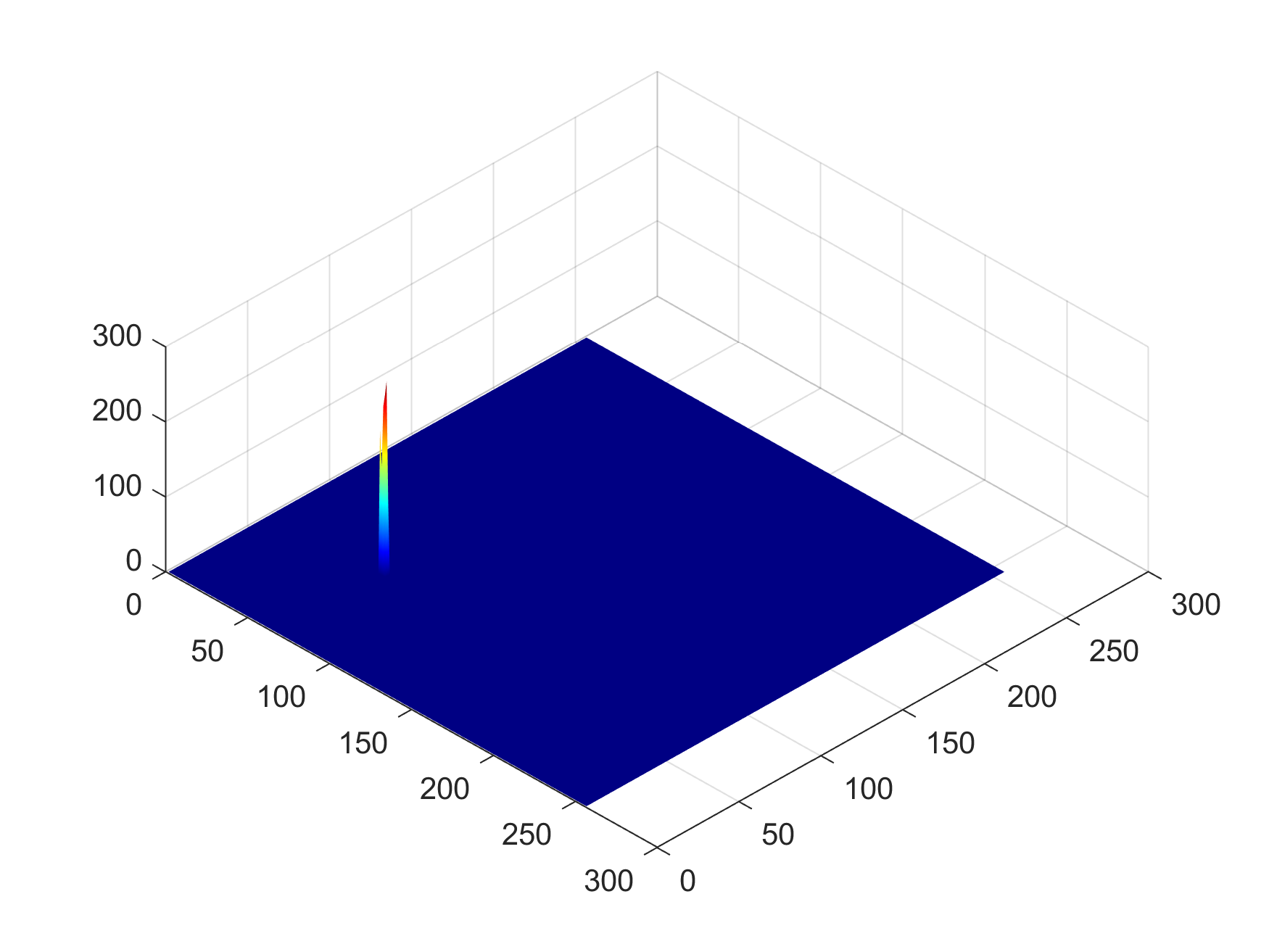} \\[-4pt]

% 第2行
\includegraphics[width=0.115\linewidth]{21.png} &
\includegraphics[width=0.115\linewidth]{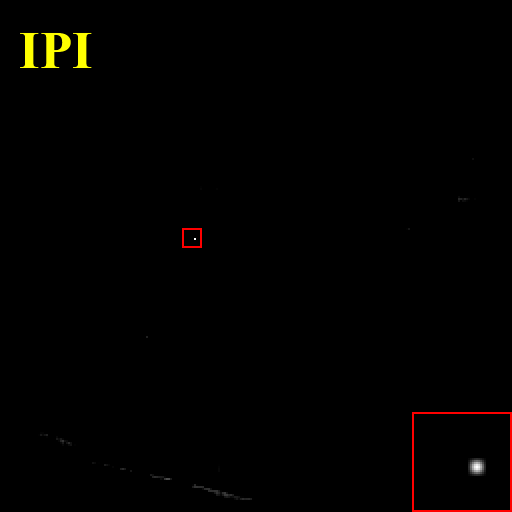} &
\includegraphics[width=0.115\linewidth]{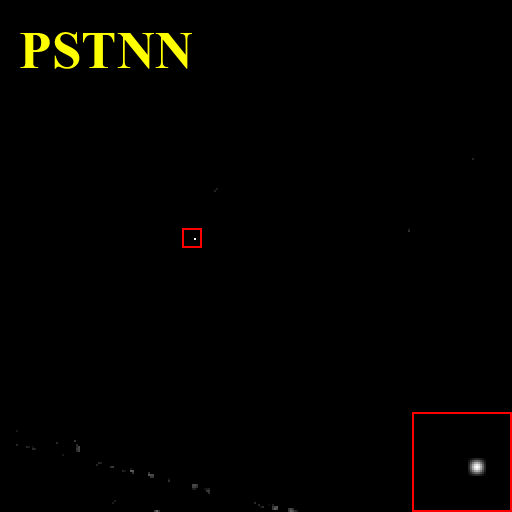} &
\includegraphics[width=0.115\linewidth]{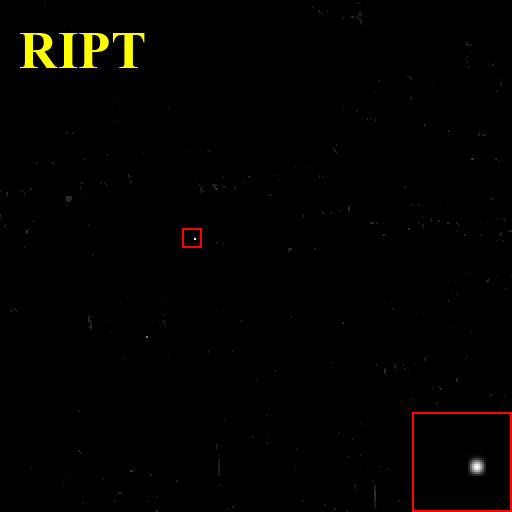} &
\includegraphics[width=0.115\linewidth]{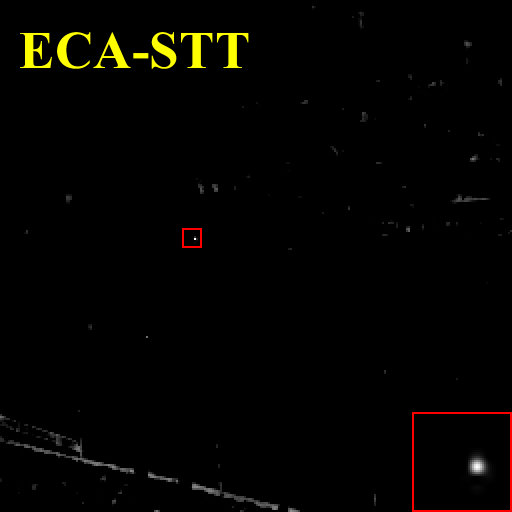} &
\includegraphics[width=0.115\linewidth]{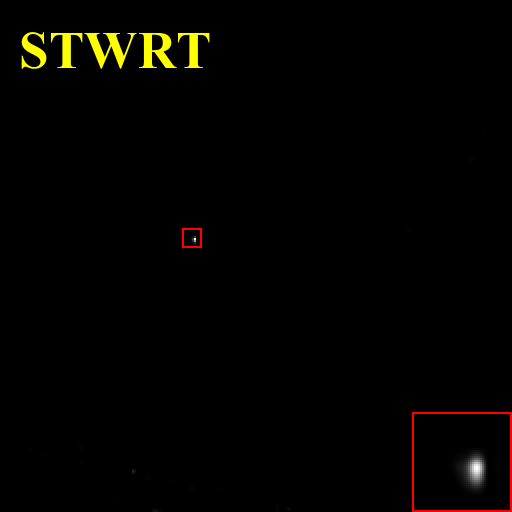} &
\includegraphics[width=0.115\linewidth]{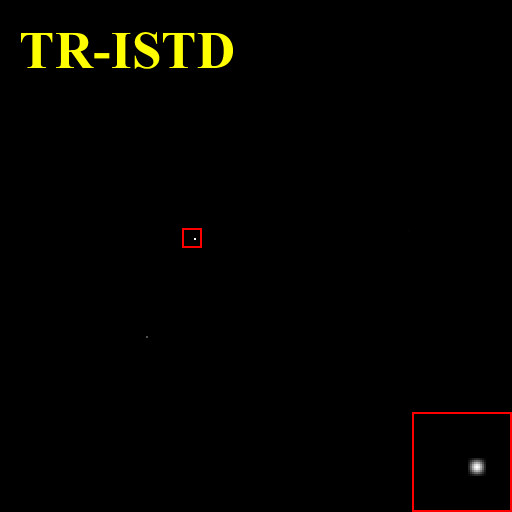} &
\includegraphics[width=0.115\linewidth]{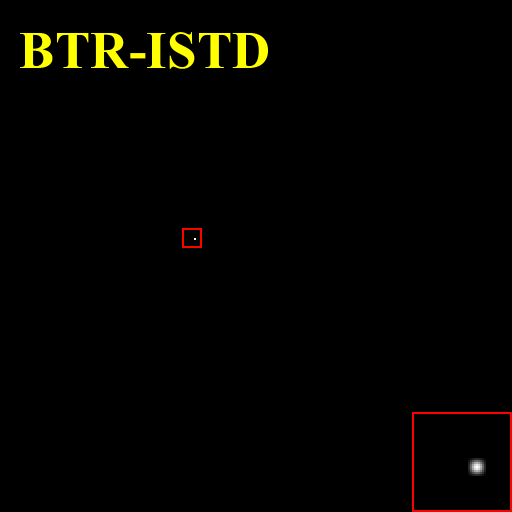} \\[-4pt]

\includegraphics[width=0.115\linewidth]{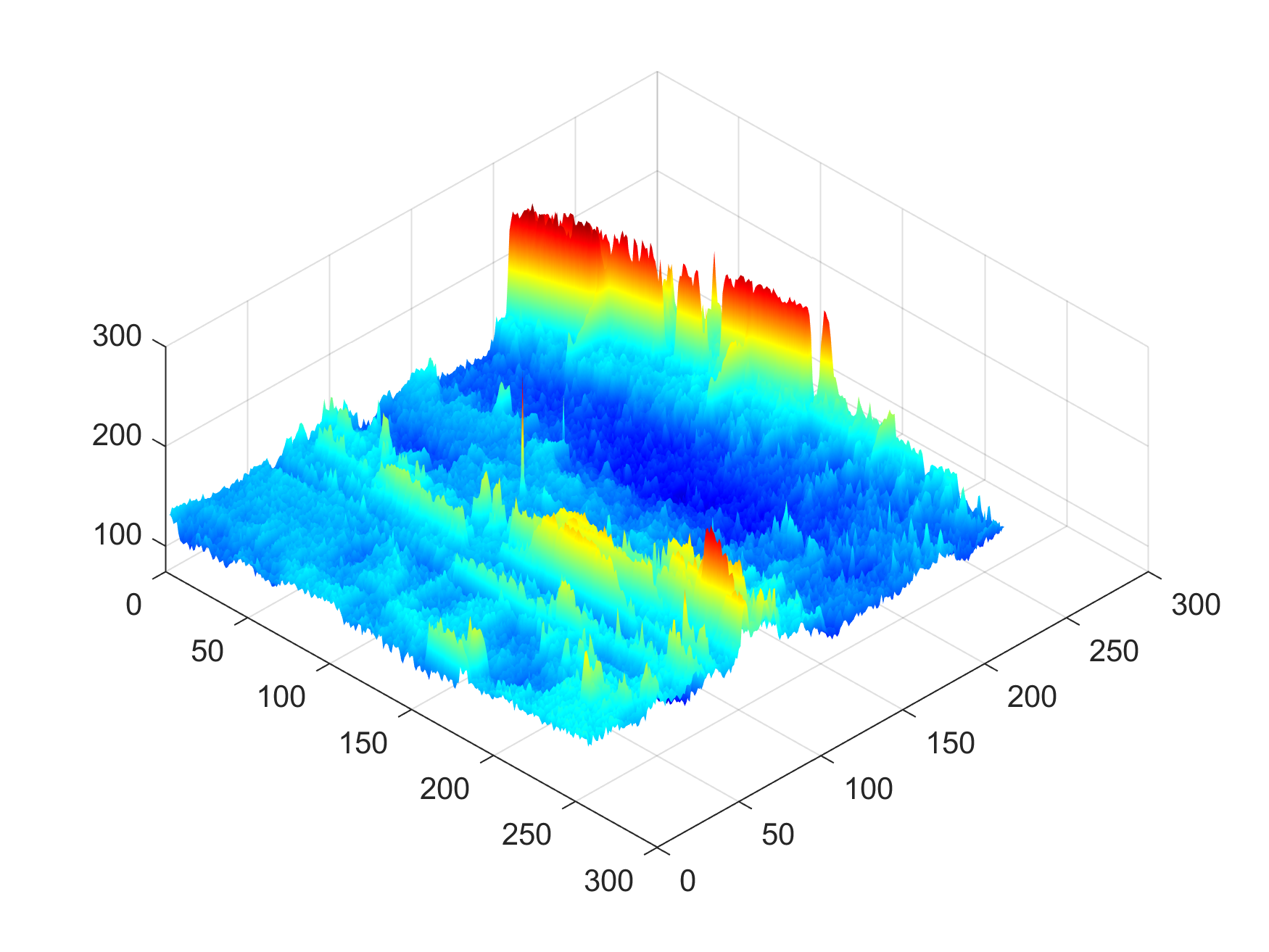} &
\includegraphics[width=0.115\linewidth]{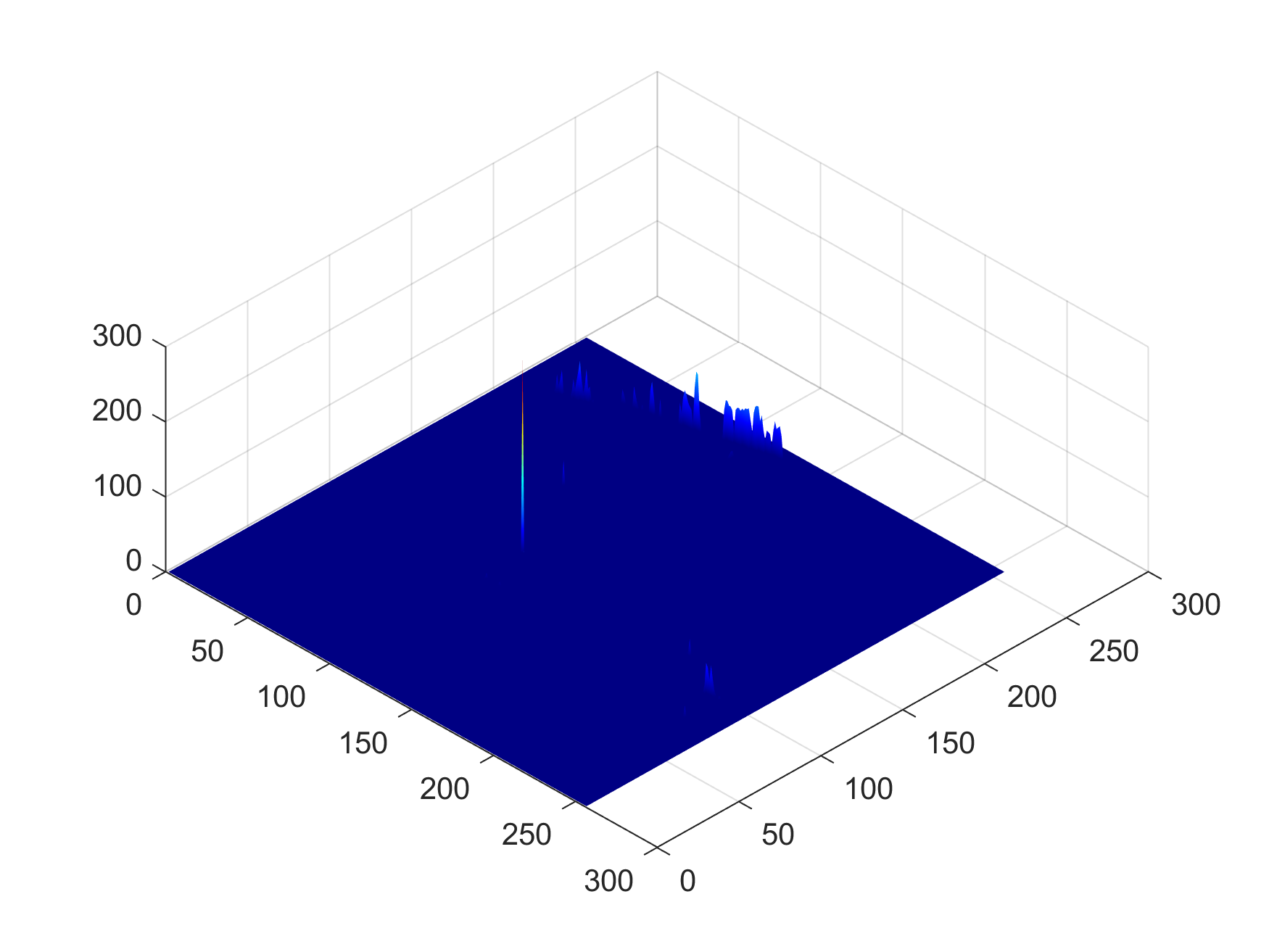} &
\includegraphics[width=0.115\linewidth]{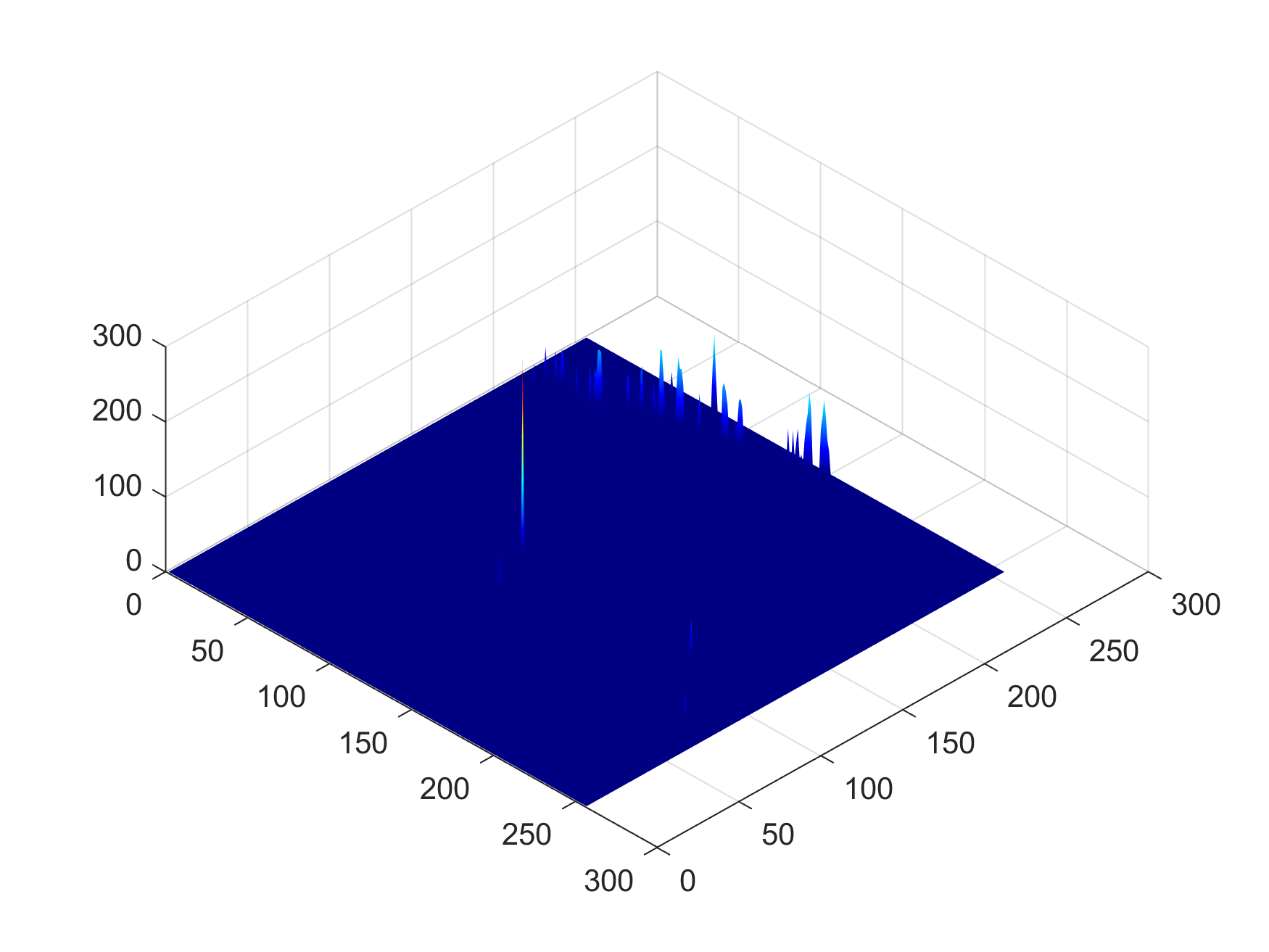} &
\includegraphics[width=0.115\linewidth]{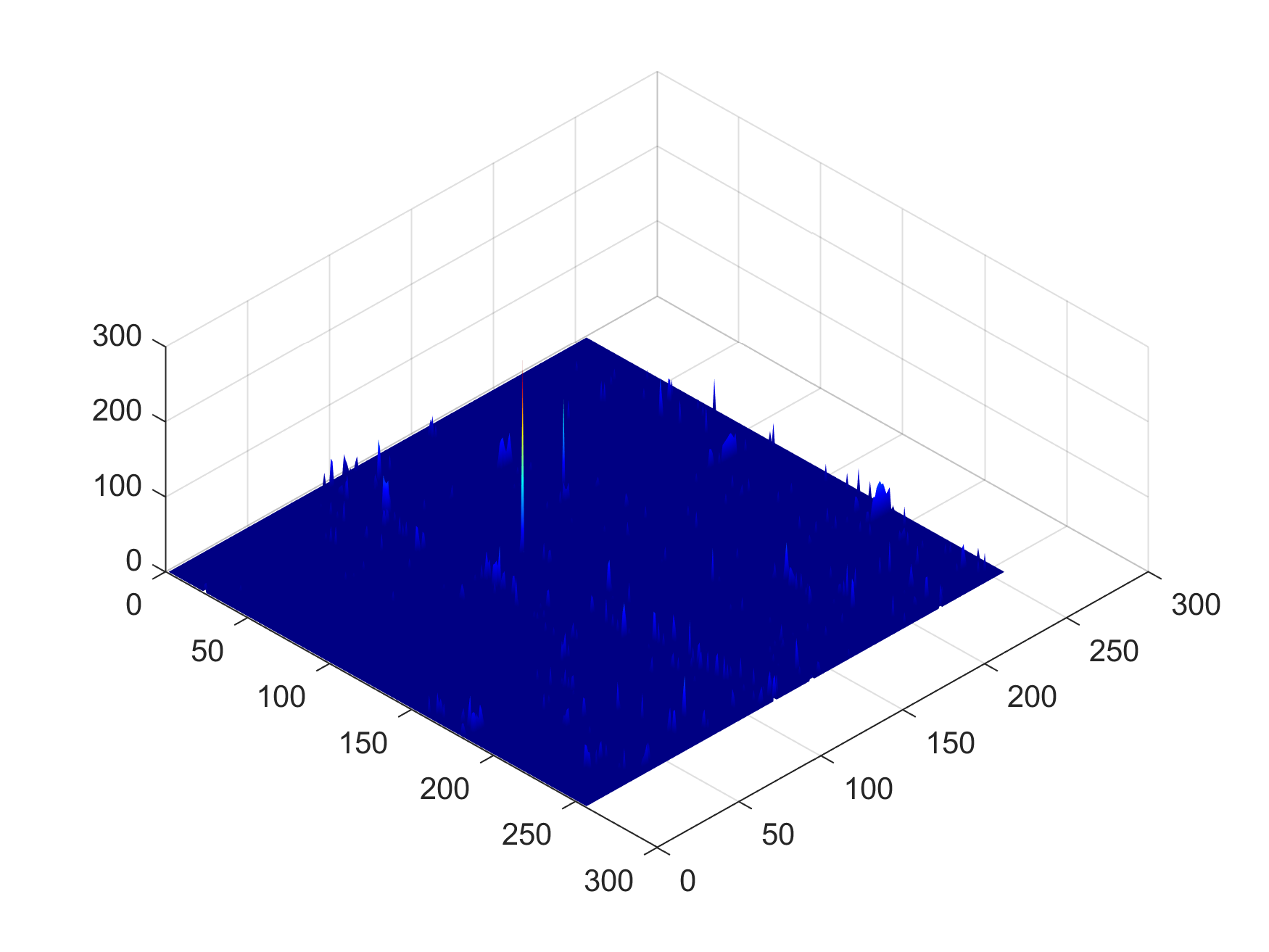} &
\includegraphics[width=0.115\linewidth]{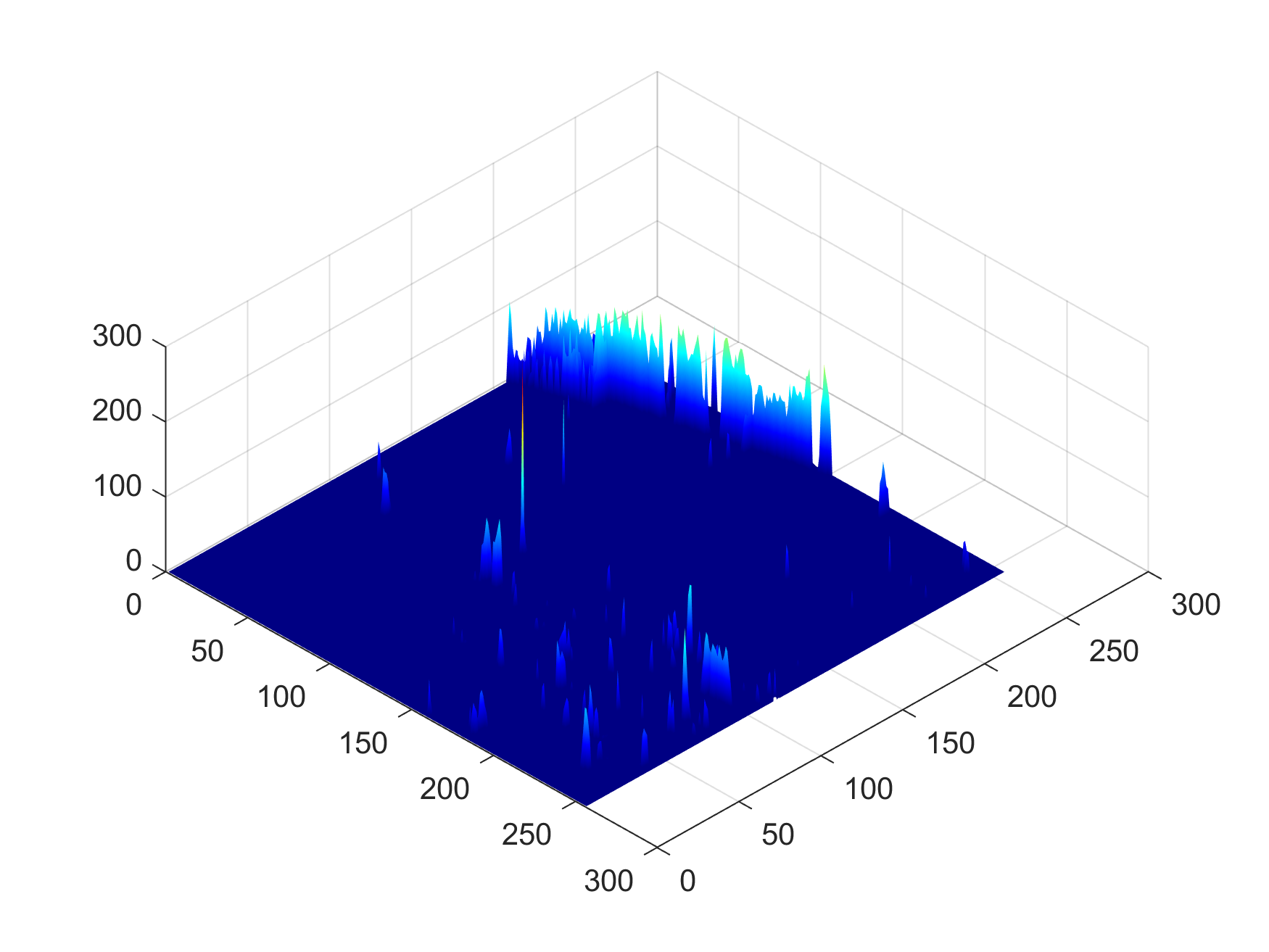} &
\includegraphics[width=0.115\linewidth]{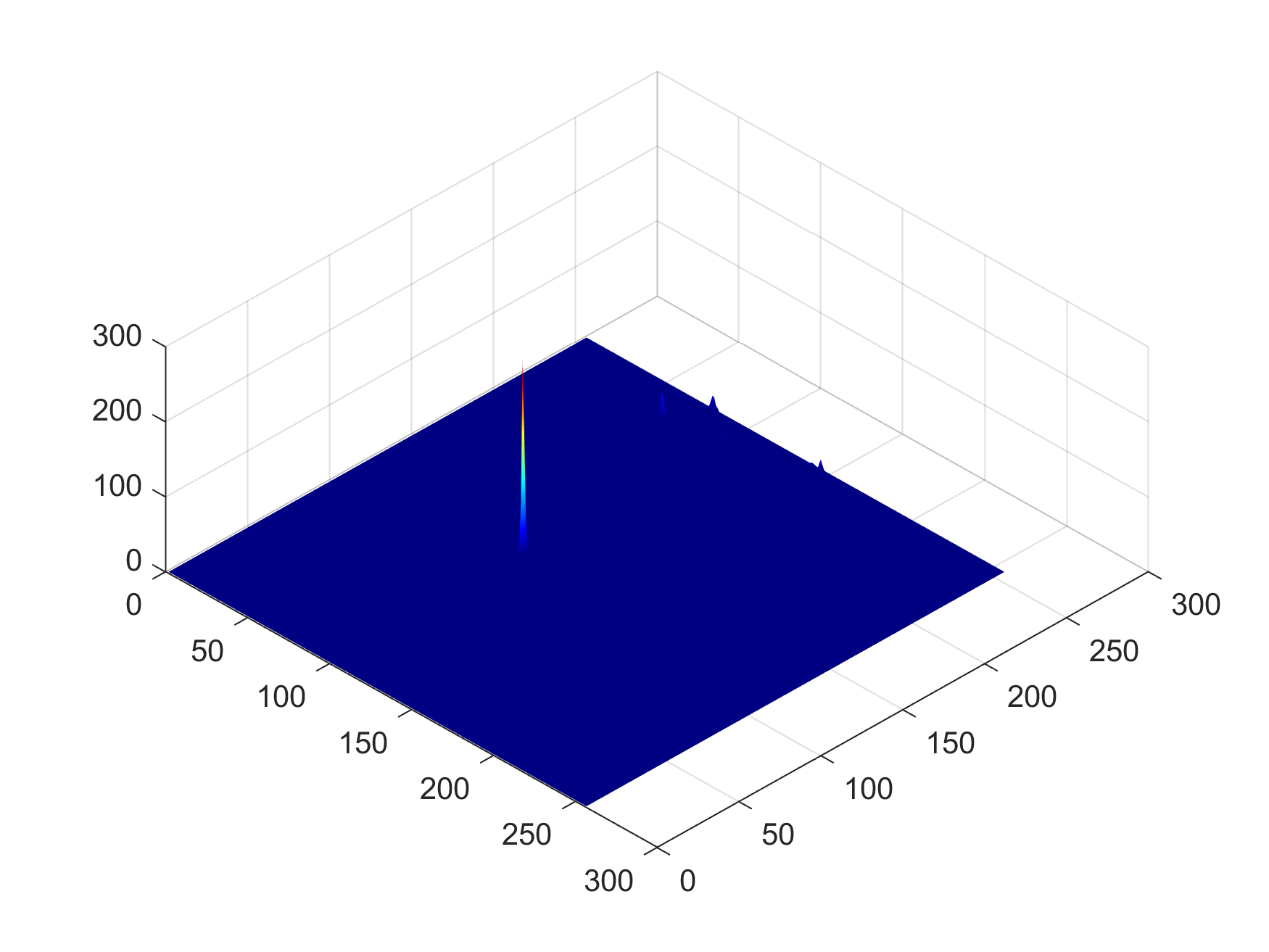} &
\includegraphics[width=0.115\linewidth]{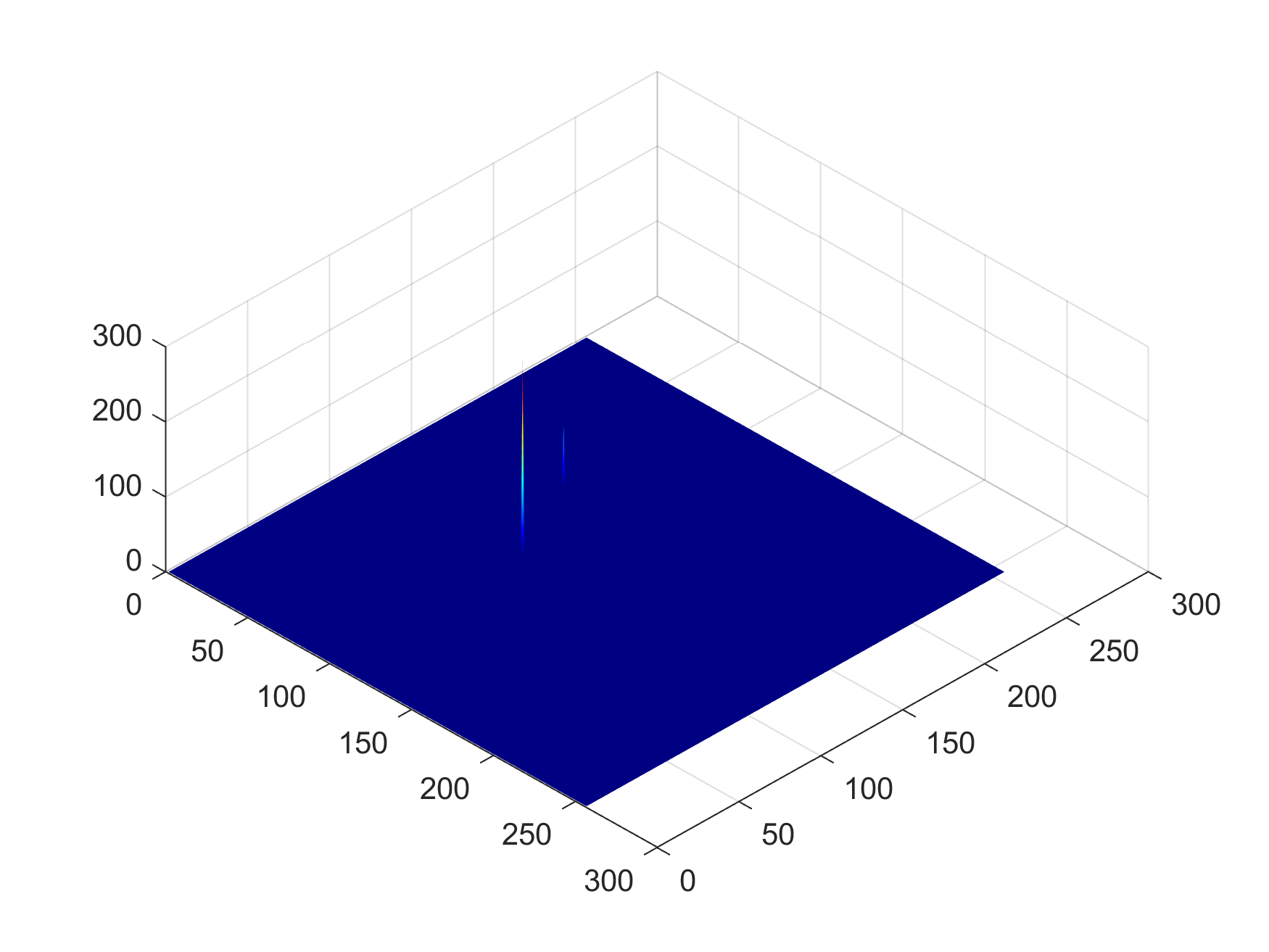} &
\includegraphics[width=0.115\linewidth]{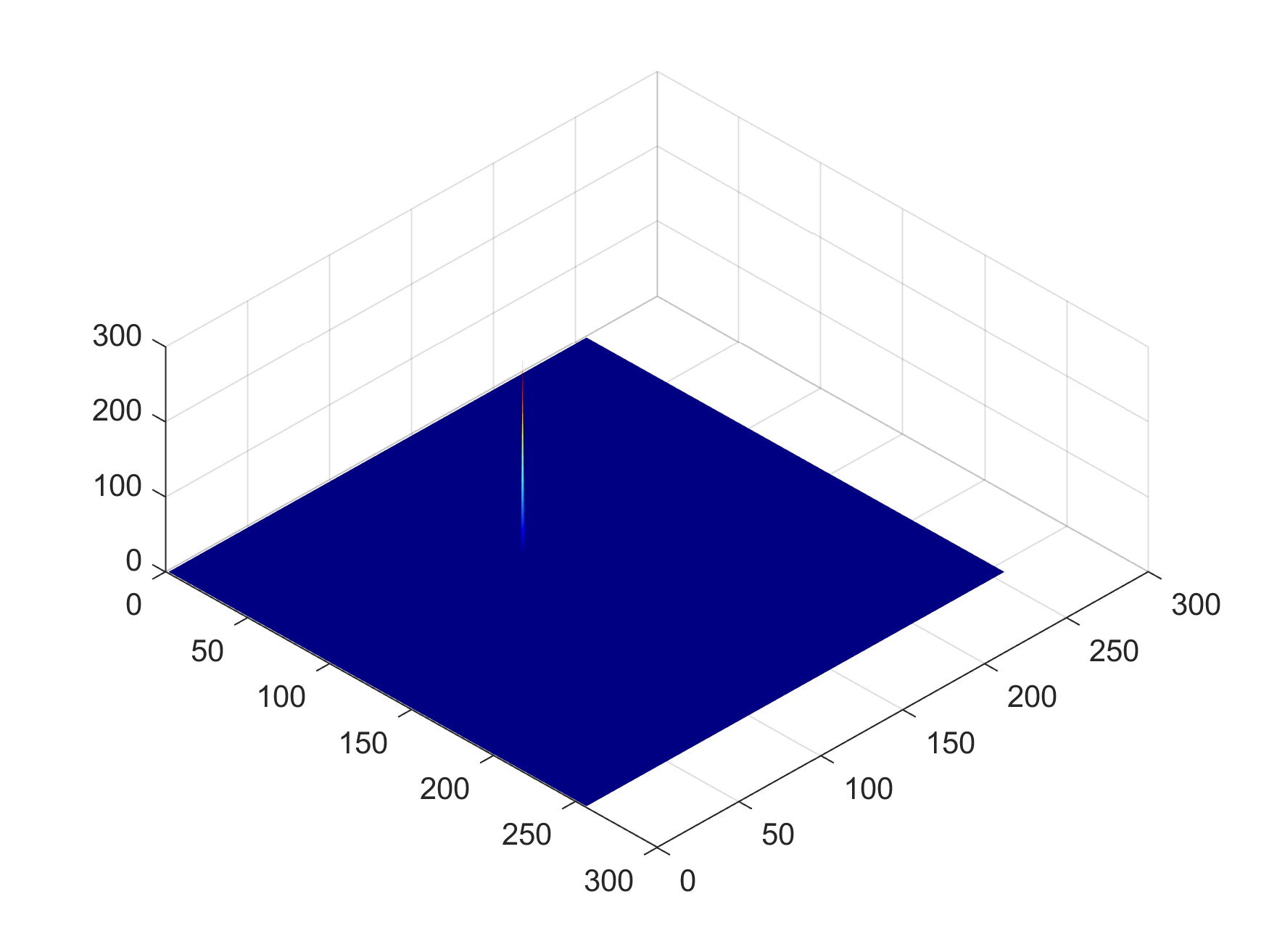} \\[-4pt]

% 第3-6行以此类推...
\includegraphics[width=0.115\linewidth]{22.png} &
\includegraphics[width=0.115\linewidth]{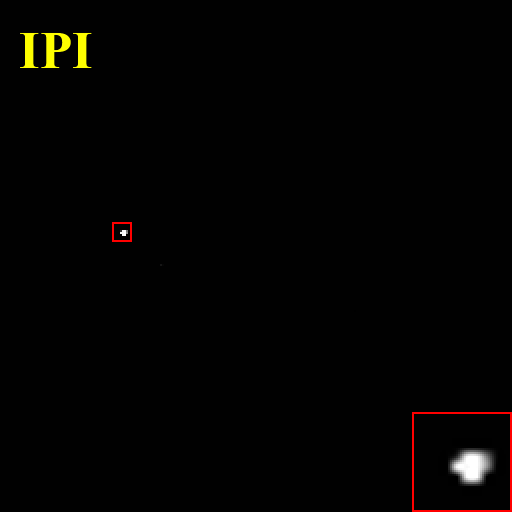} &
\includegraphics[width=0.115\linewidth]{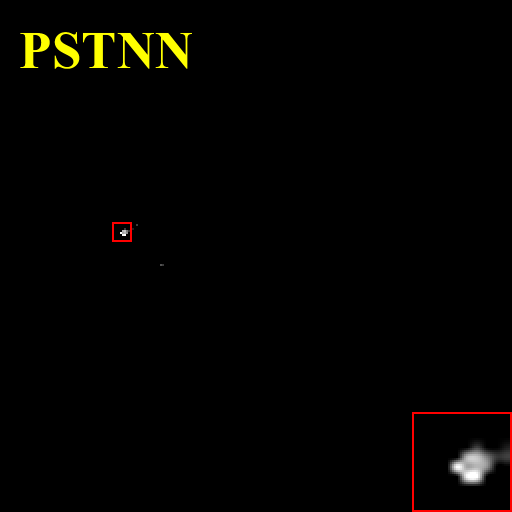} &
\includegraphics[width=0.115\linewidth]{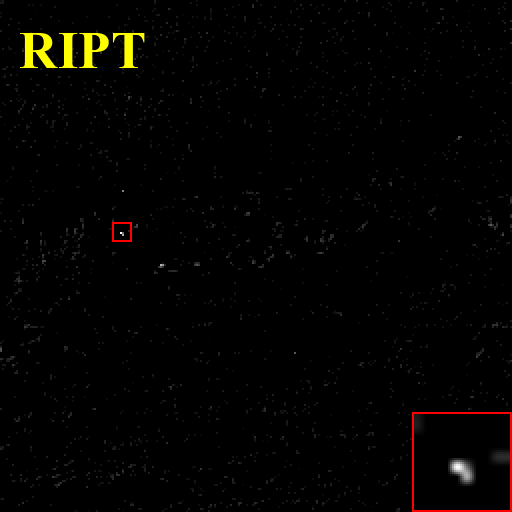} &
\includegraphics[width=0.115\linewidth]{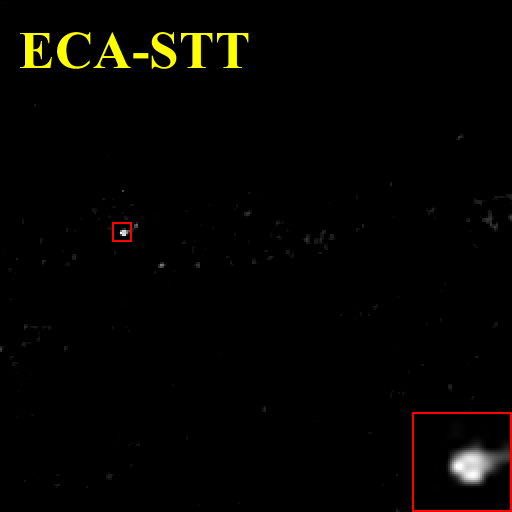} &
\includegraphics[width=0.115\linewidth]{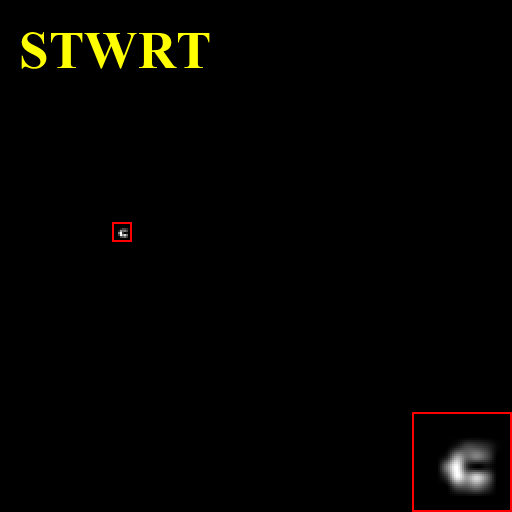} &
\includegraphics[width=0.115\linewidth]{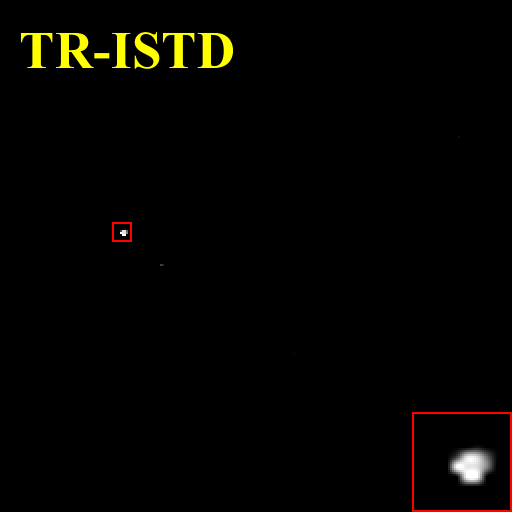} &
\includegraphics[width=0.115\linewidth]{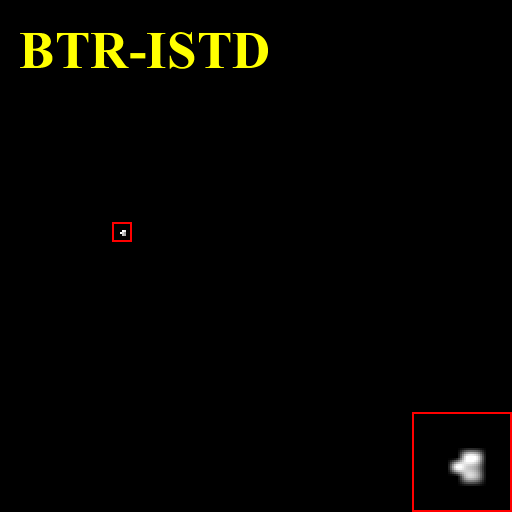} \\[-4pt]

\includegraphics[width=0.115\linewidth]{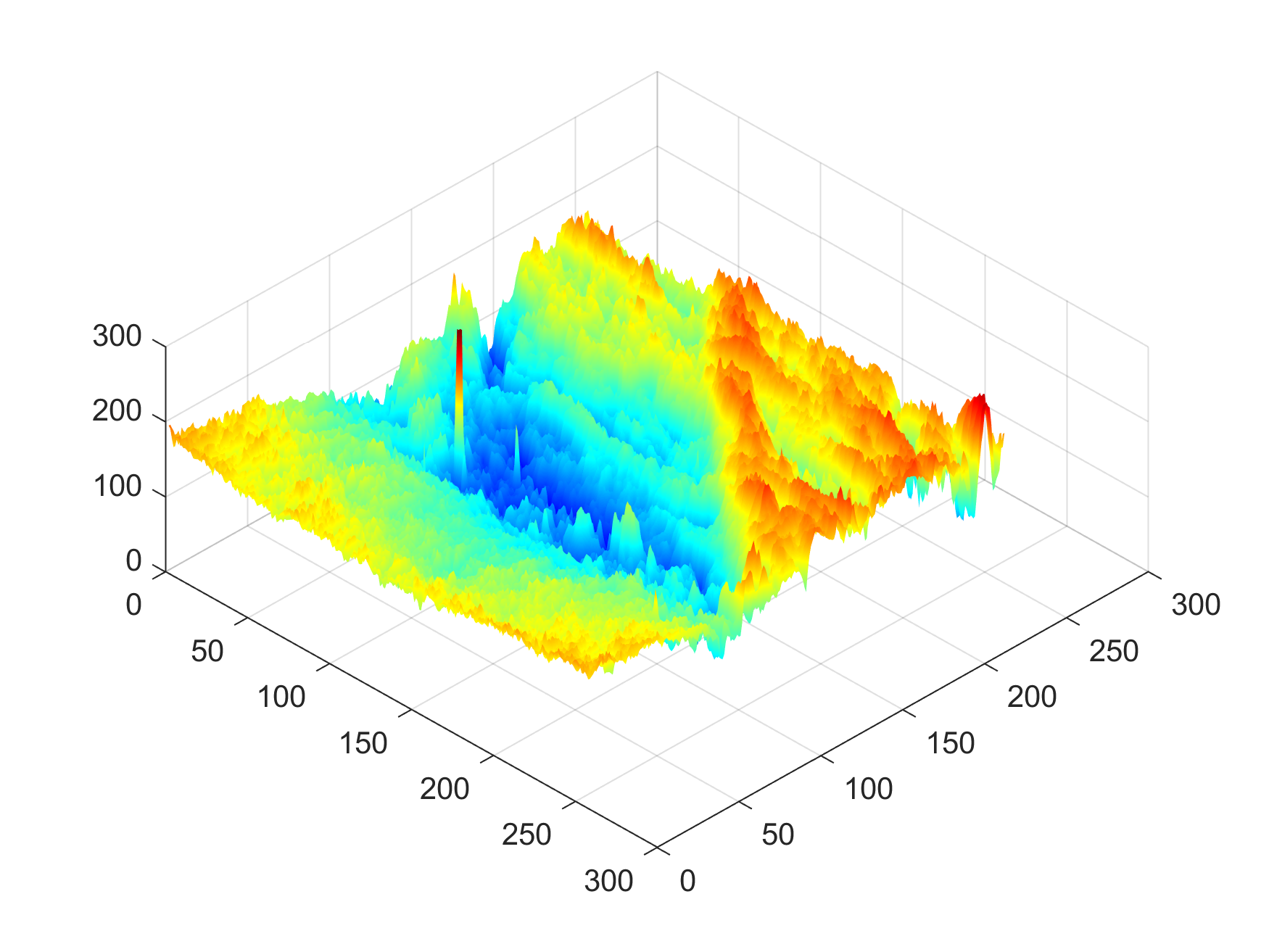} &
\includegraphics[width=0.115\linewidth]{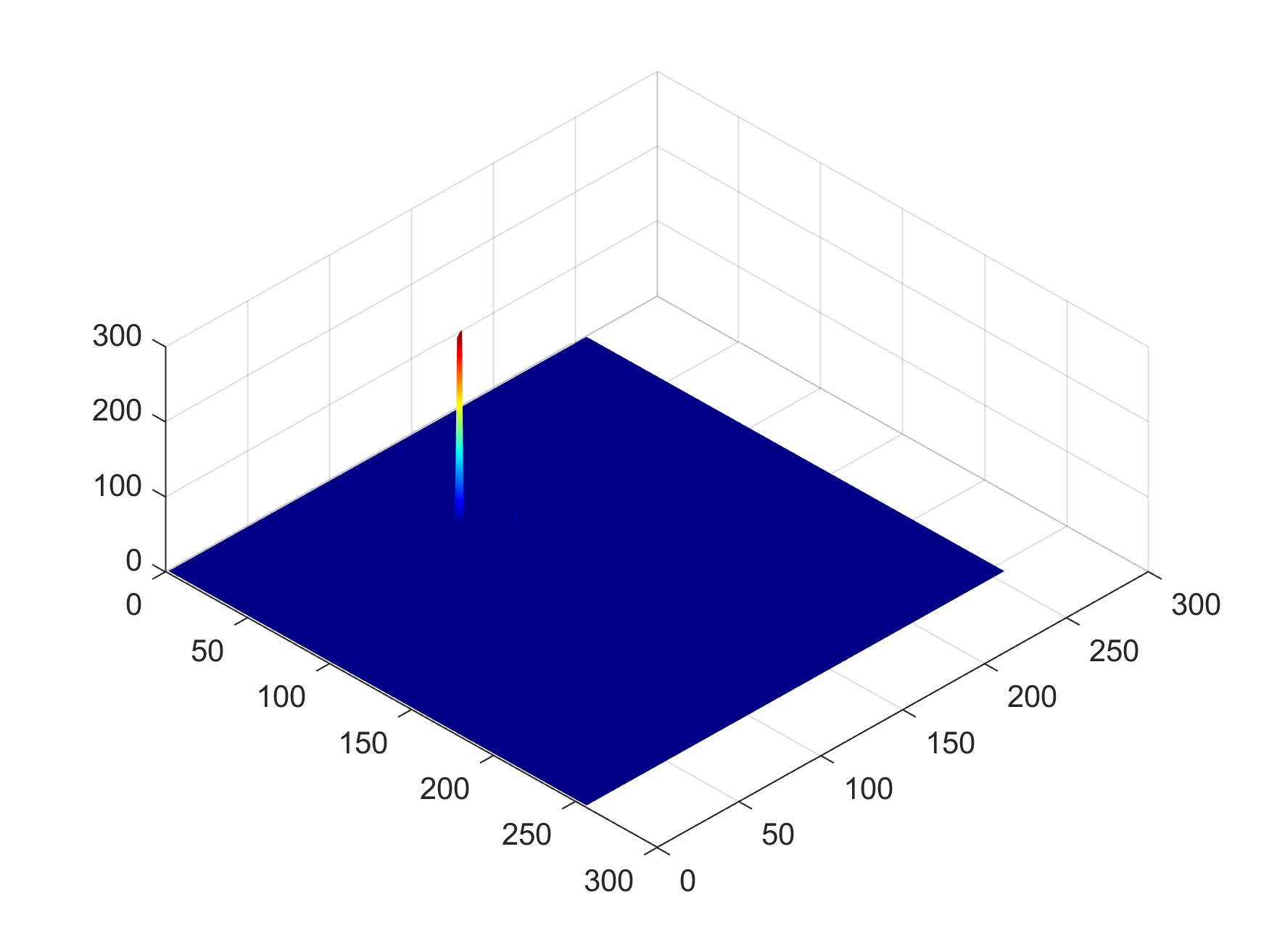} &
\includegraphics[width=0.115\linewidth]{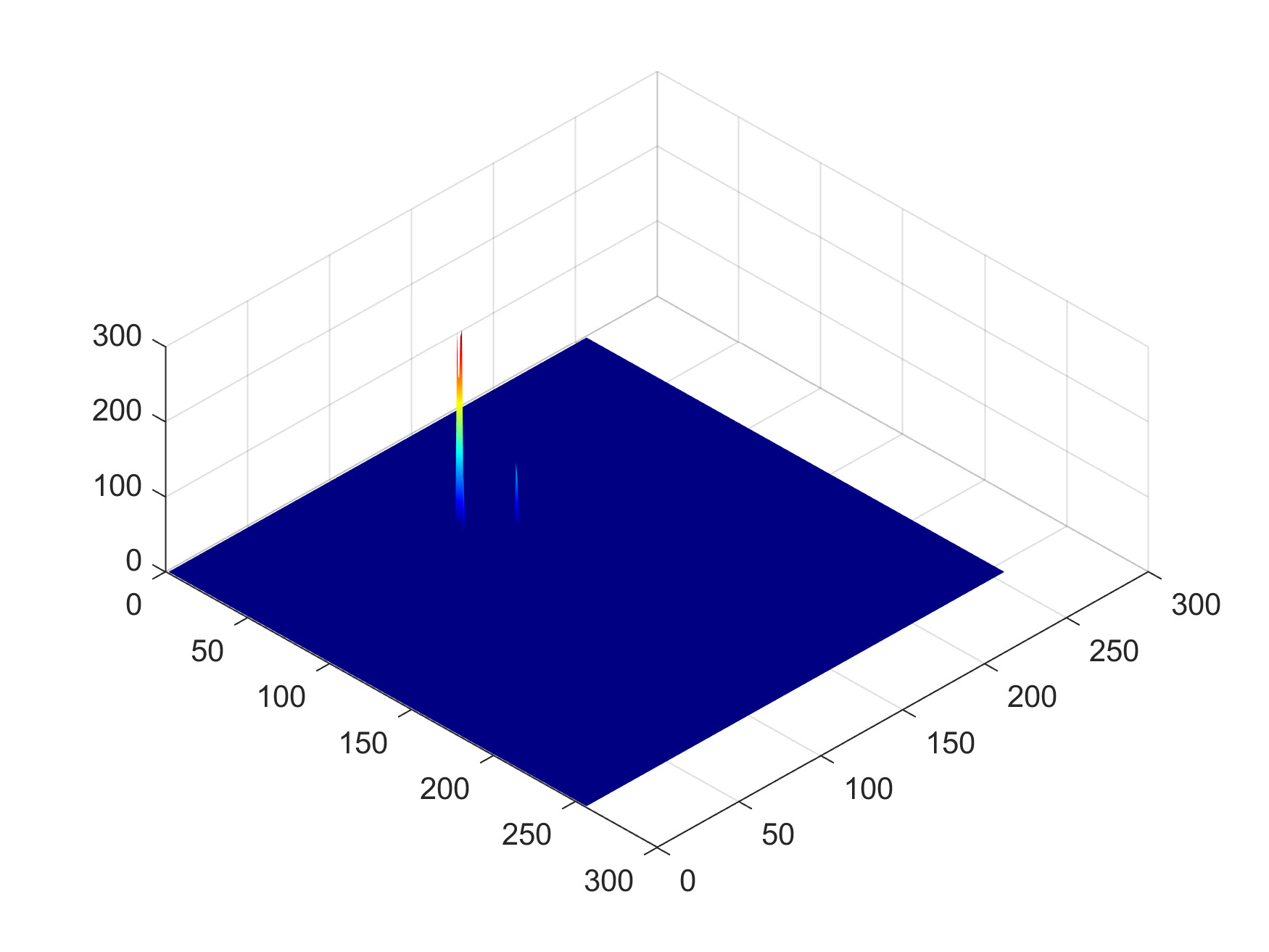} &
\includegraphics[width=0.115\linewidth]{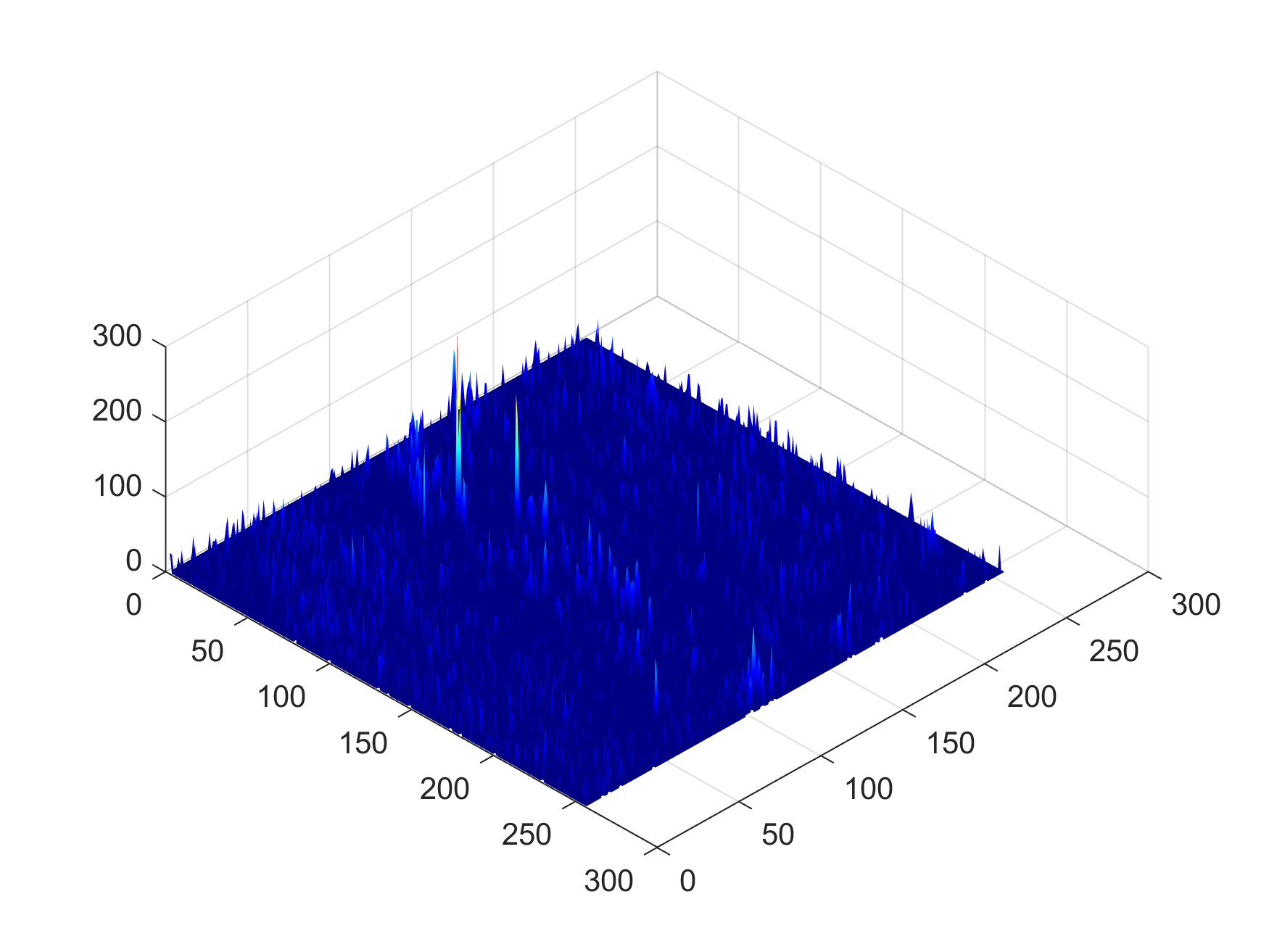} &
\includegraphics[width=0.115\linewidth]{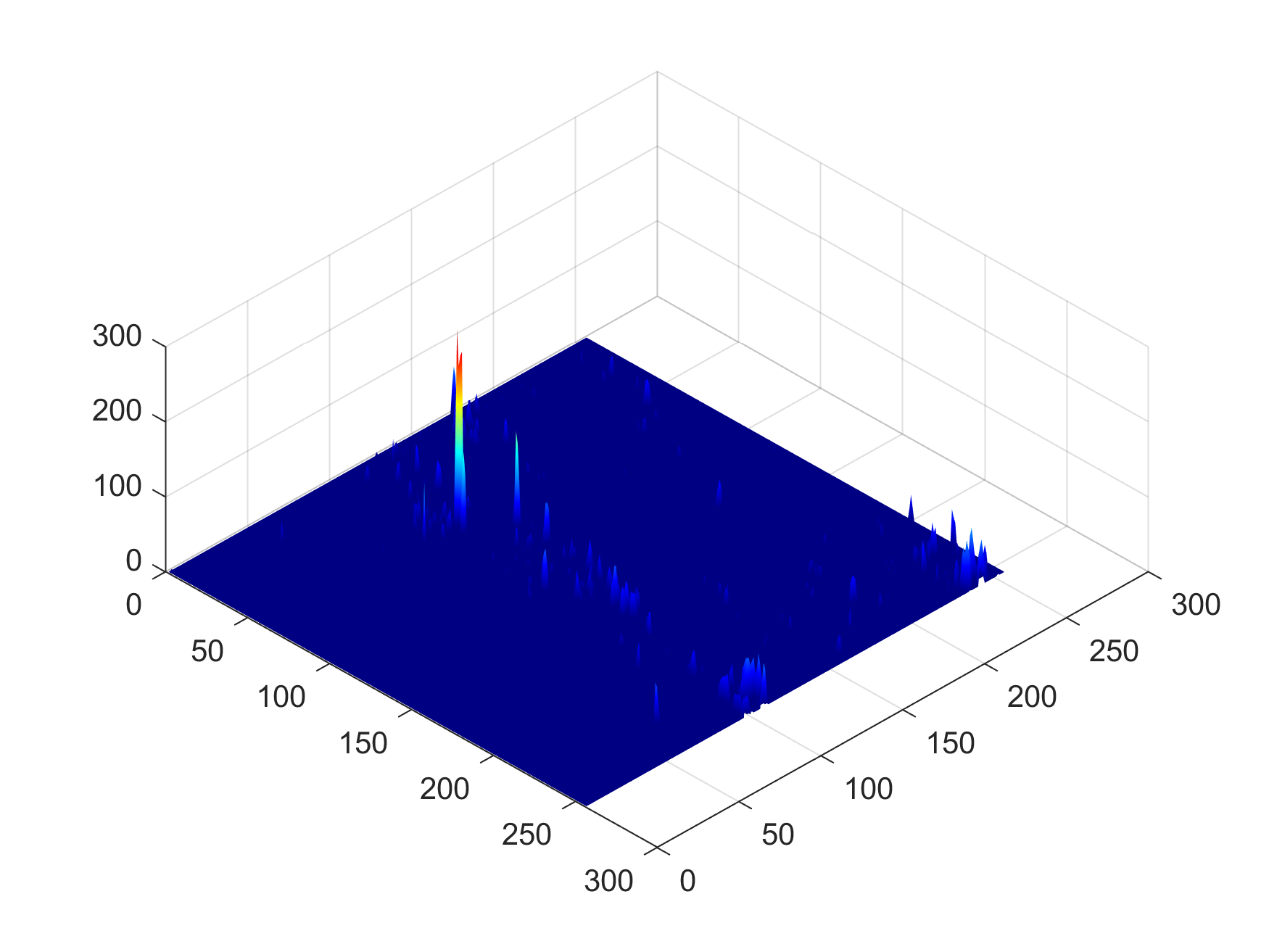} &
\includegraphics[width=0.115\linewidth]{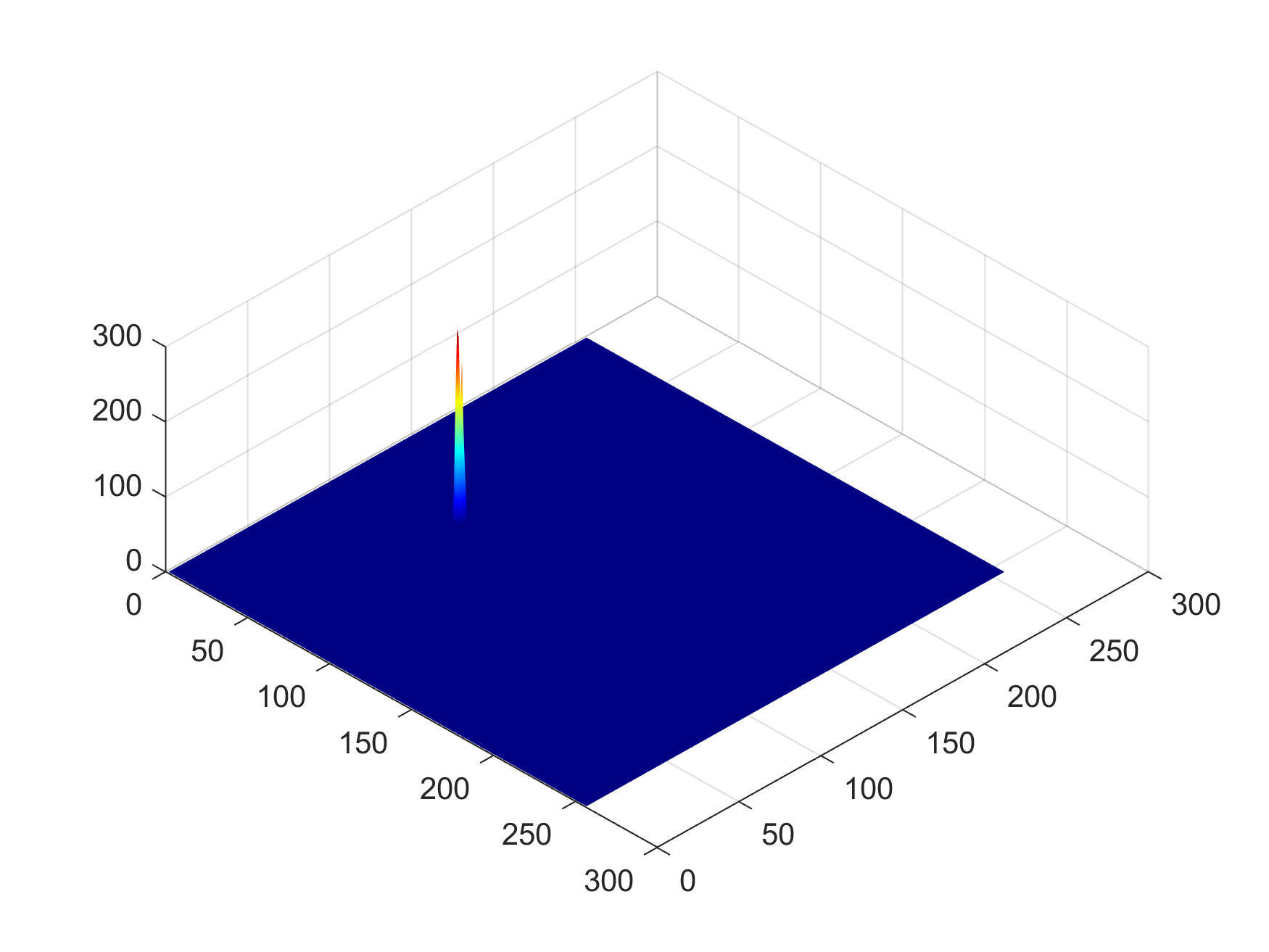} &
\includegraphics[width=0.115\linewidth]{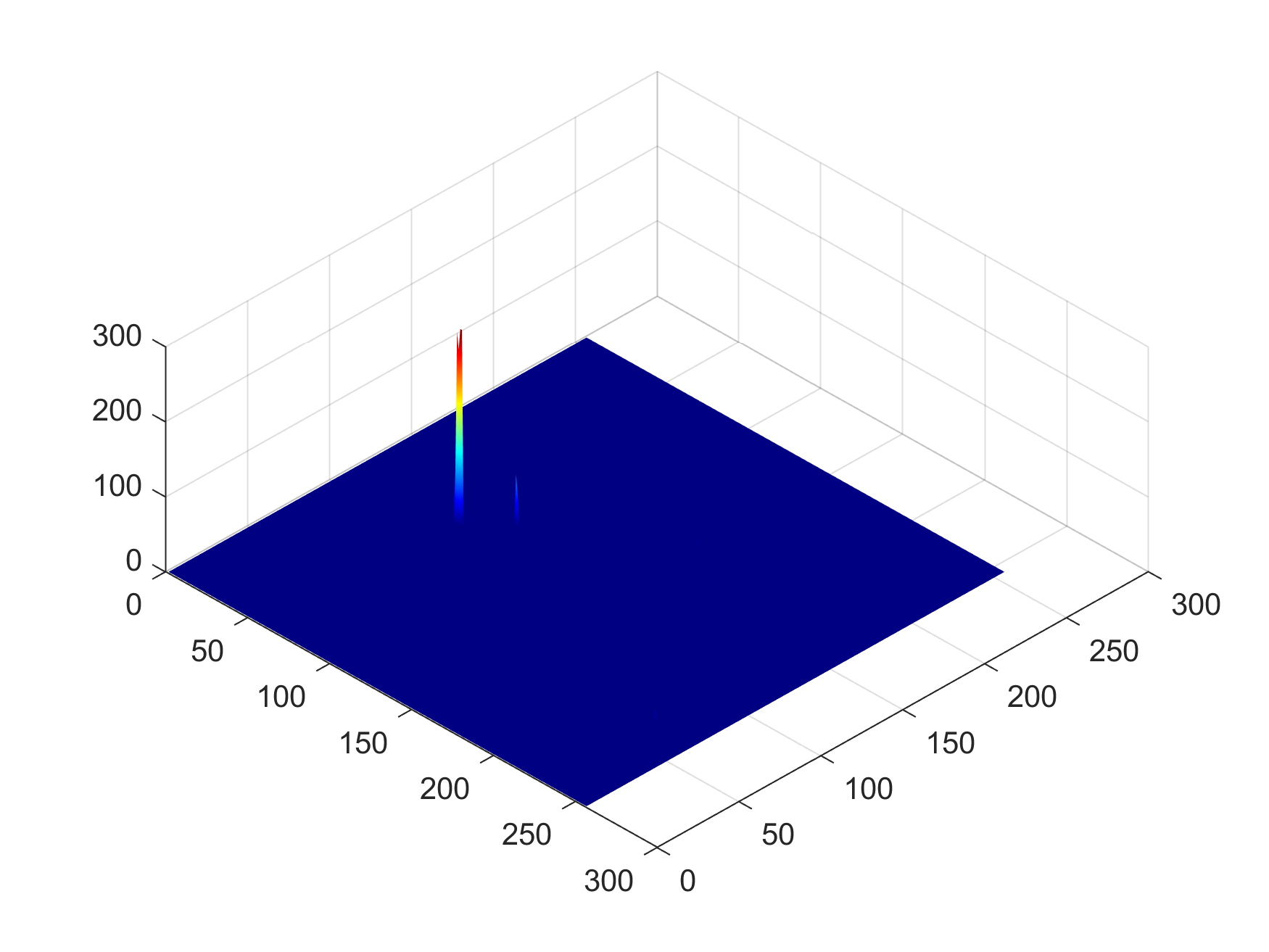} &
\includegraphics[width=0.115\linewidth]{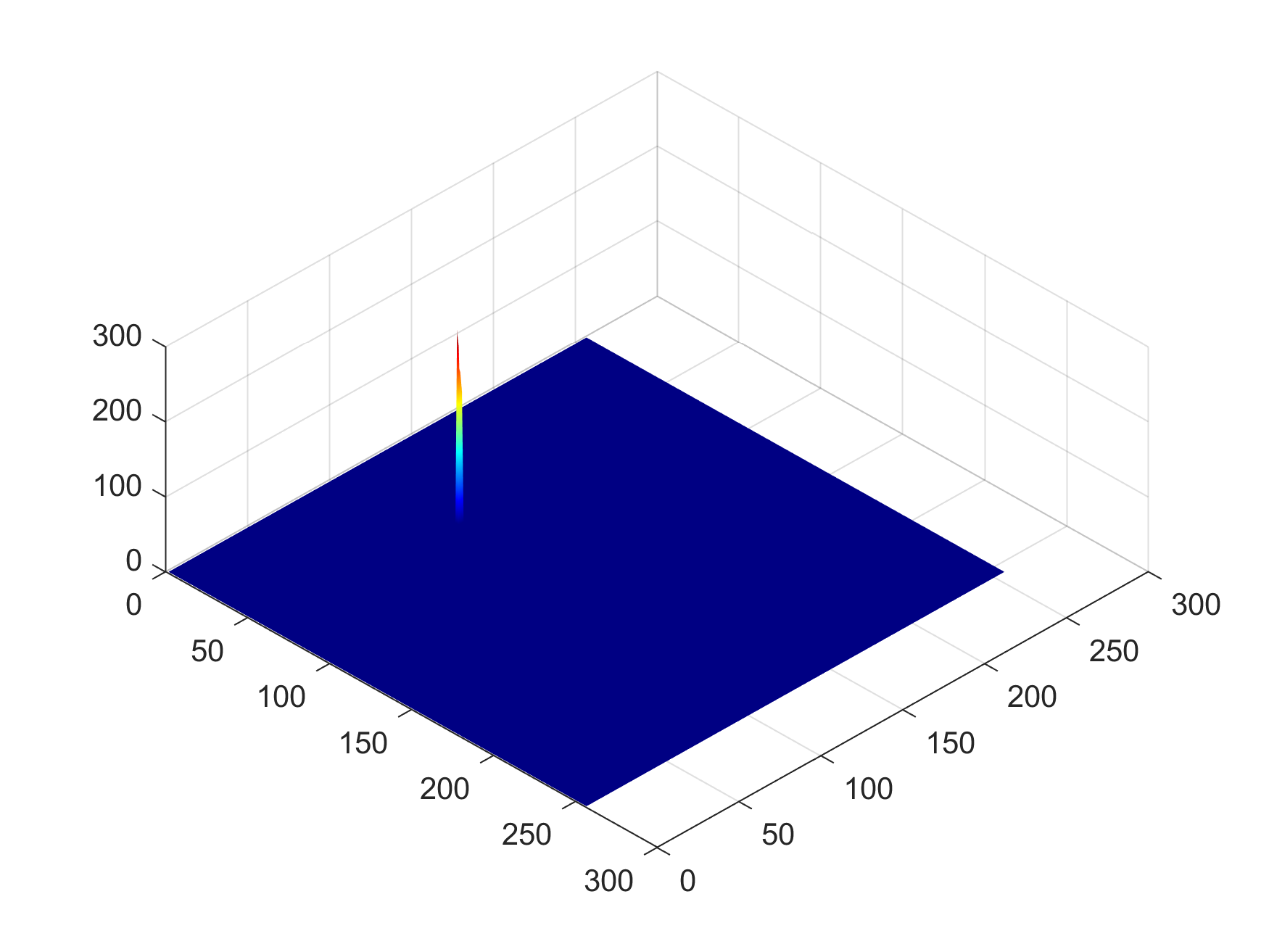} \\[-4pt]

\includegraphics[width=0.115\linewidth]{23.png} &
\includegraphics[width=0.115\linewidth]{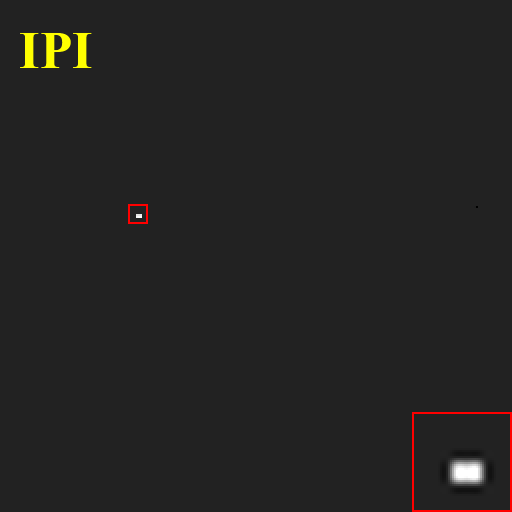} &
\includegraphics[width=0.115\linewidth]{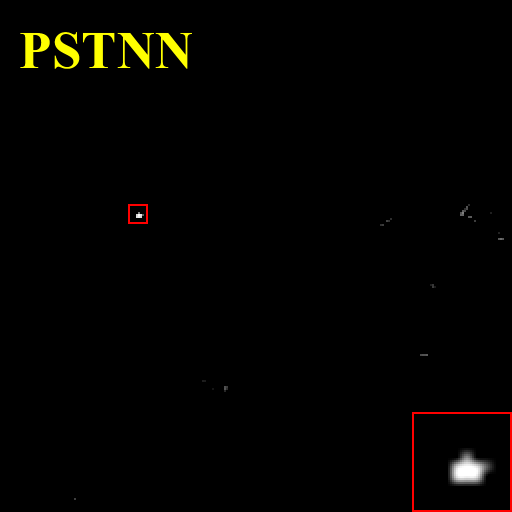} &
\includegraphics[width=0.115\linewidth]{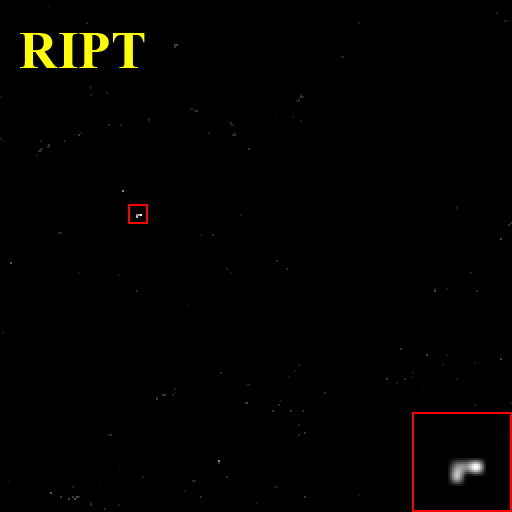} &
\includegraphics[width=0.115\linewidth]{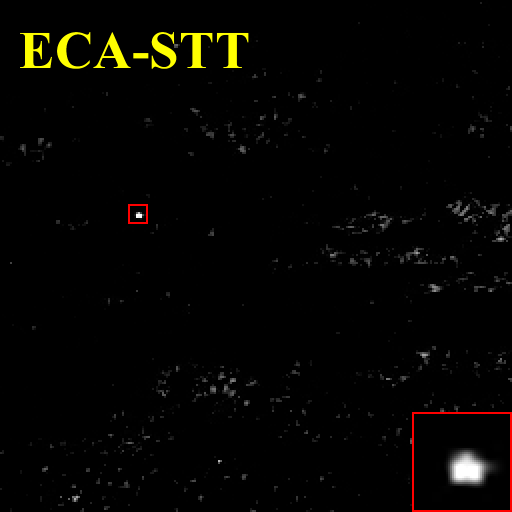} &
\includegraphics[width=0.115\linewidth]{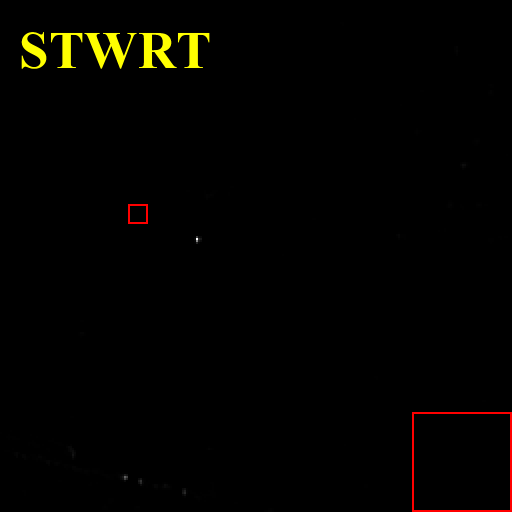} &
\includegraphics[width=0.115\linewidth]{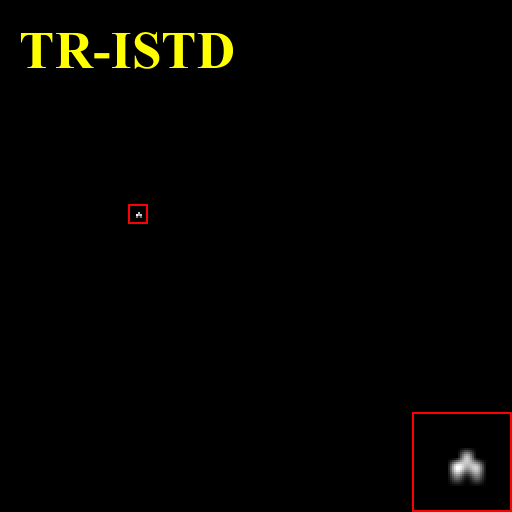} &
\includegraphics[width=0.115\linewidth]{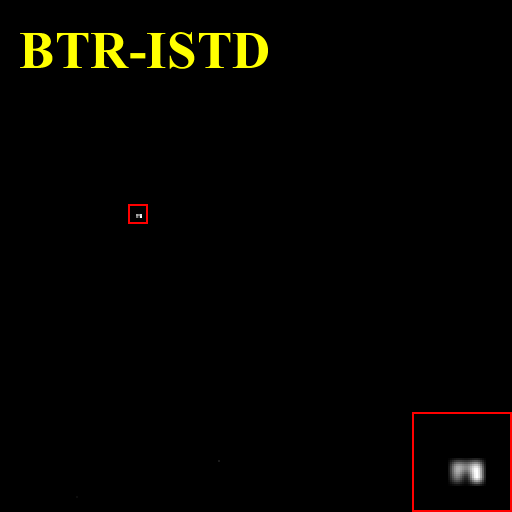} \\[-4pt]

\includegraphics[width=0.115\linewidth]{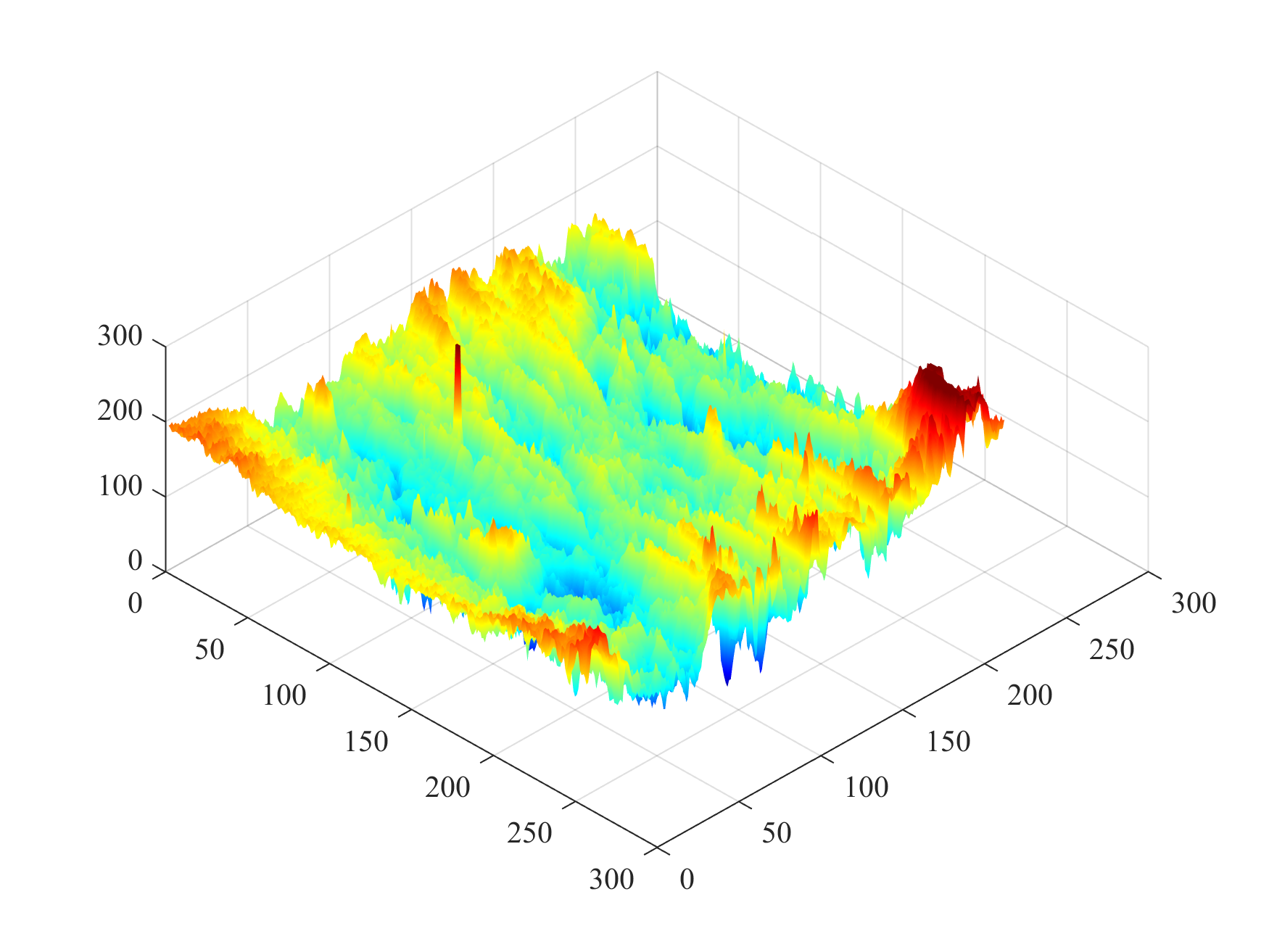} &
\includegraphics[width=0.115\linewidth]{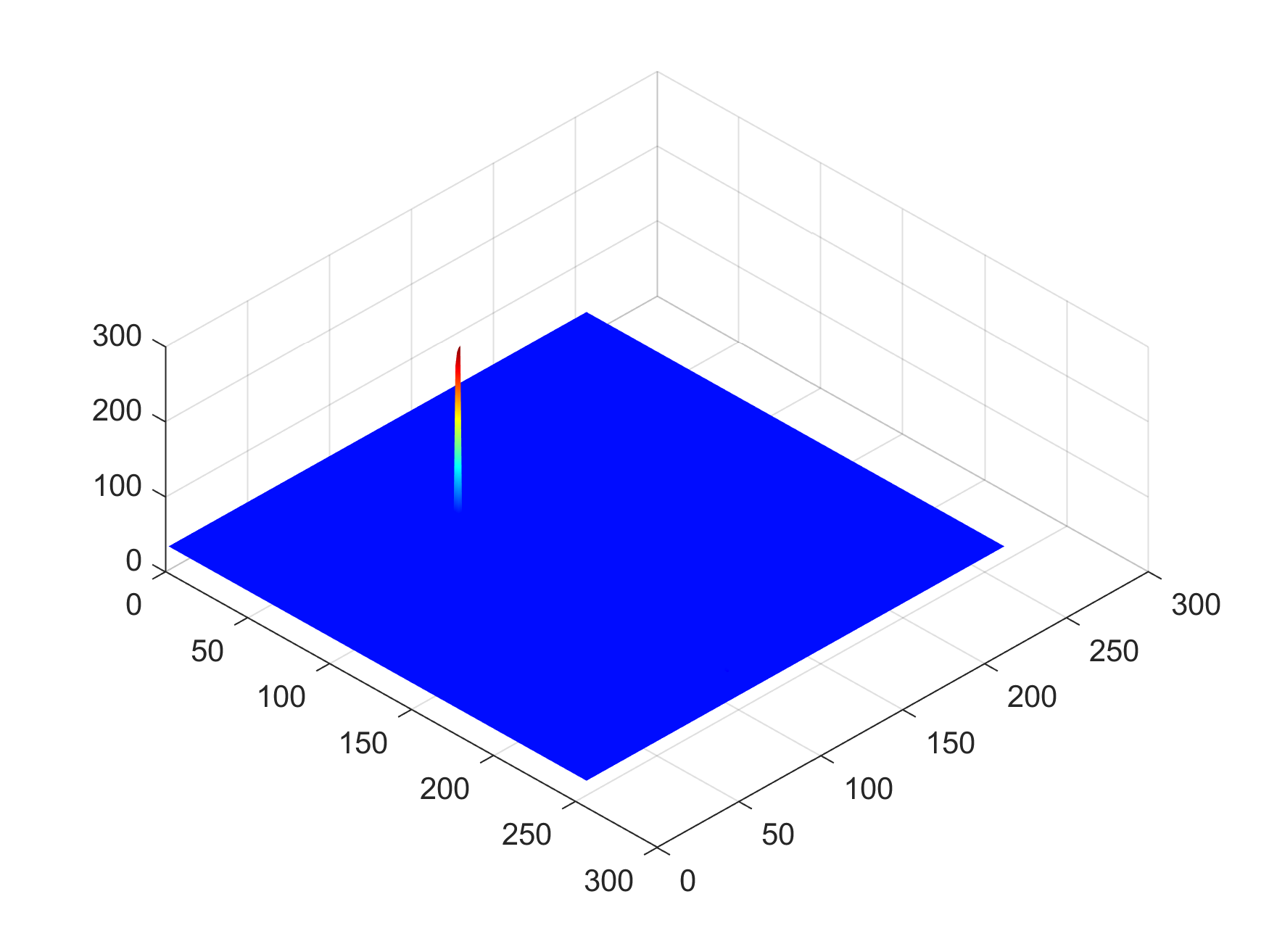} &
\includegraphics[width=0.115\linewidth]{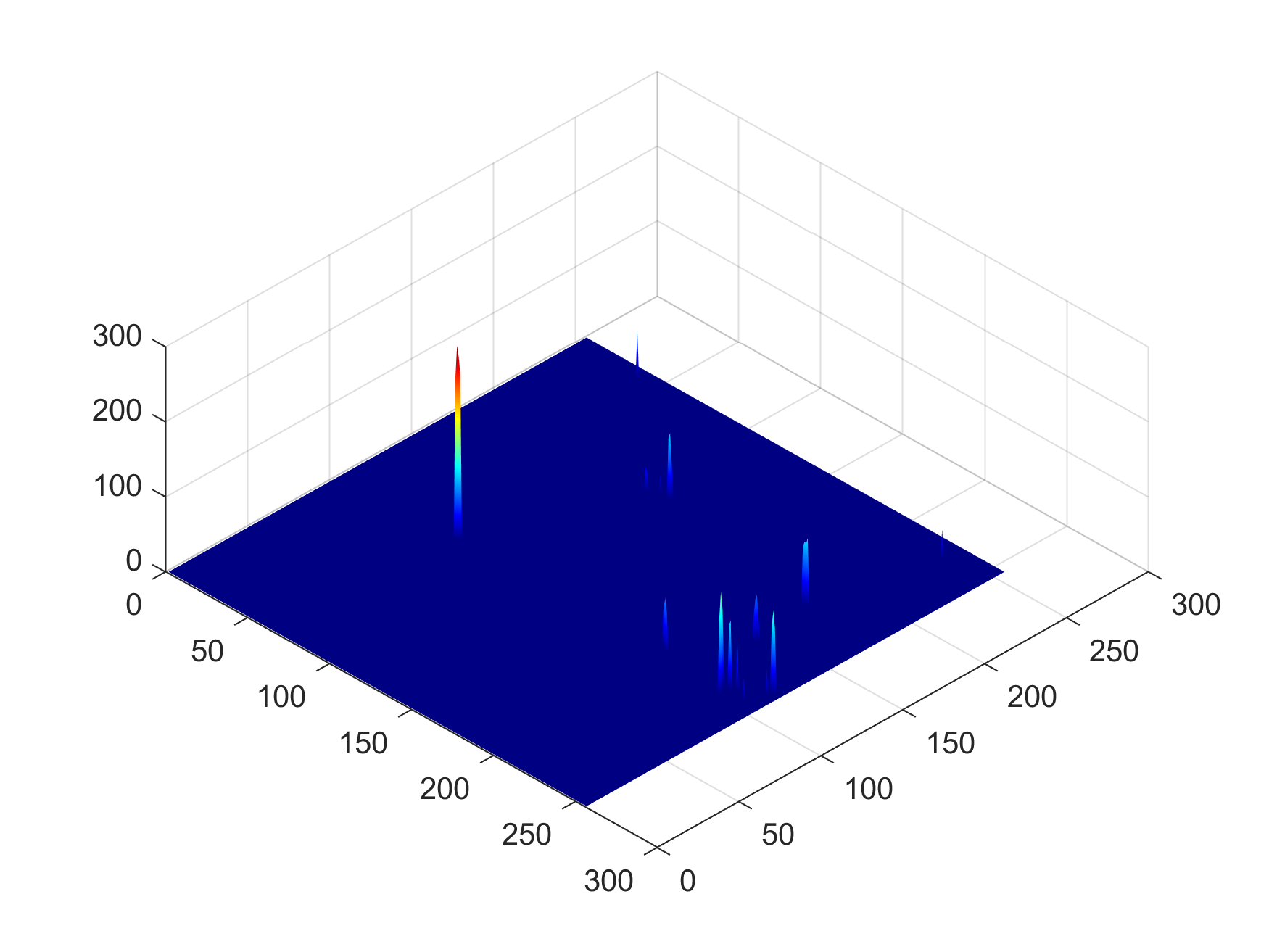} &
\includegraphics[width=0.115\linewidth]{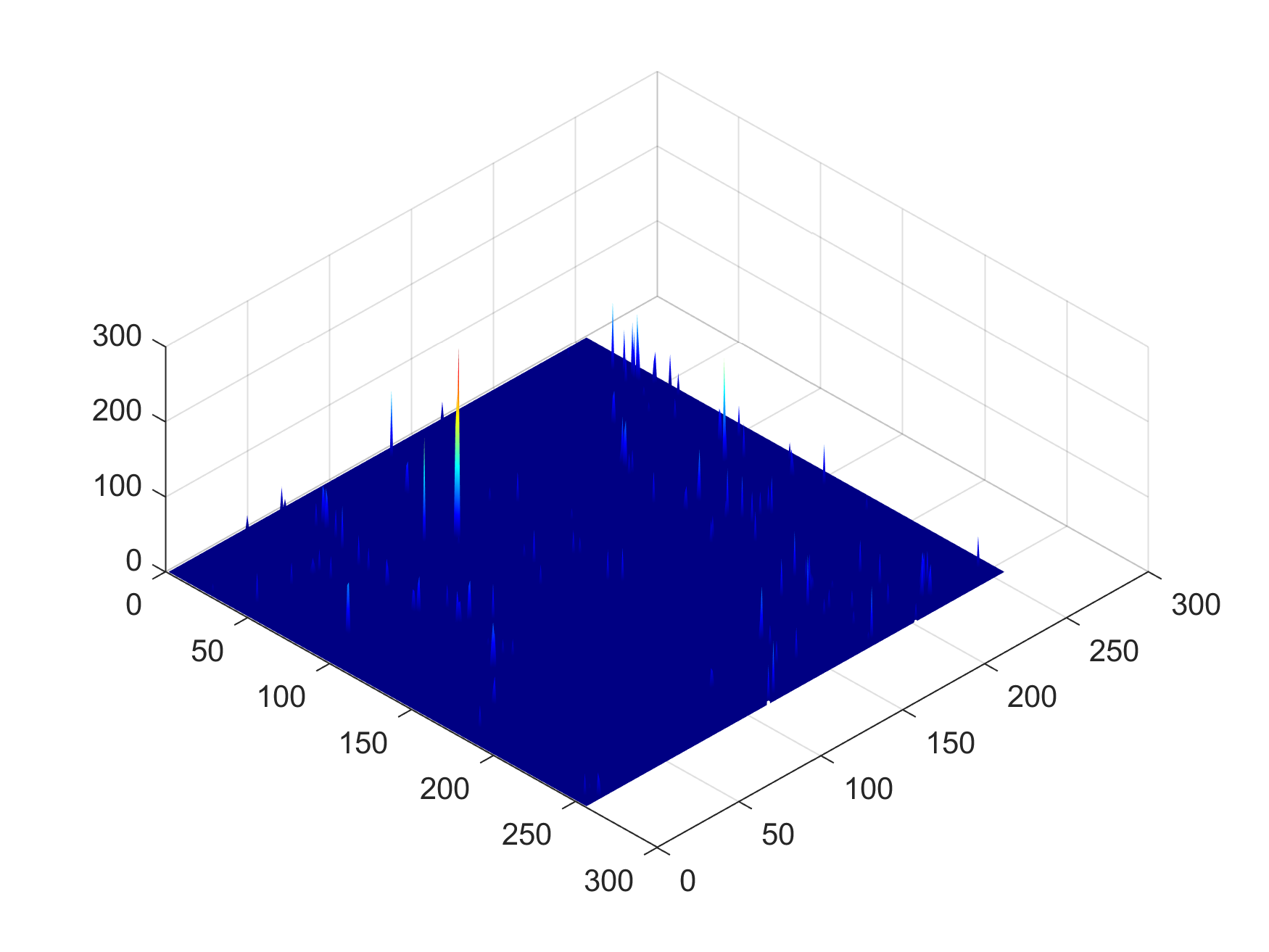} &
\includegraphics[width=0.115\linewidth]{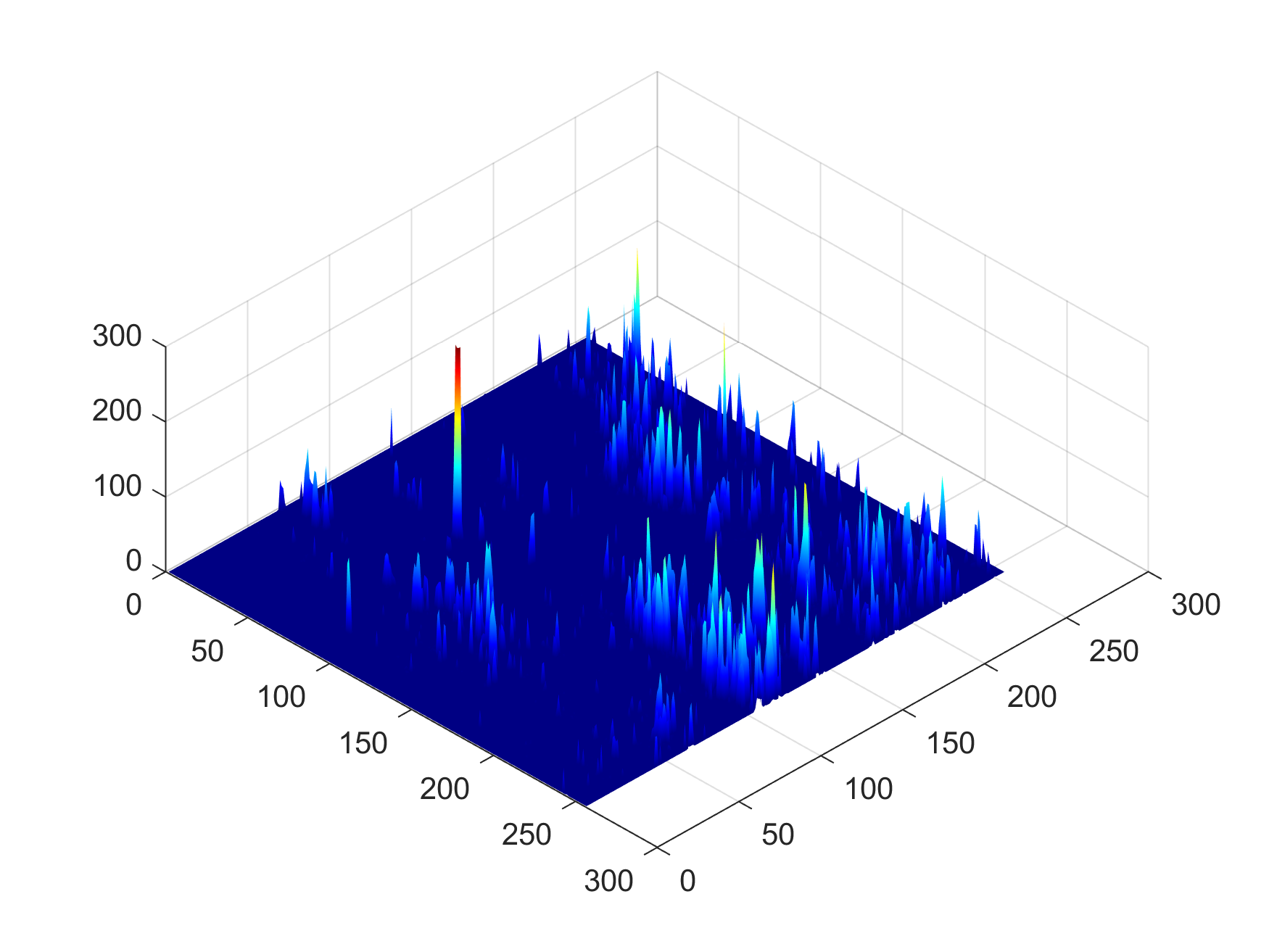} &
\includegraphics[width=0.115\linewidth]{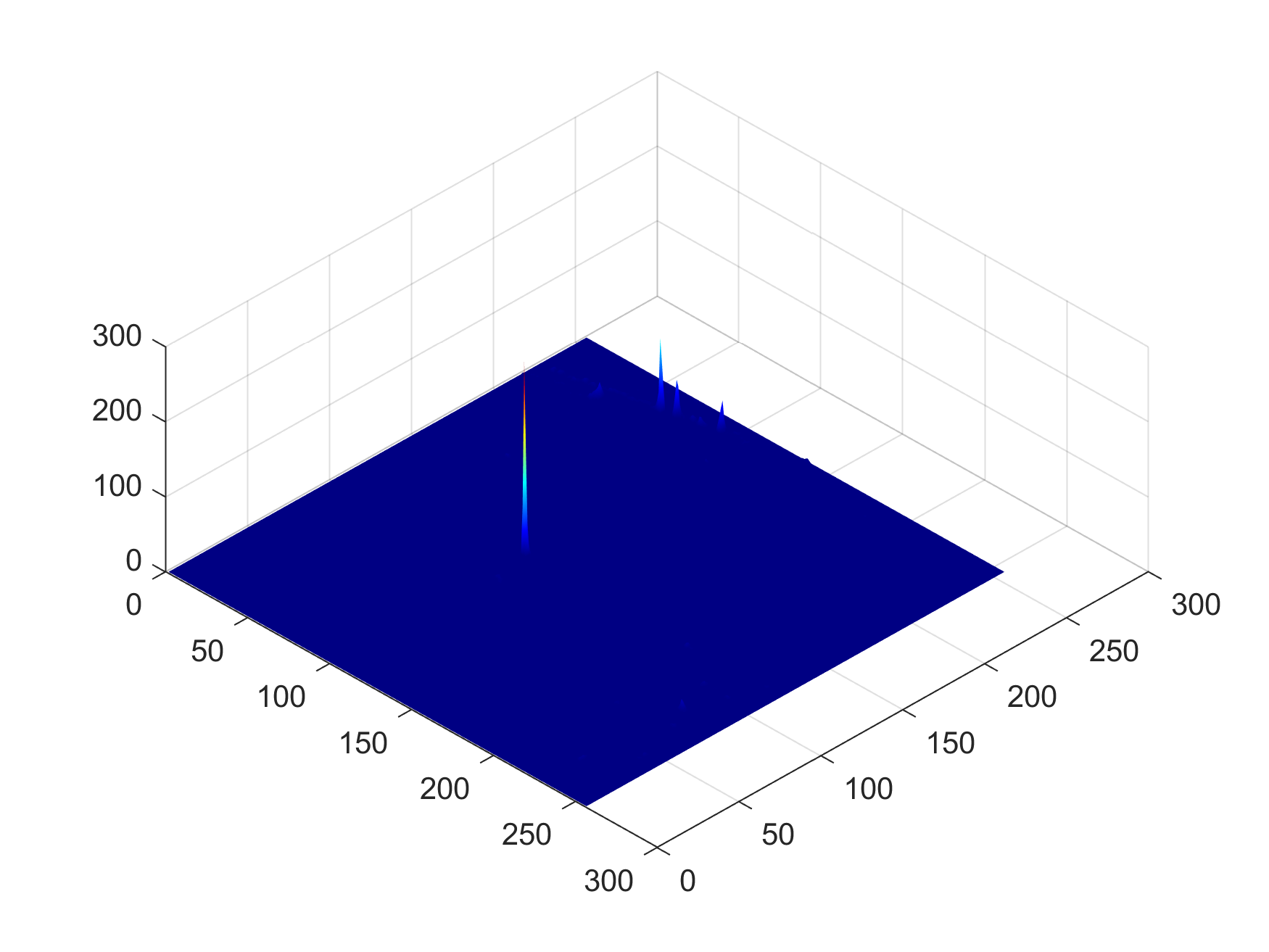} &
\includegraphics[width=0.115\linewidth]{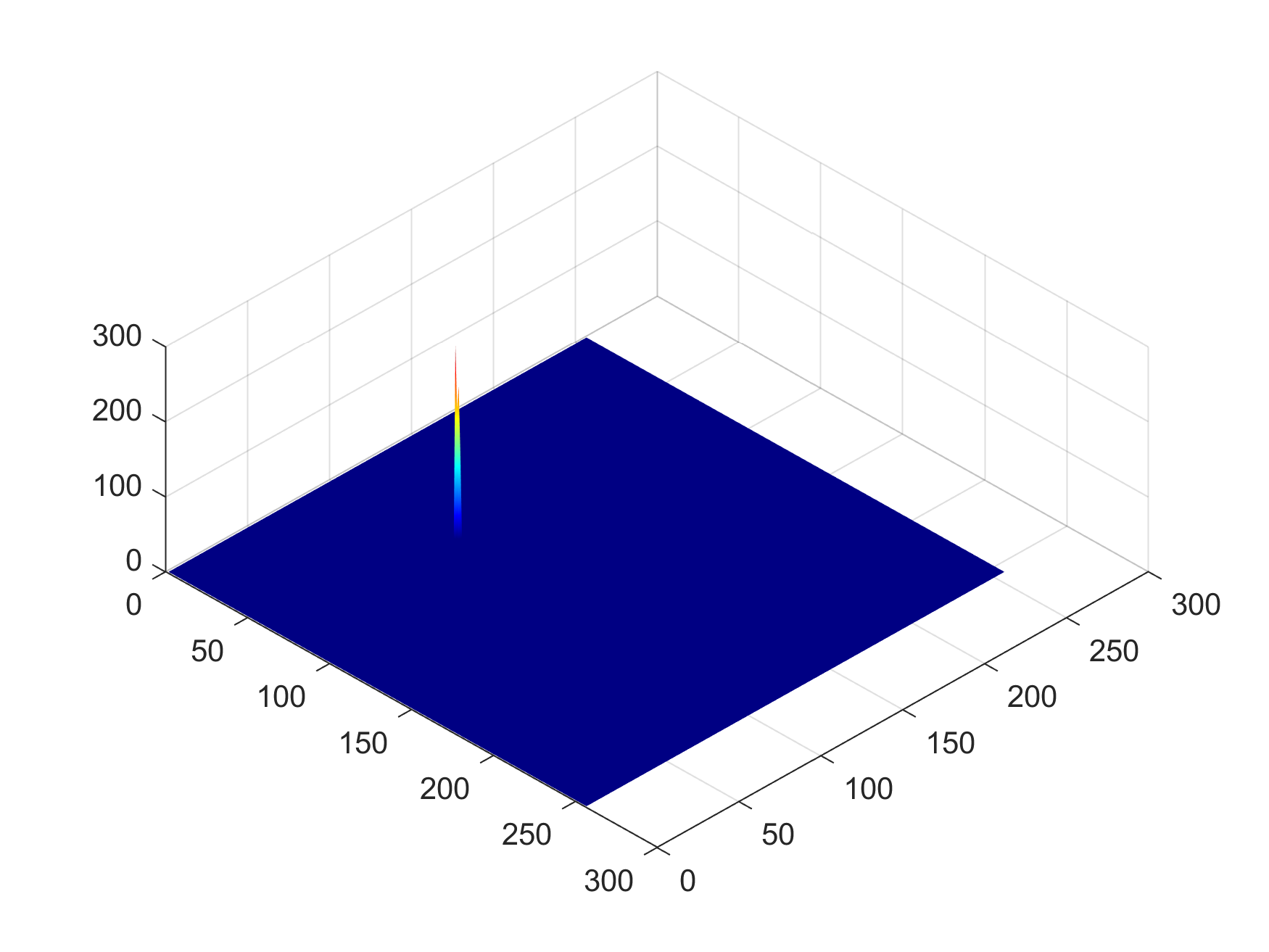} &
\includegraphics[width=0.115\linewidth]{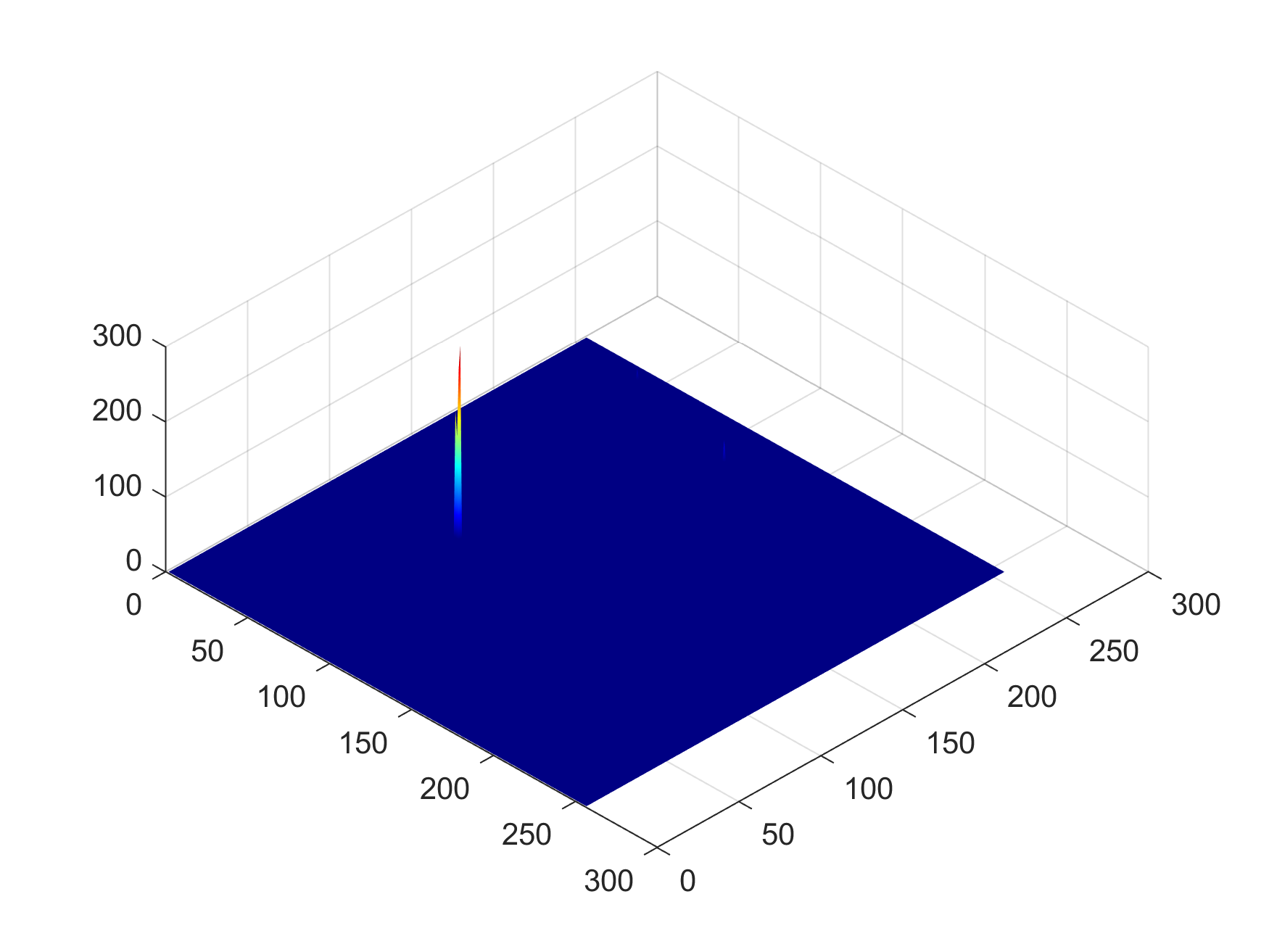} \\[-4pt]

\includegraphics[width=0.115\linewidth]{25.png} &
\includegraphics[width=0.115\linewidth]{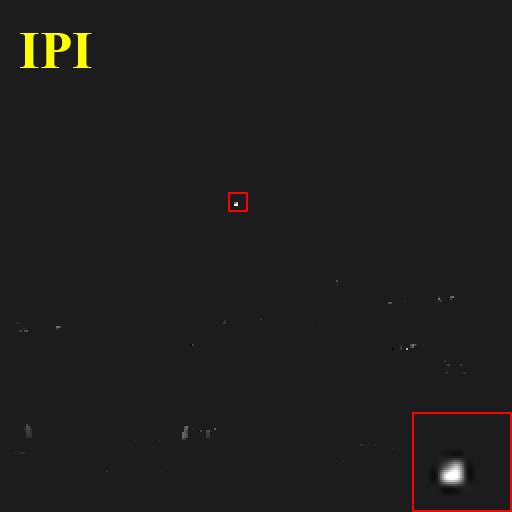} &
\includegraphics[width=0.115\linewidth]{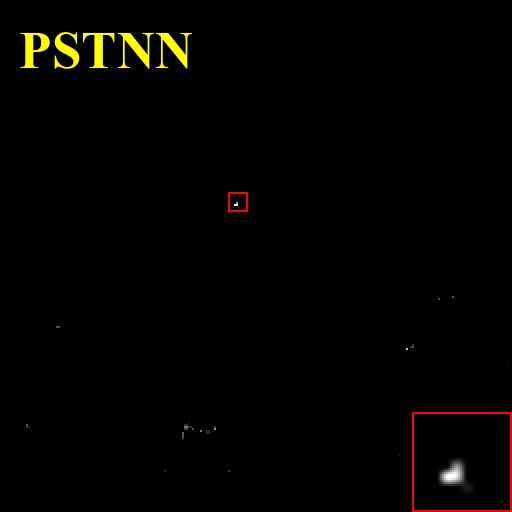} &
\includegraphics[width=0.115\linewidth]{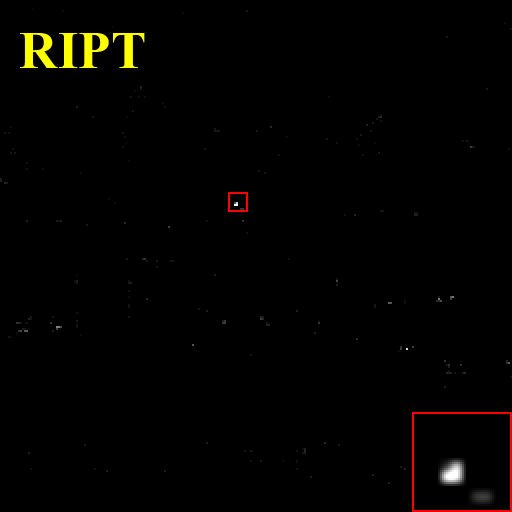} &
\includegraphics[width=0.115\linewidth]{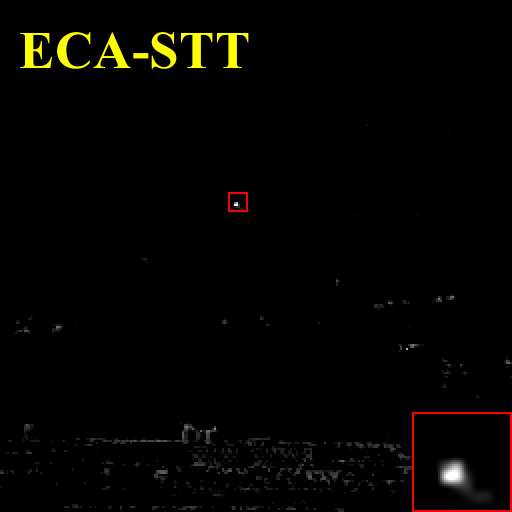} &
\includegraphics[width=0.115\linewidth]{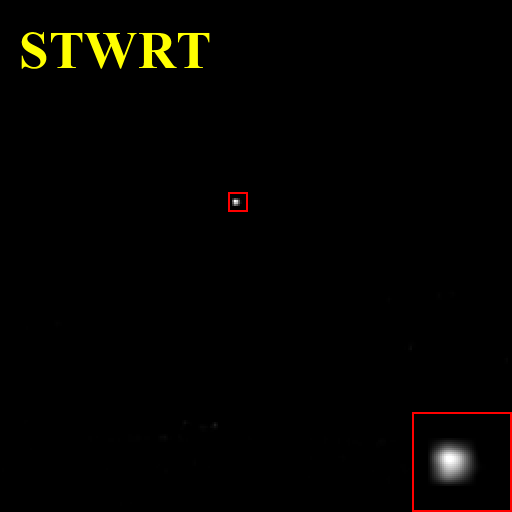} &
\includegraphics[width=0.115\linewidth]{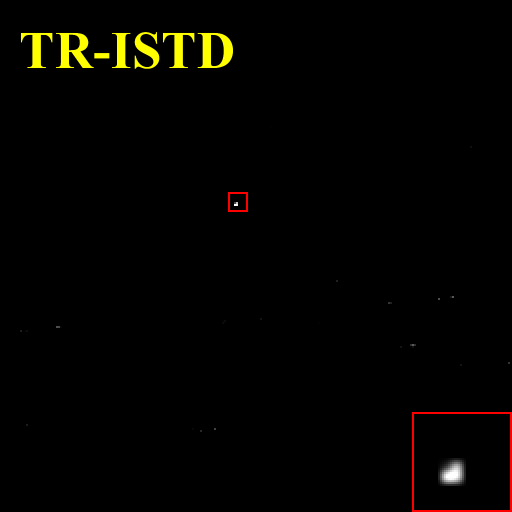} &
\includegraphics[width=0.115\linewidth]{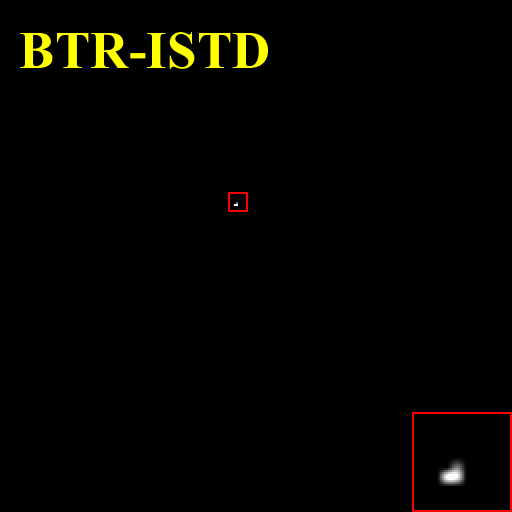}\\[-4pt]

\includegraphics[width=0.115\linewidth]{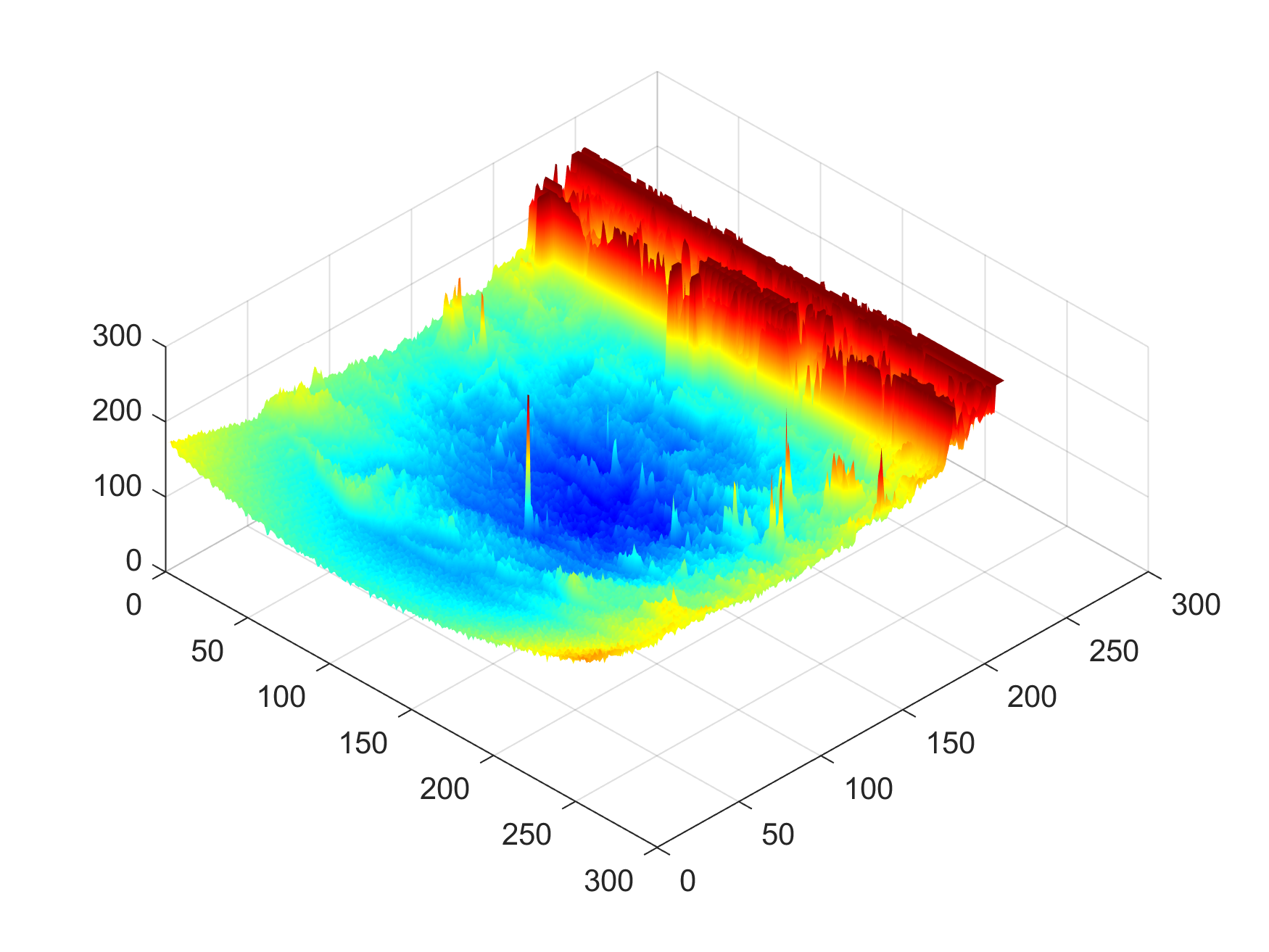} &
\includegraphics[width=0.115\linewidth]{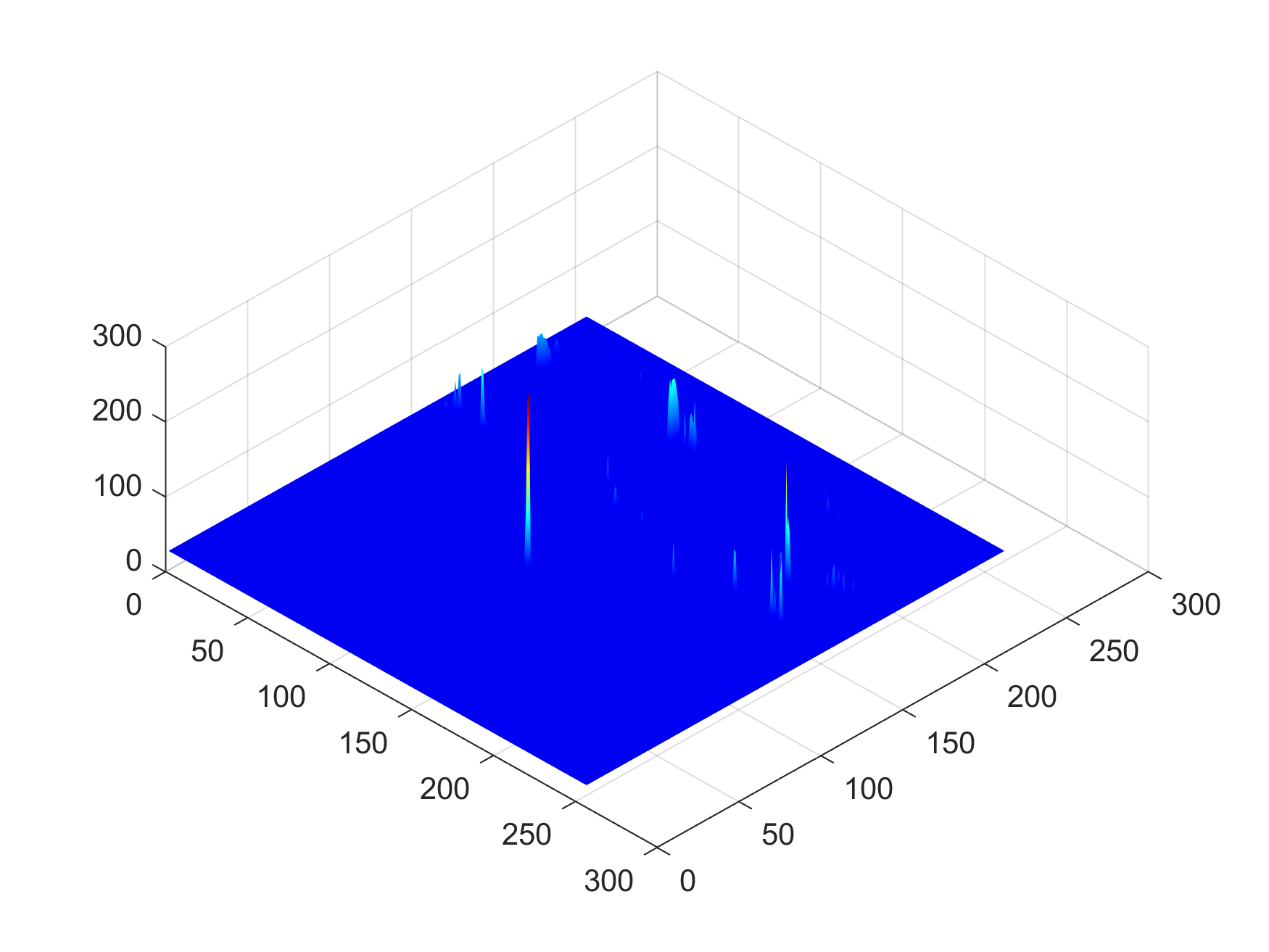} &
\includegraphics[width=0.115\linewidth]{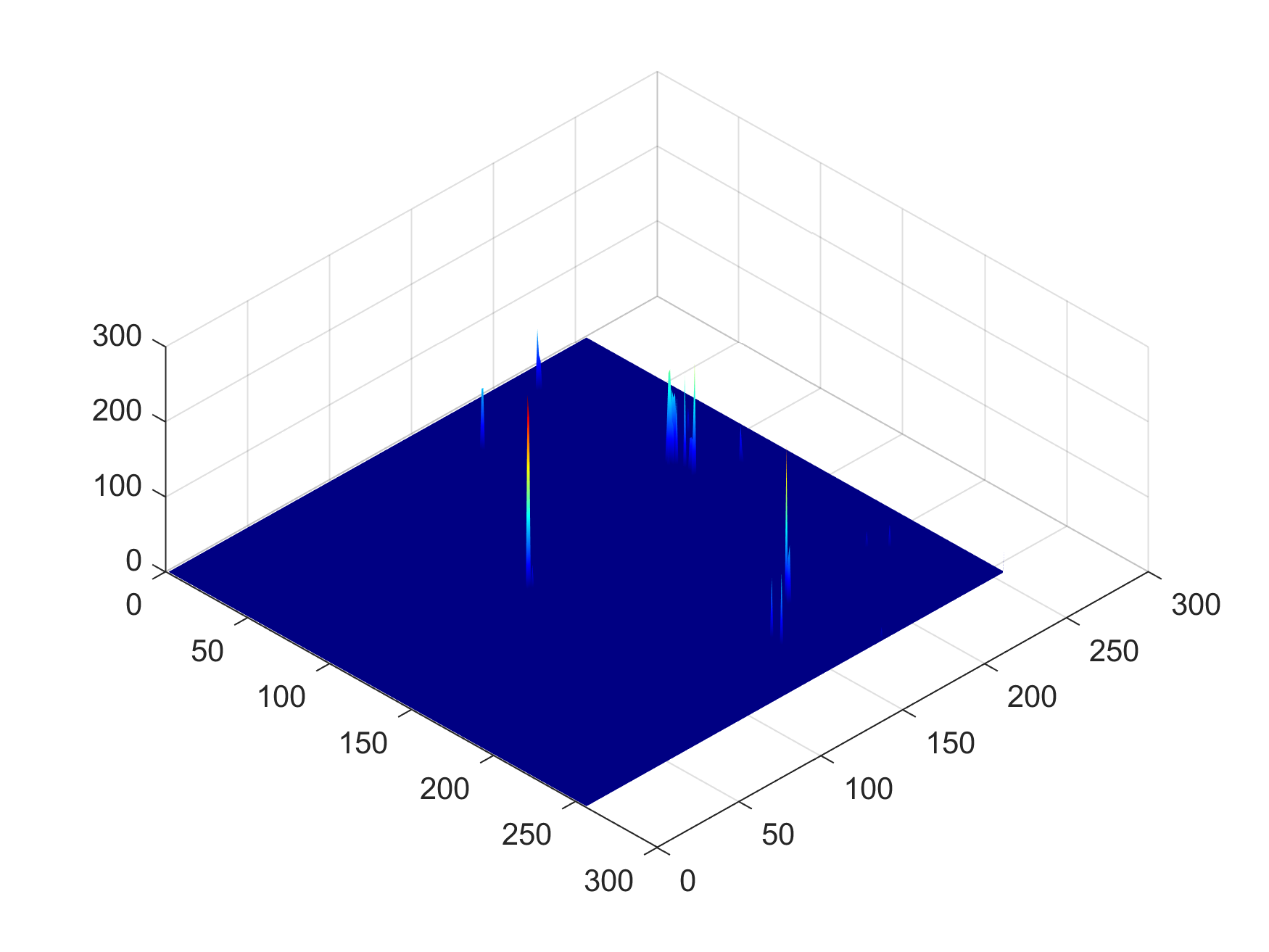} &
\includegraphics[width=0.115\linewidth]{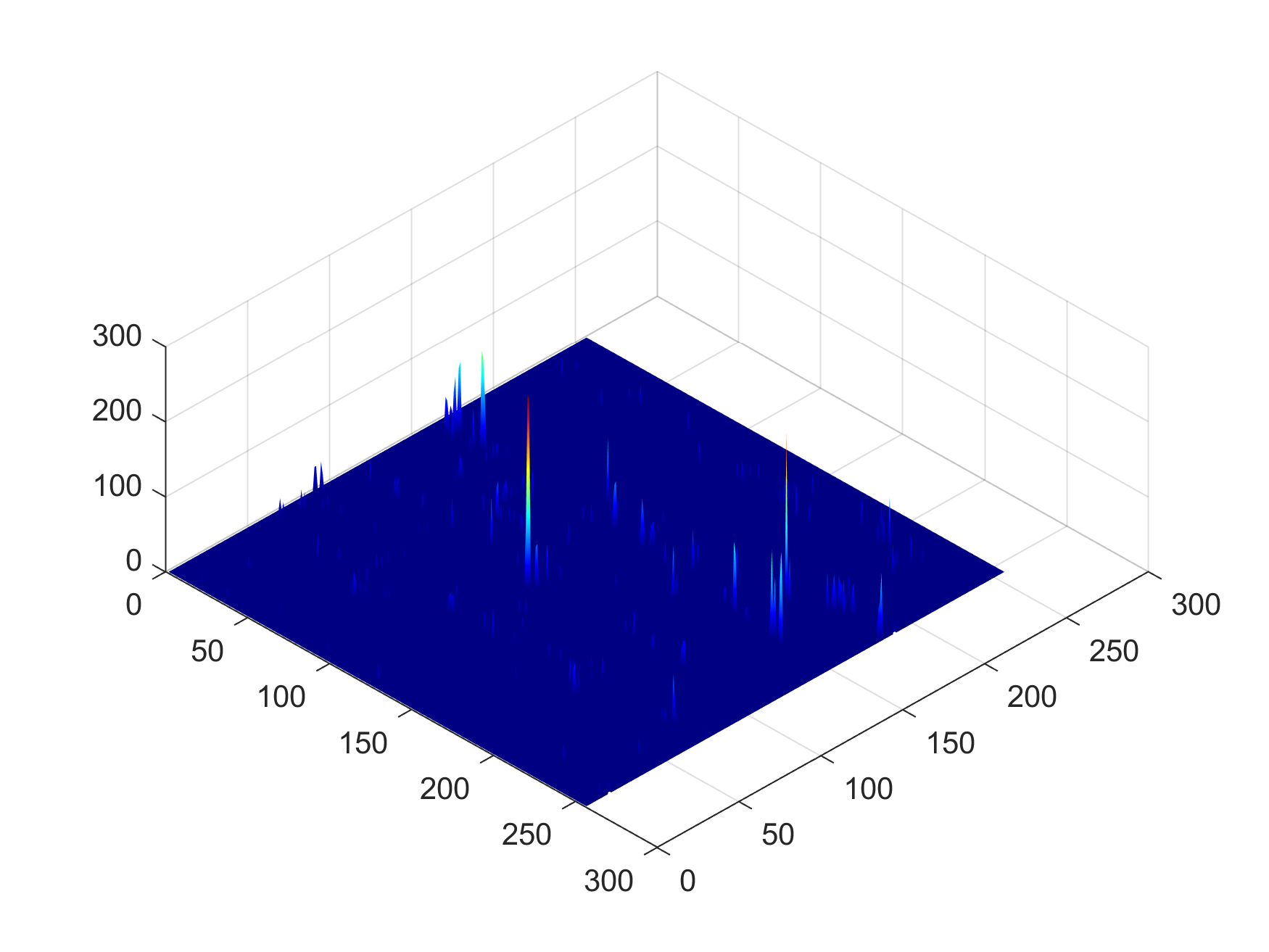} &
\includegraphics[width=0.115\linewidth]{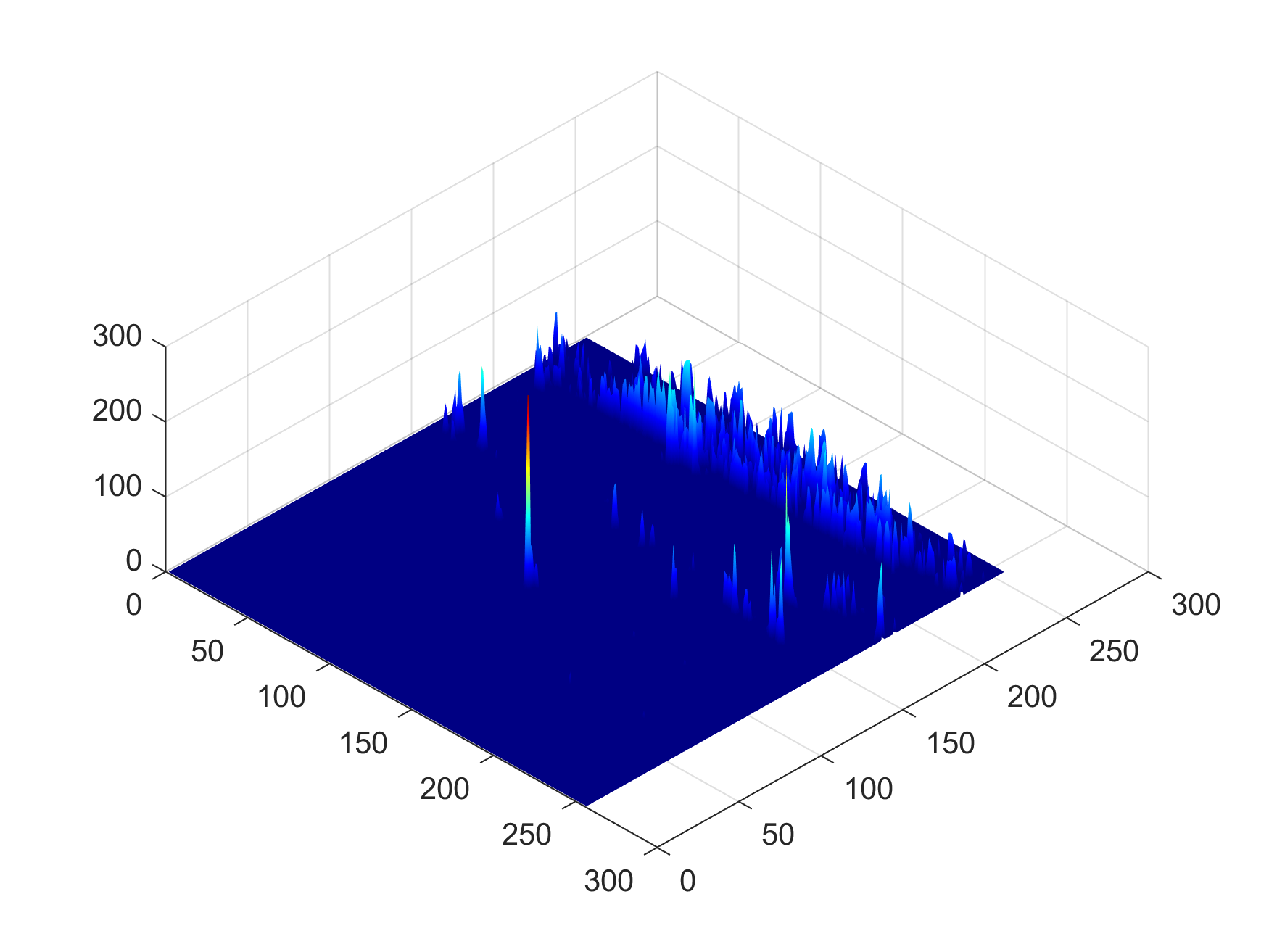} &
\includegraphics[width=0.115\linewidth]{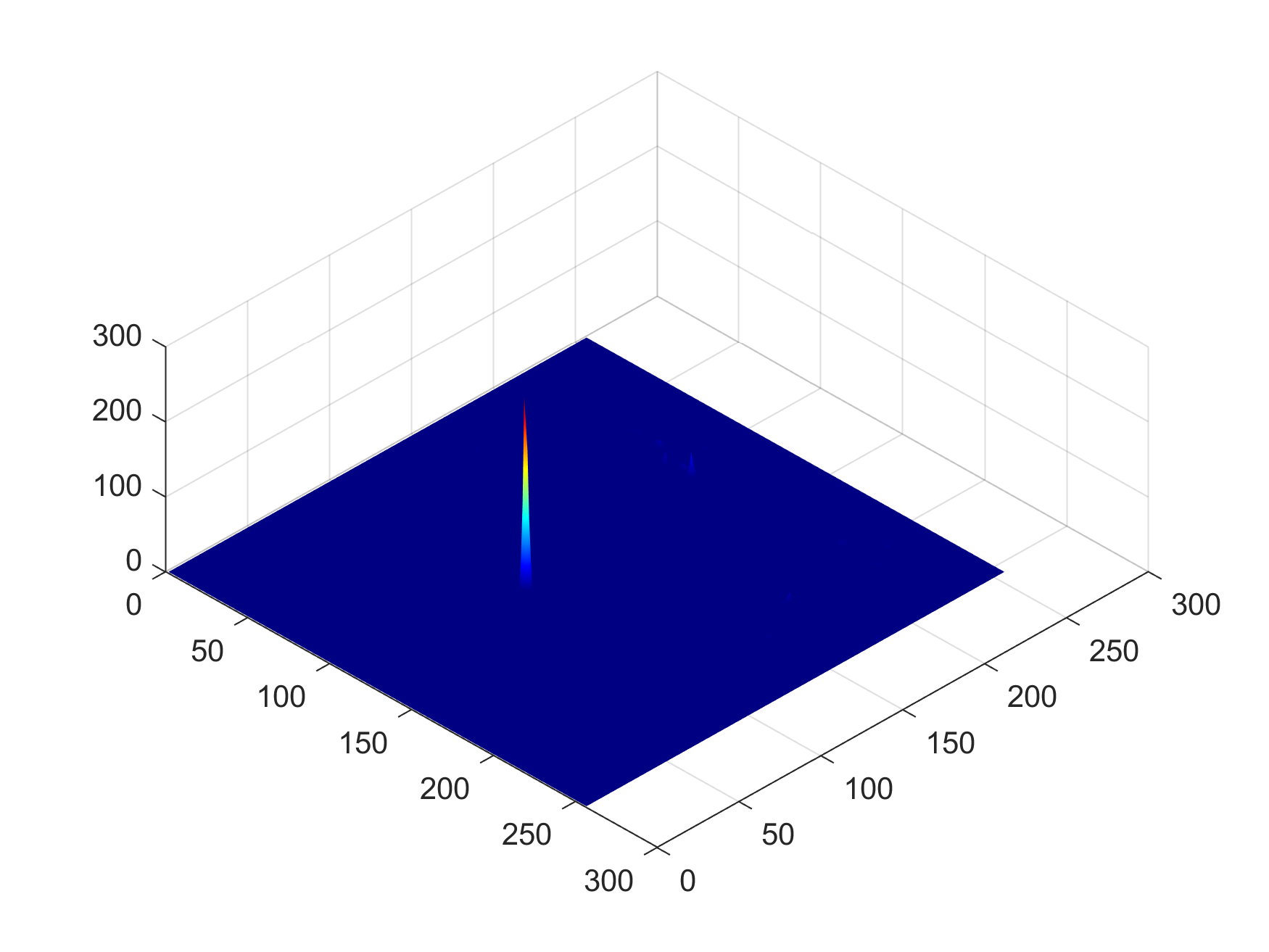} &
\includegraphics[width=0.115\linewidth]{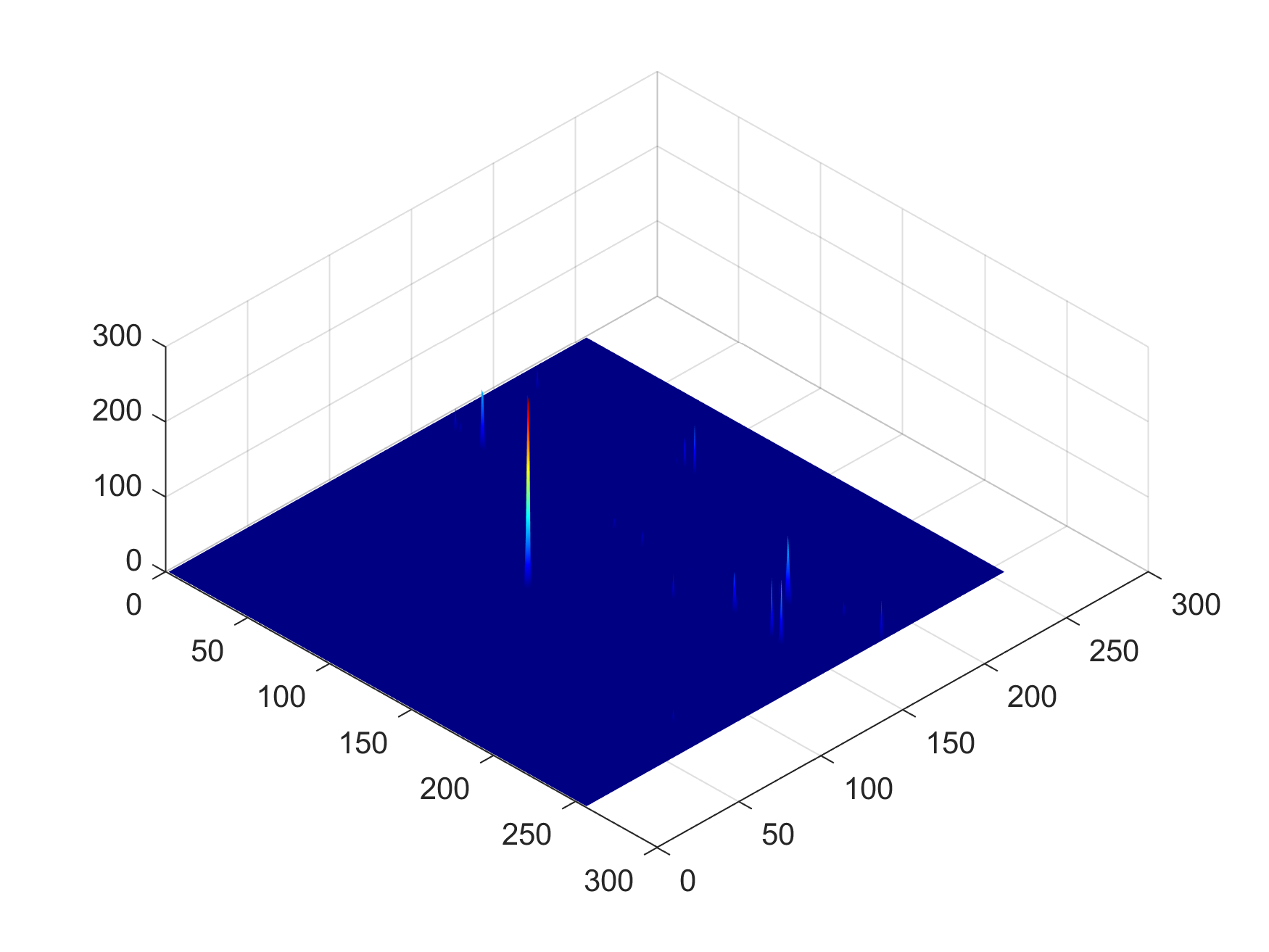} &
\includegraphics[width=0.115\linewidth]{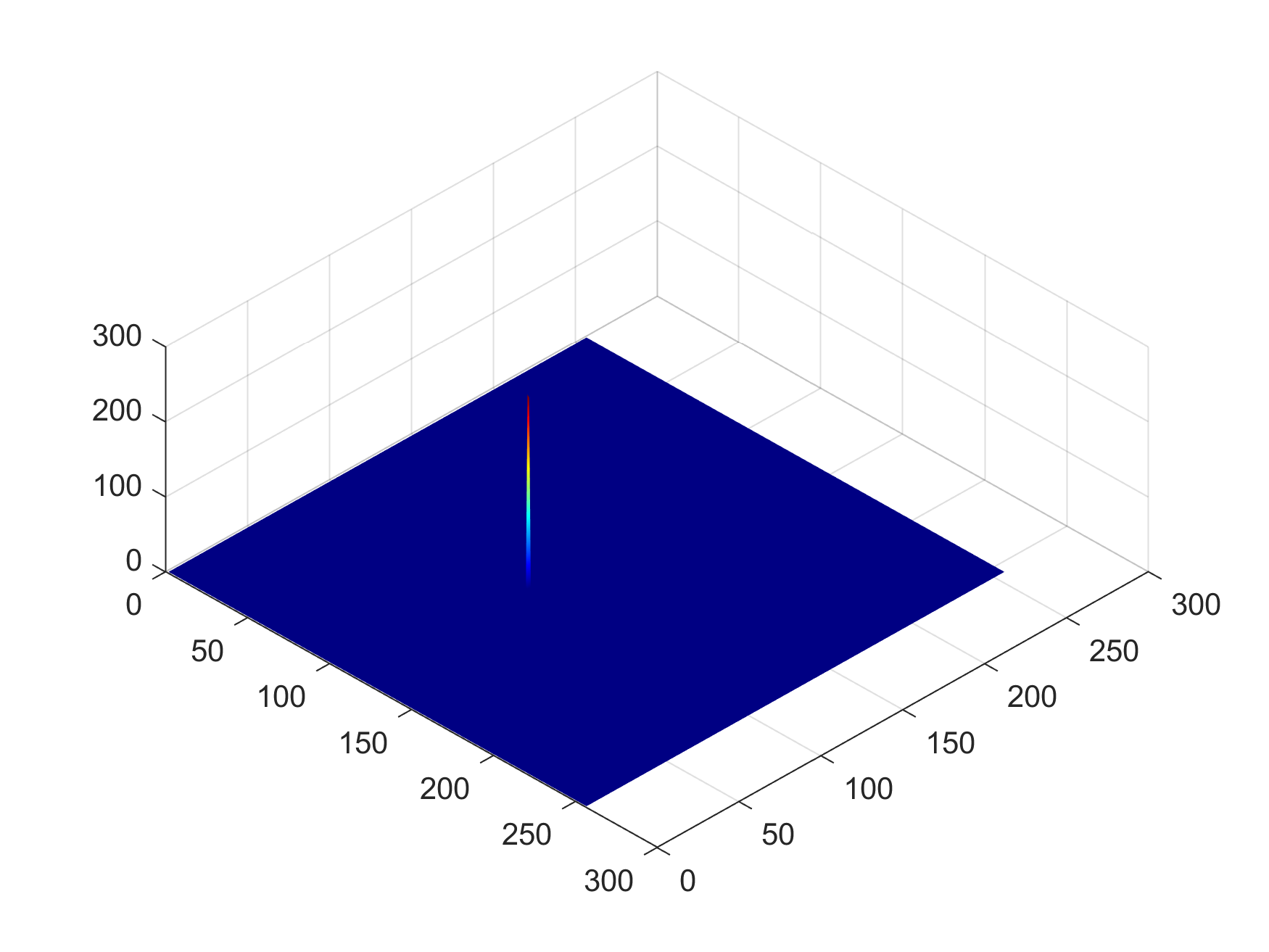}\\[-4pt]

\includegraphics[width=0.115\linewidth]{266.png} &
\includegraphics[width=0.115\linewidth]{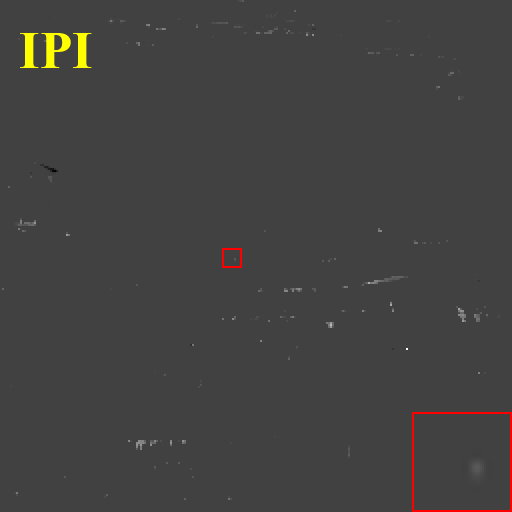} &
\includegraphics[width=0.115\linewidth]{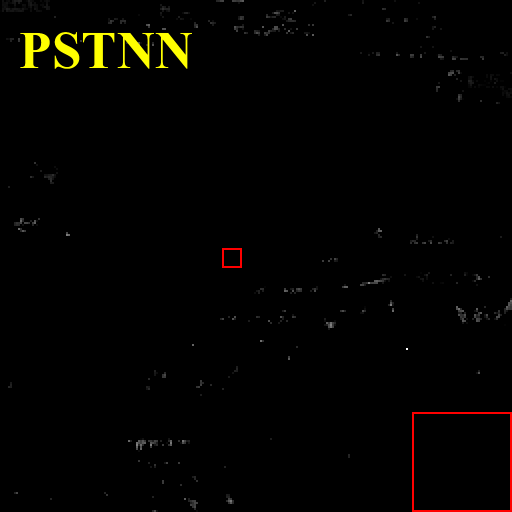} &
\includegraphics[width=0.115\linewidth]{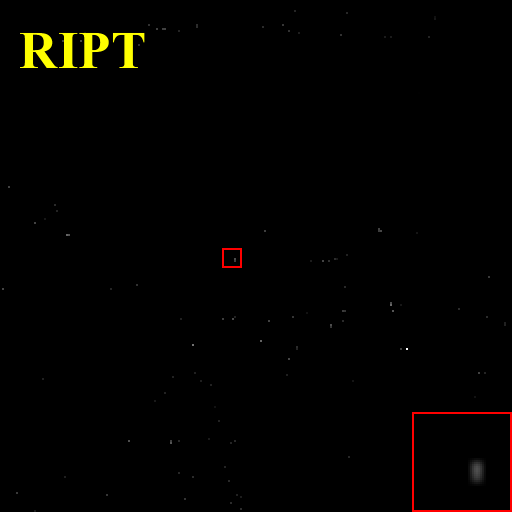} &
\includegraphics[width=0.115\linewidth]{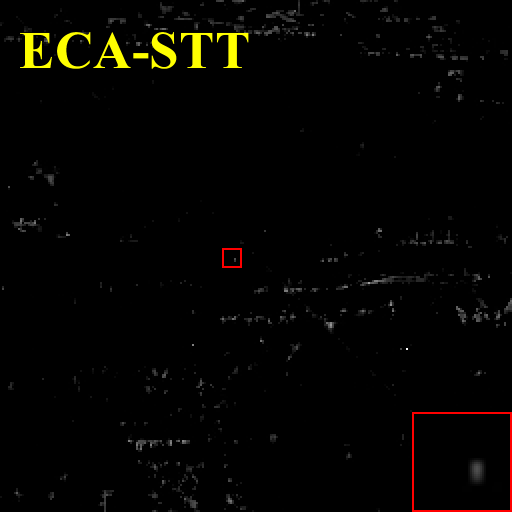} &
\includegraphics[width=0.115\linewidth]{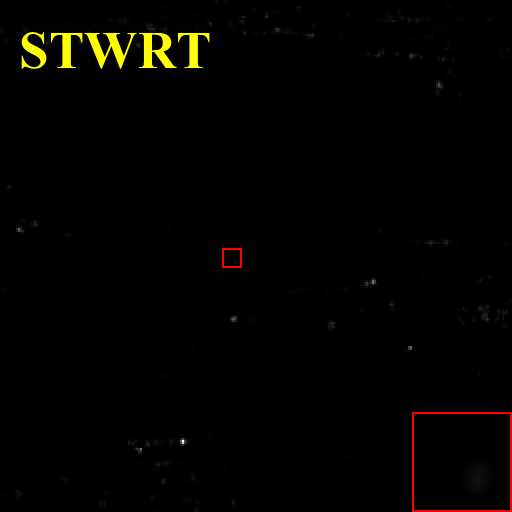} &
\includegraphics[width=0.115\linewidth]{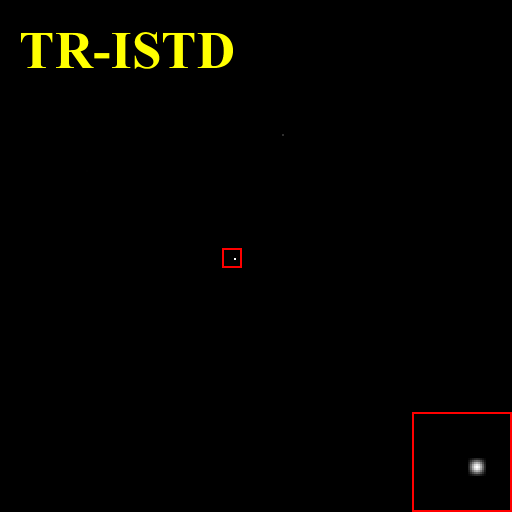} &
\includegraphics[width=0.115\linewidth]{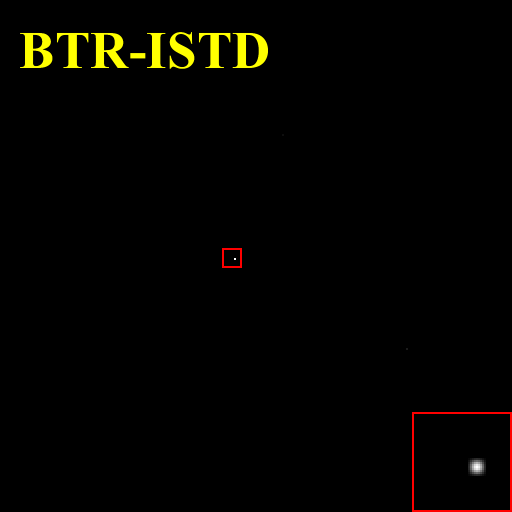} \\[-4pt]

\includegraphics[width=0.115\linewidth]{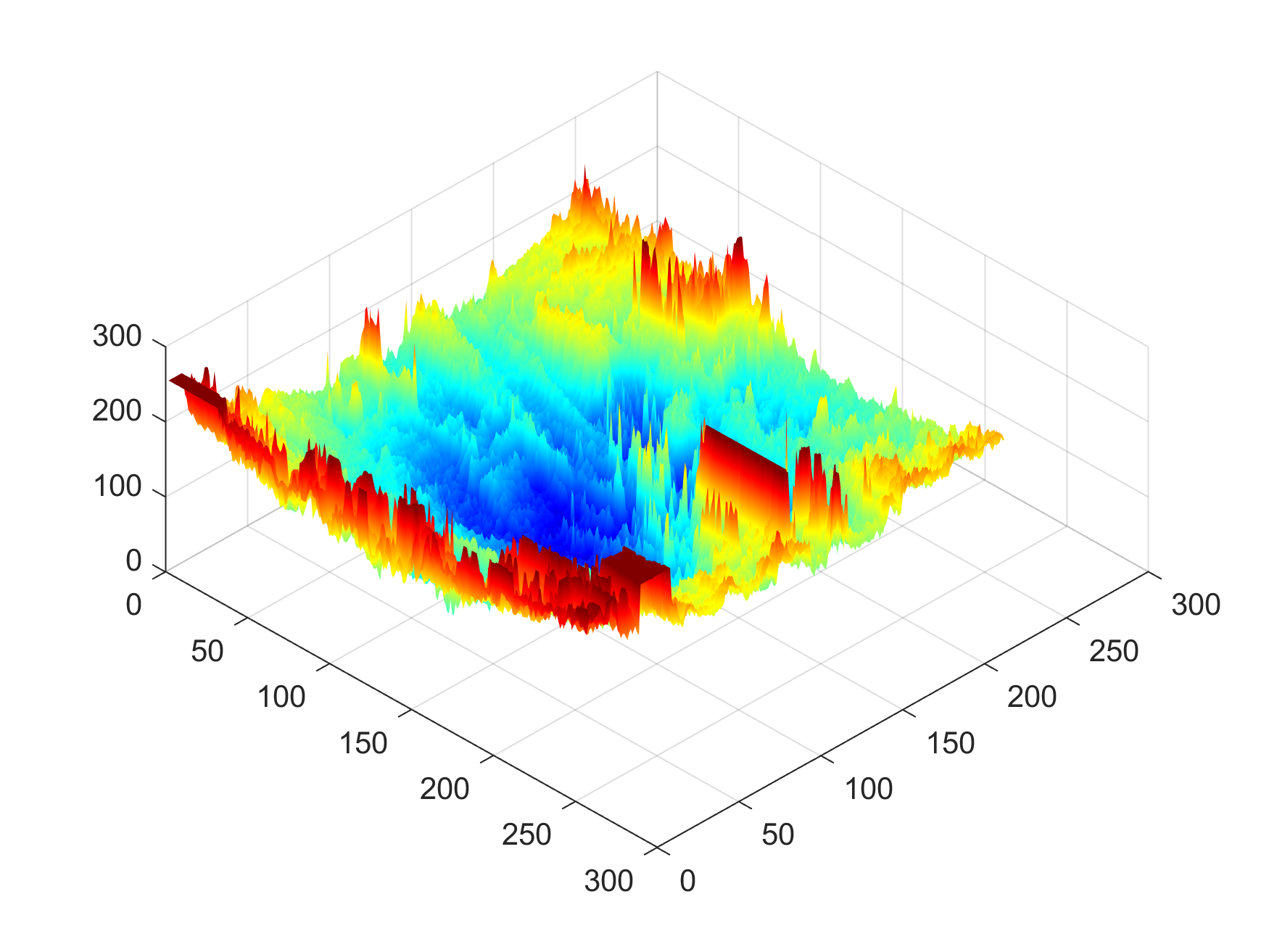} &
\includegraphics[width=0.115\linewidth]{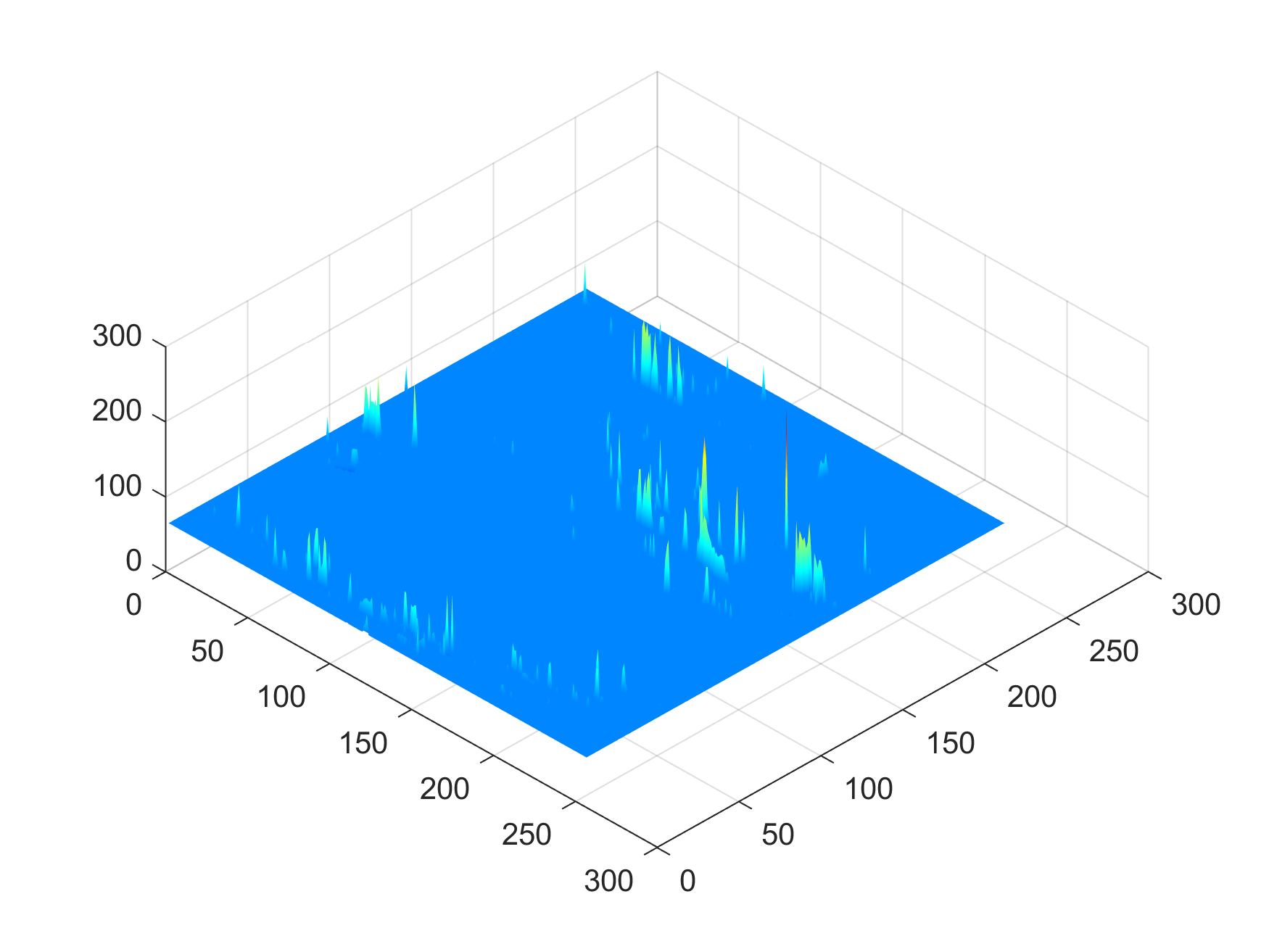} &
\includegraphics[width=0.115\linewidth]{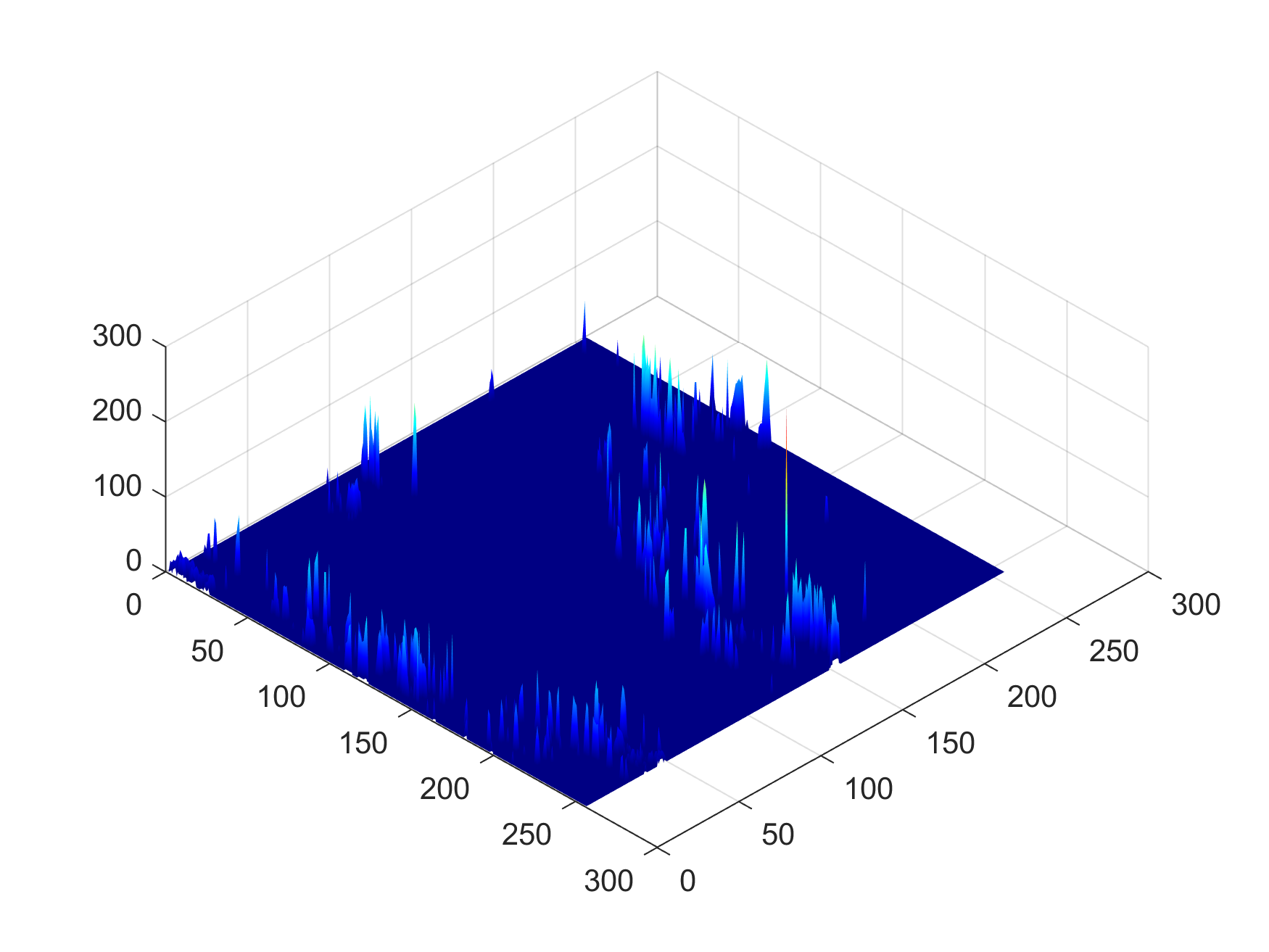} &
\includegraphics[width=0.115\linewidth]{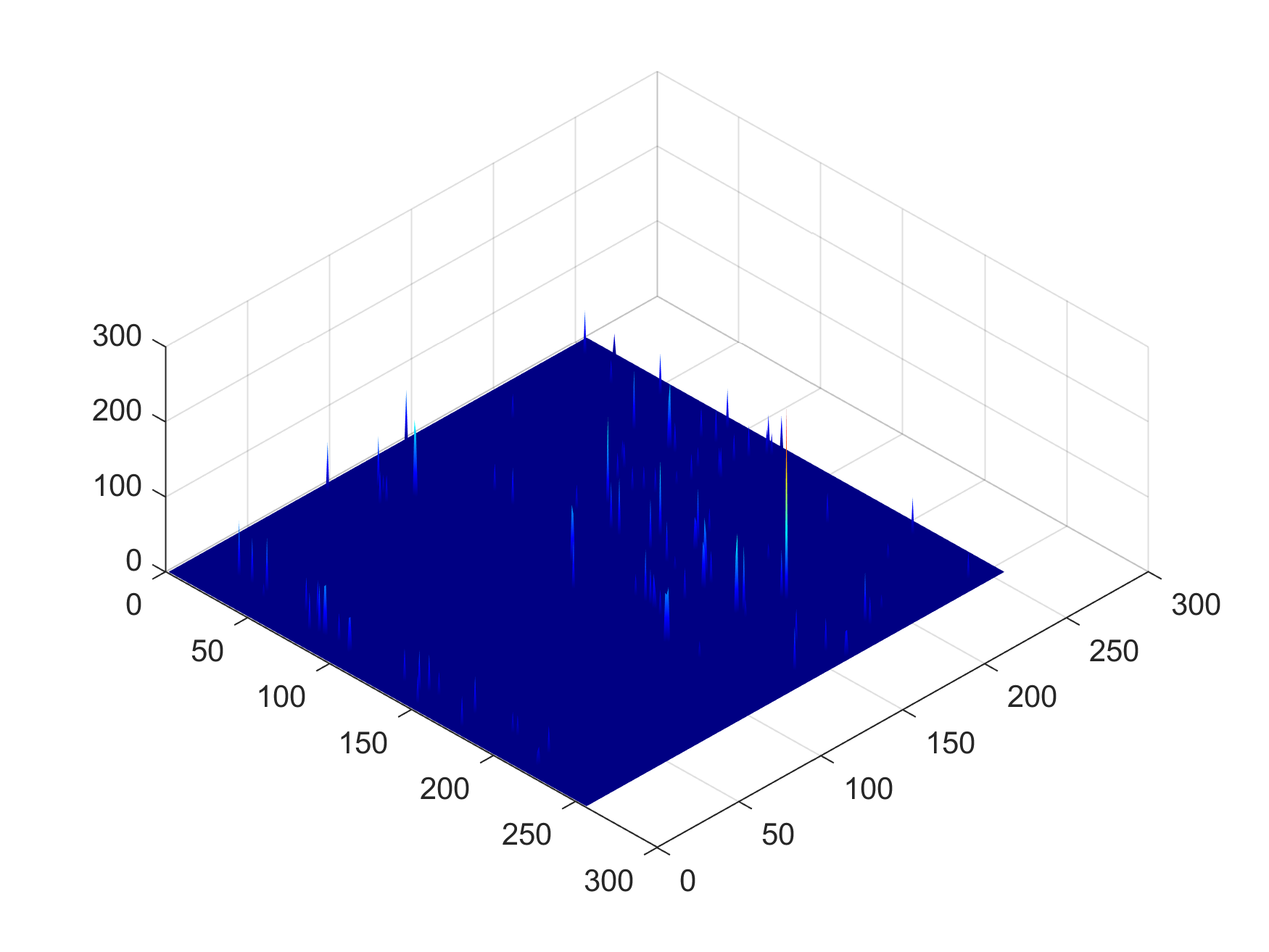} &
\includegraphics[width=0.115\linewidth]{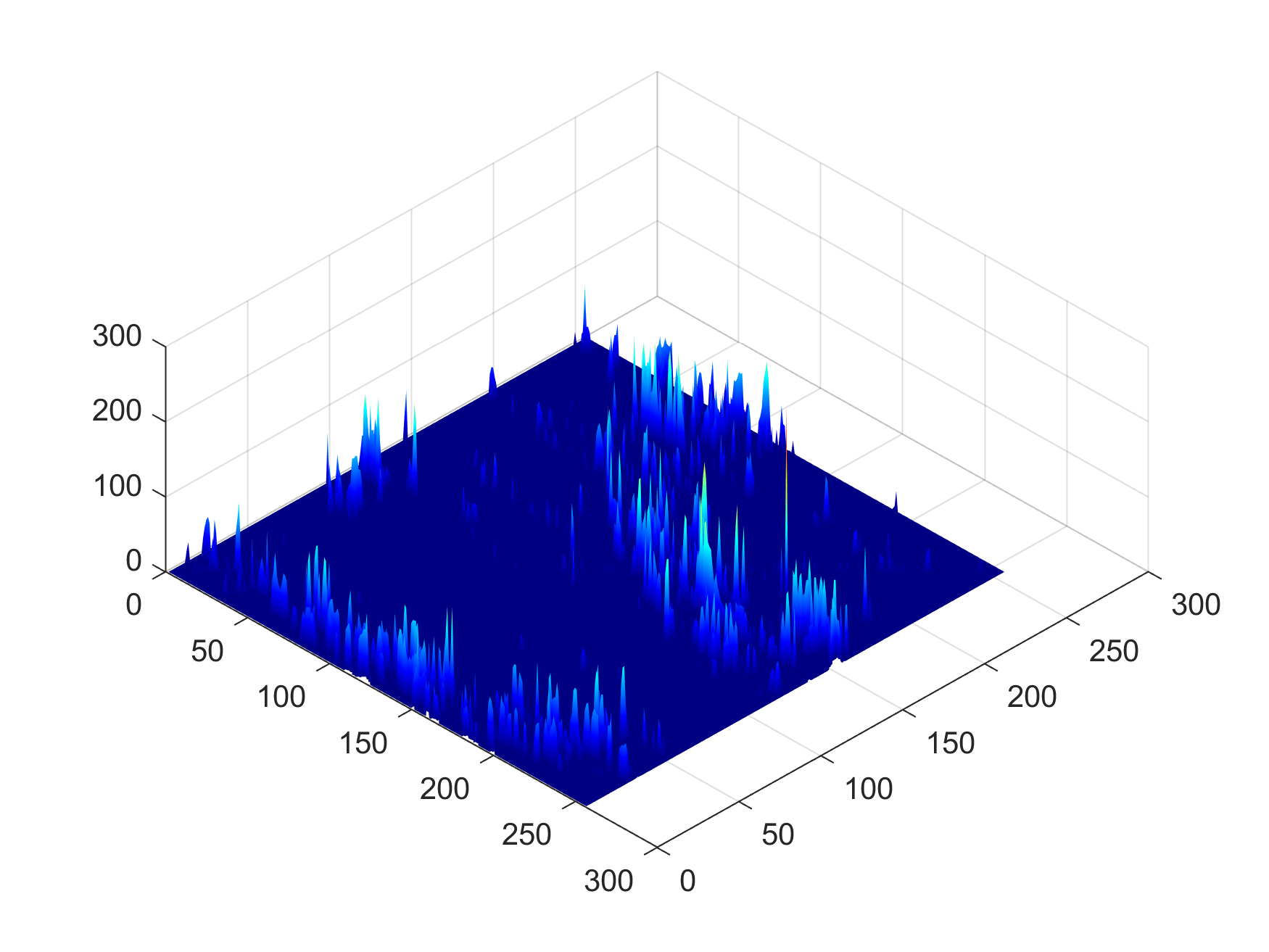} &
\includegraphics[width=0.115\linewidth]{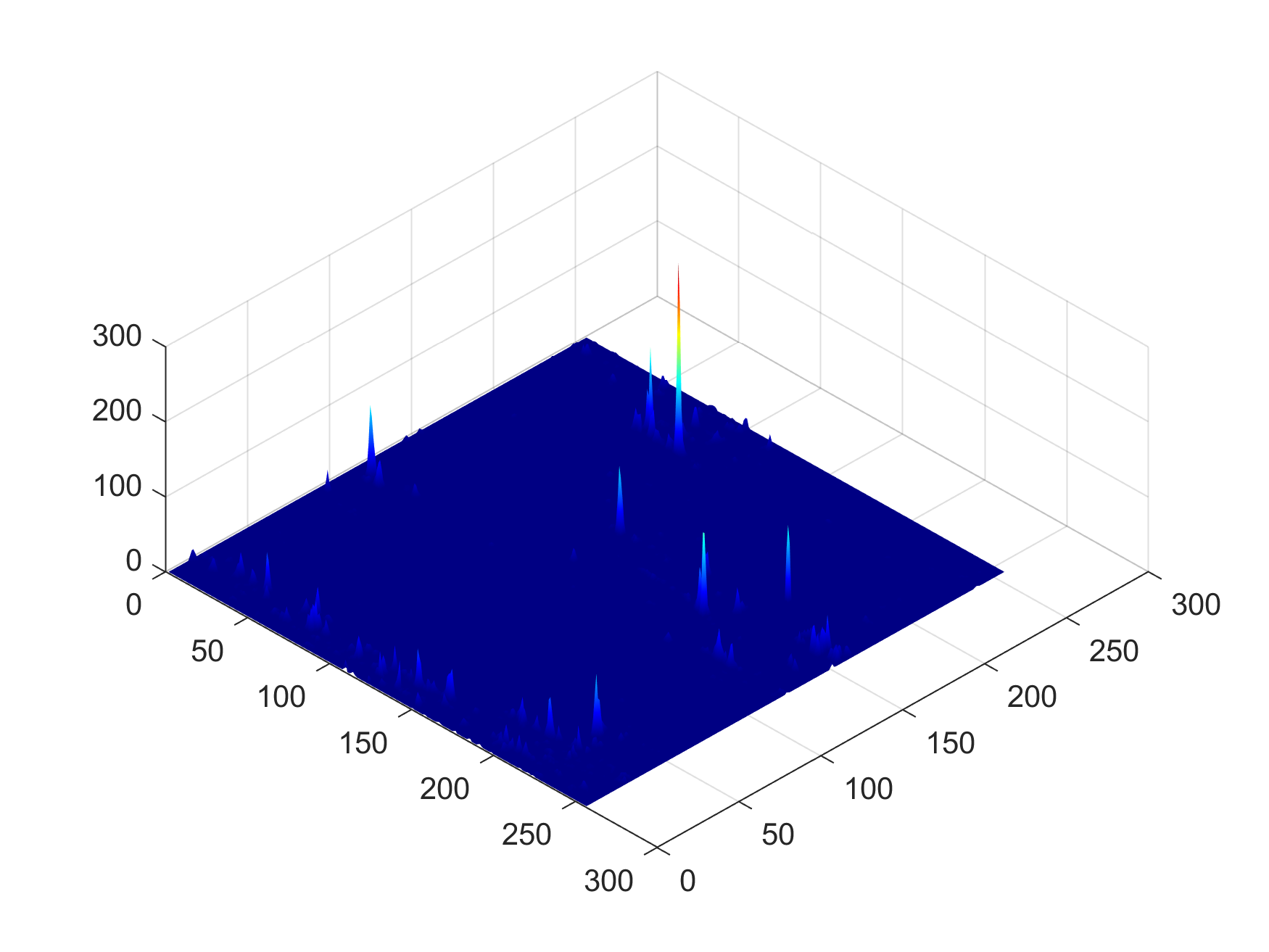} &
\includegraphics[width=0.115\linewidth]{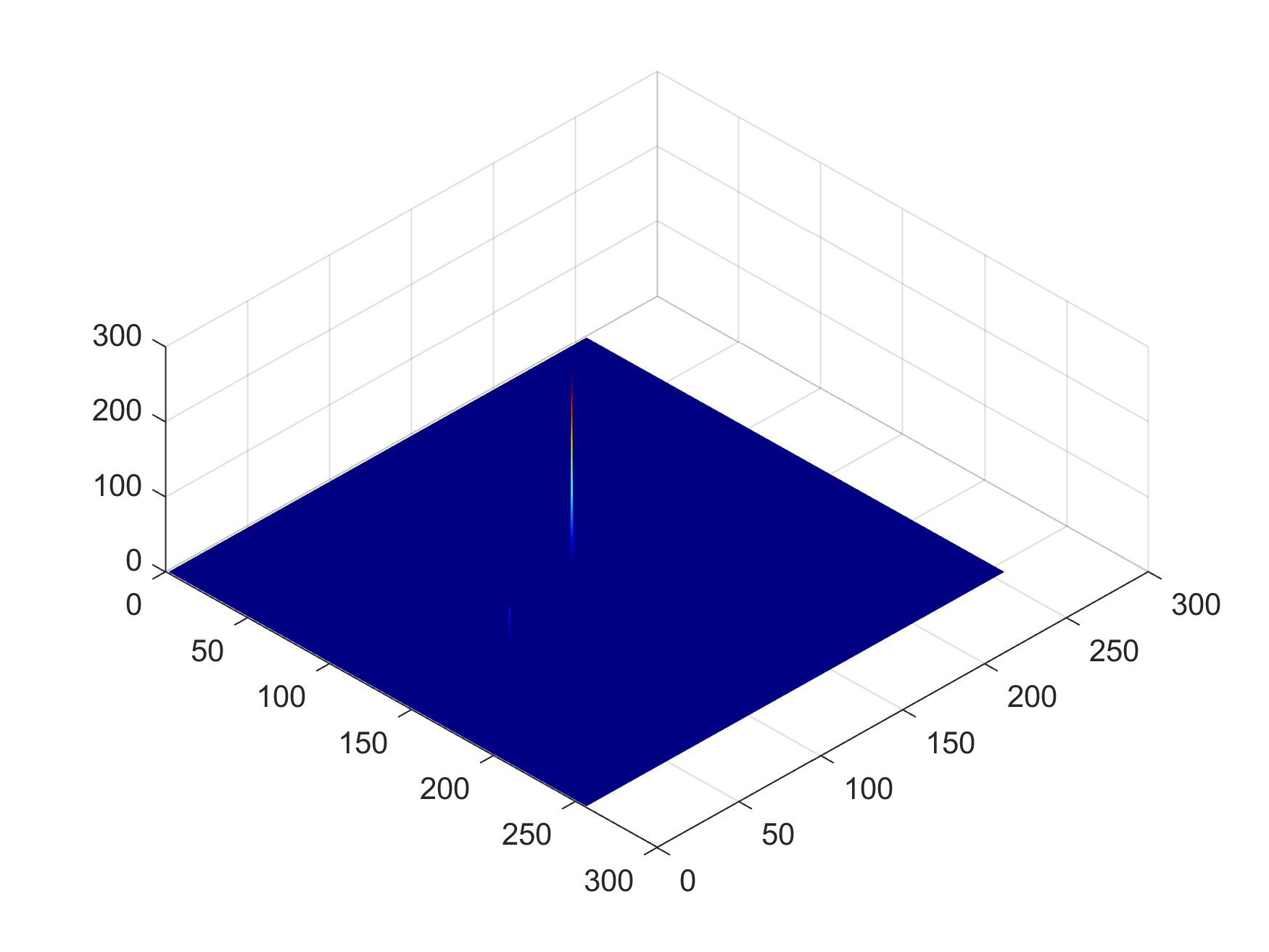} &
\includegraphics[width=0.115\linewidth]{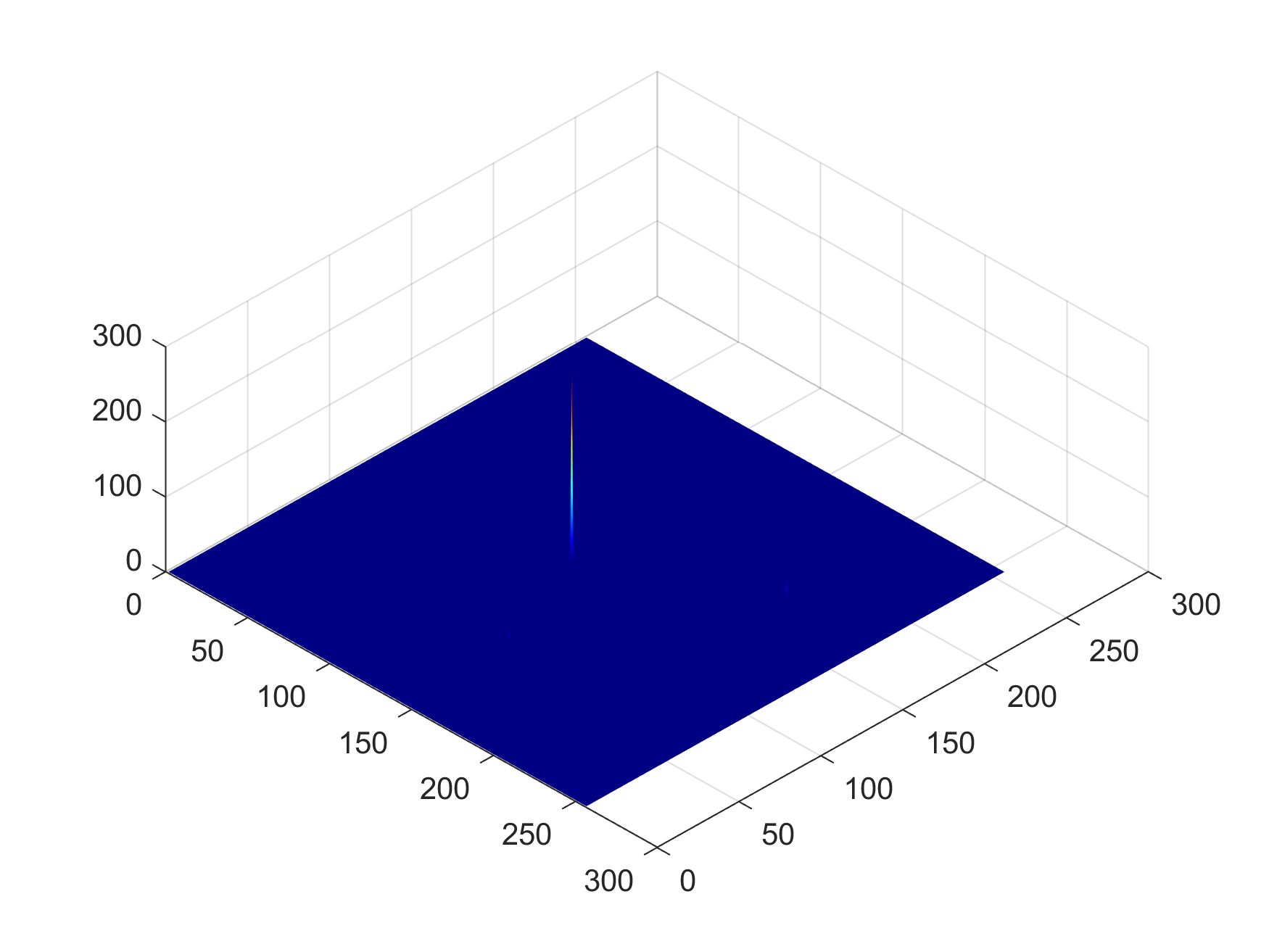} \\
\end{tabular}
\caption{Qualitative comparison and 3-D intuitive visualization of the seven competing methods for sequences 1–6. The target regions are enlarged for clarity.}
\label{fig:6x8_grid}
\end{figure*}

\begin{figure*}[!t]
\centering
\includegraphics[height=0.15\textheight,width=0.22\linewidth]{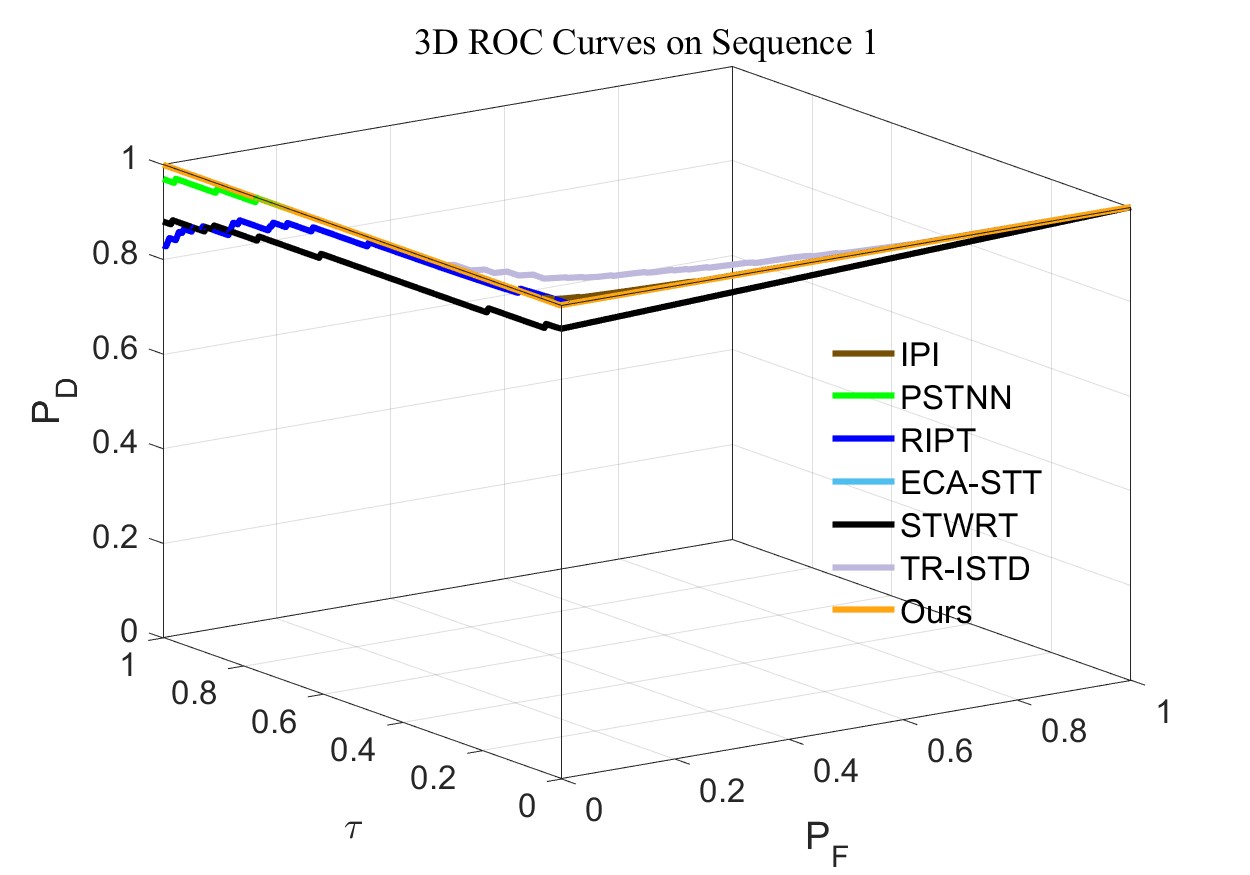}%
\hspace{0.01\linewidth}%
\includegraphics[height=0.15\textheight,width=0.22\linewidth]{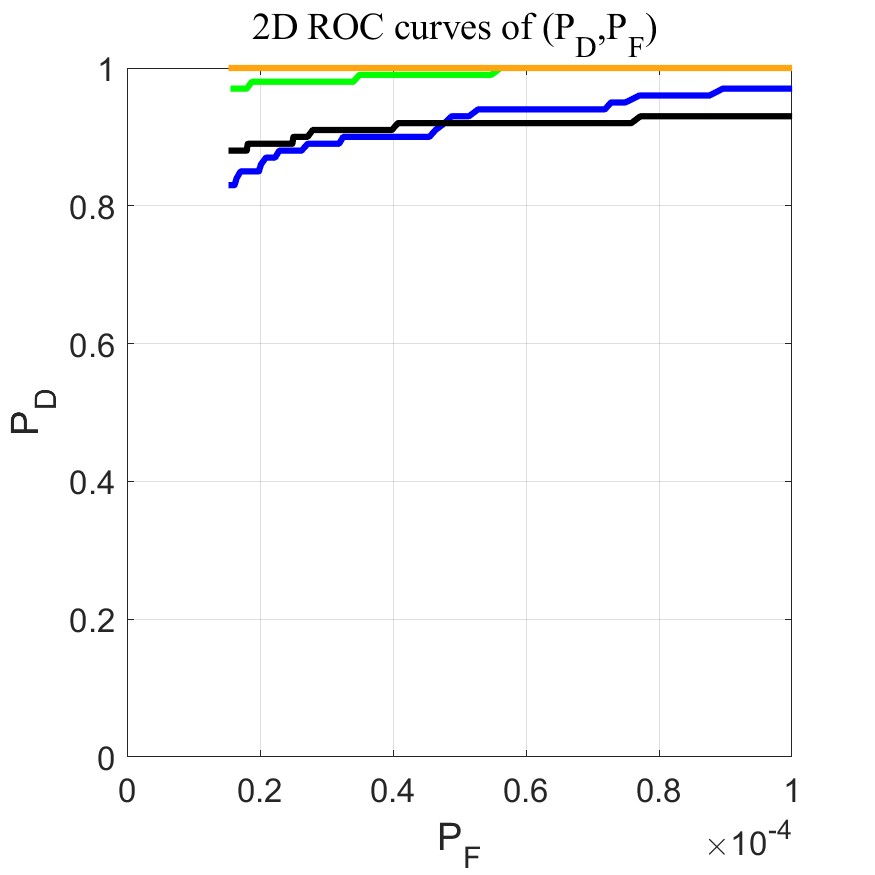}%
\hspace{0.01\linewidth}%
\includegraphics[height=0.15\textheight,width=0.22\linewidth]{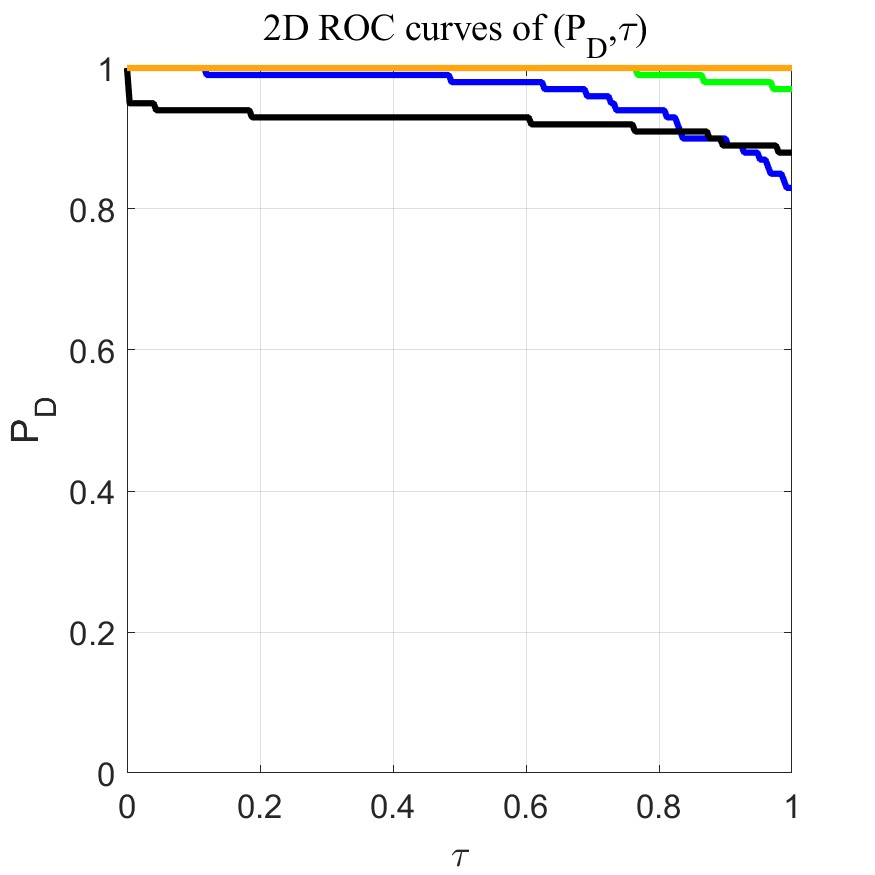}%
\hspace{0.01\linewidth}%
\includegraphics[height=0.15\textheight,width=0.22\linewidth]{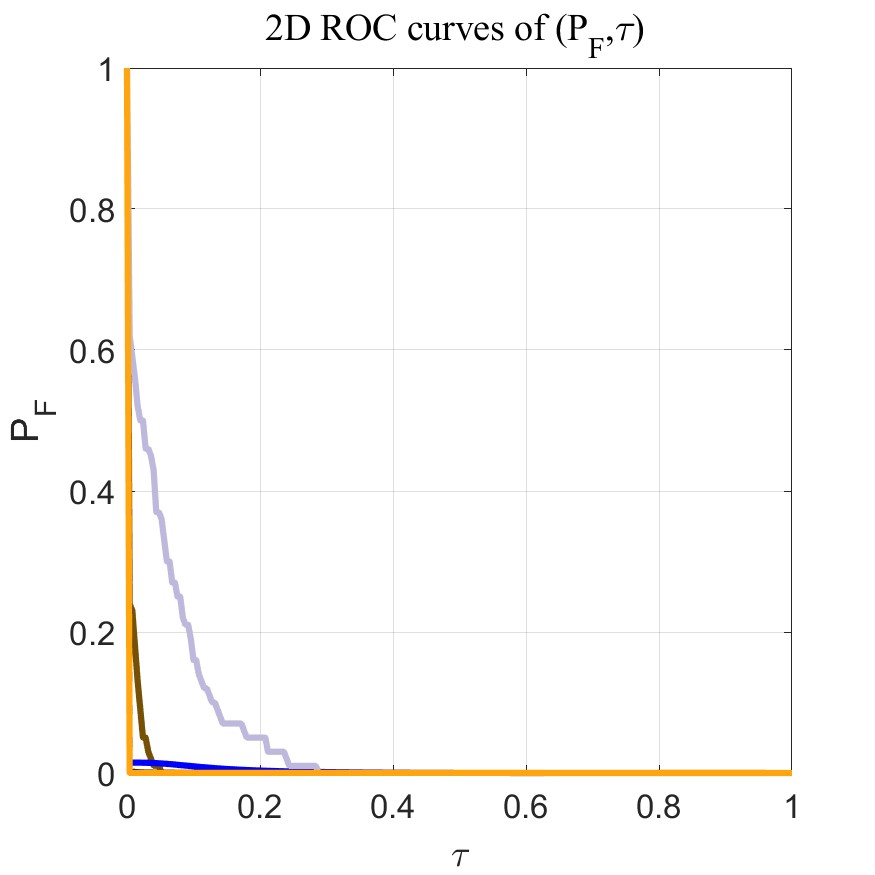}
\caption{Three-dimensional ROC curves of nine competing methods and their resulting 2-D ROC curves in sequence 1.}
\label{fig_row1}
\end{figure*}

\begin{figure*}[!t]
\centering
\includegraphics[height=0.15\textheight,width=0.22\linewidth]{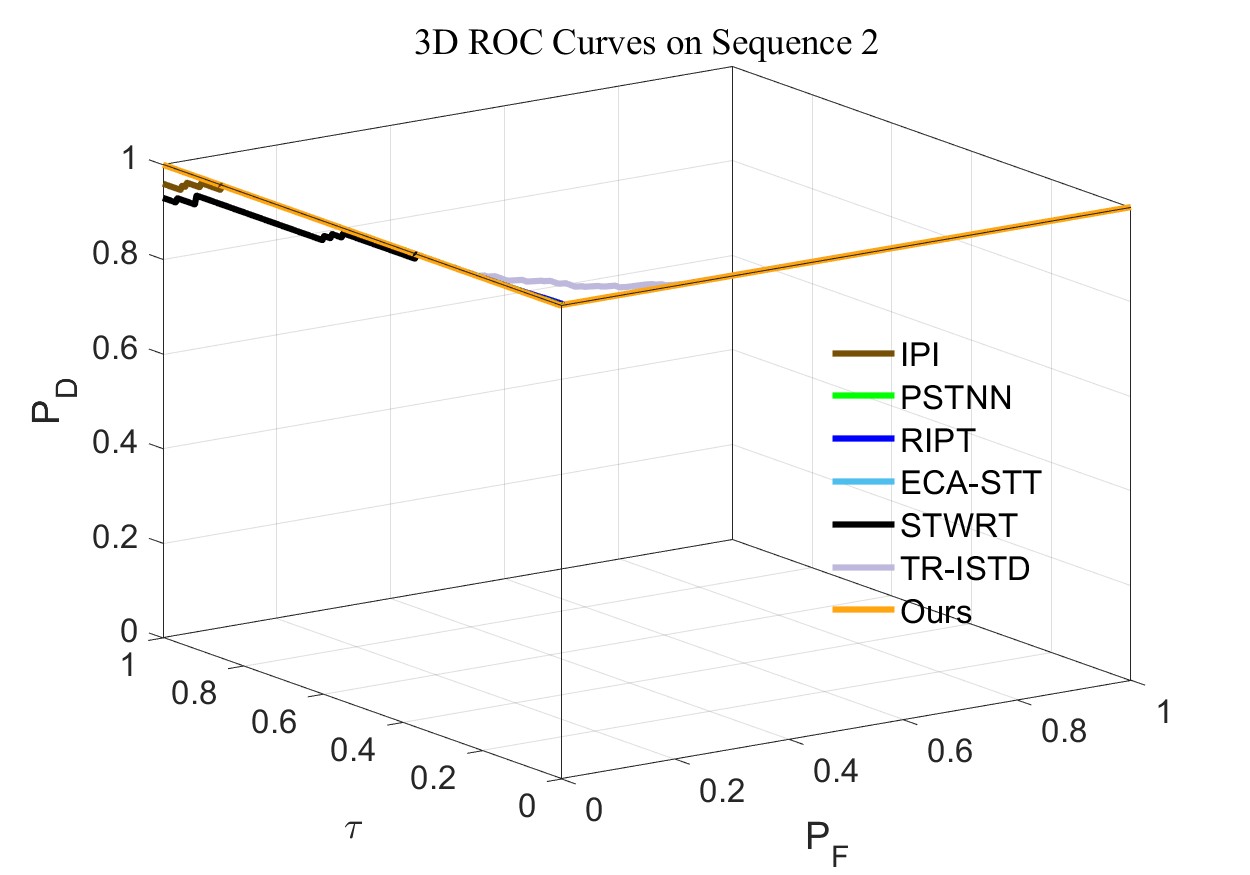}%
\hspace{0.01\linewidth}%
\includegraphics[height=0.15\textheight,width=0.22\linewidth]{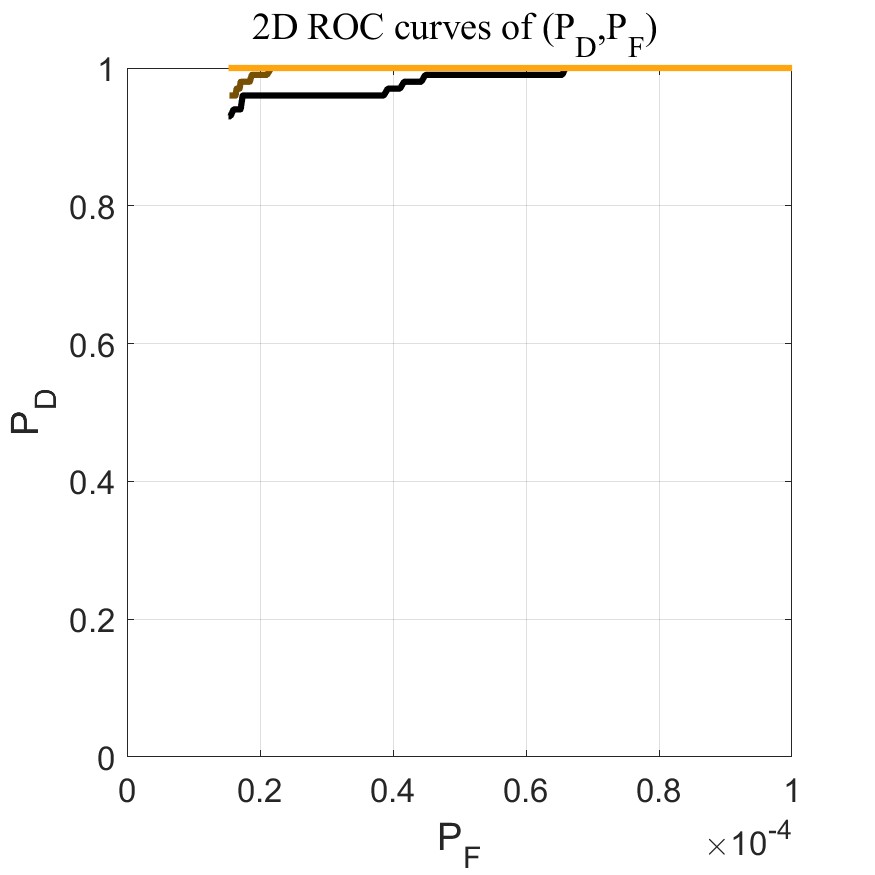}%
\hspace{0.01\linewidth}%
\includegraphics[height=0.15\textheight,width=0.22\linewidth]{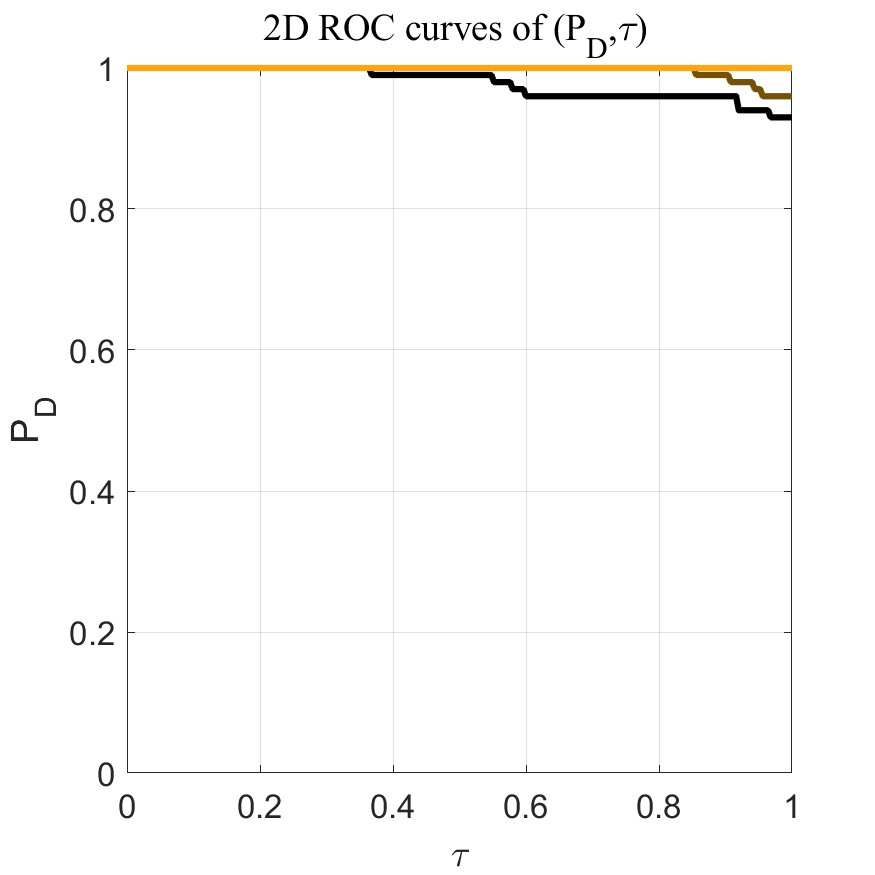}%
\hspace{0.01\linewidth}%
\includegraphics[height=0.15\textheight,width=0.22\linewidth]{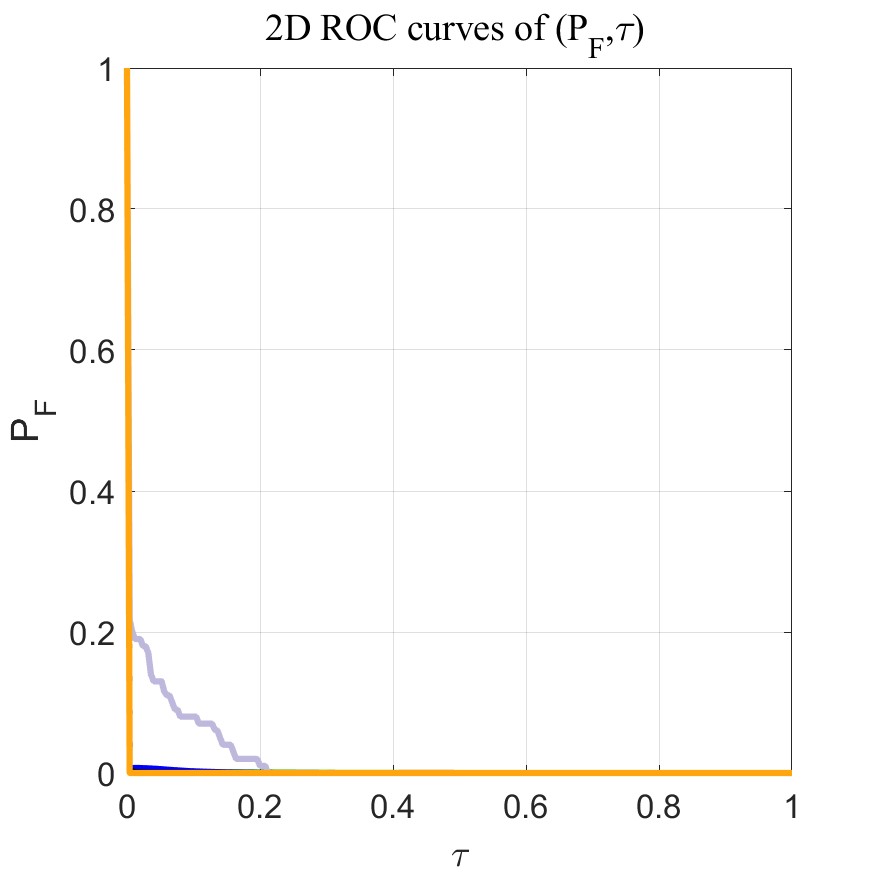}
\caption{Three-dimensional ROC curves of nine competing methods and their resulting 2-D ROC curves in sequence 2.}
\label{fig_row2}
\end{figure*}

\begin{figure*}[!t]
\centering
\includegraphics[height=0.15\textheight,width=0.22\linewidth]{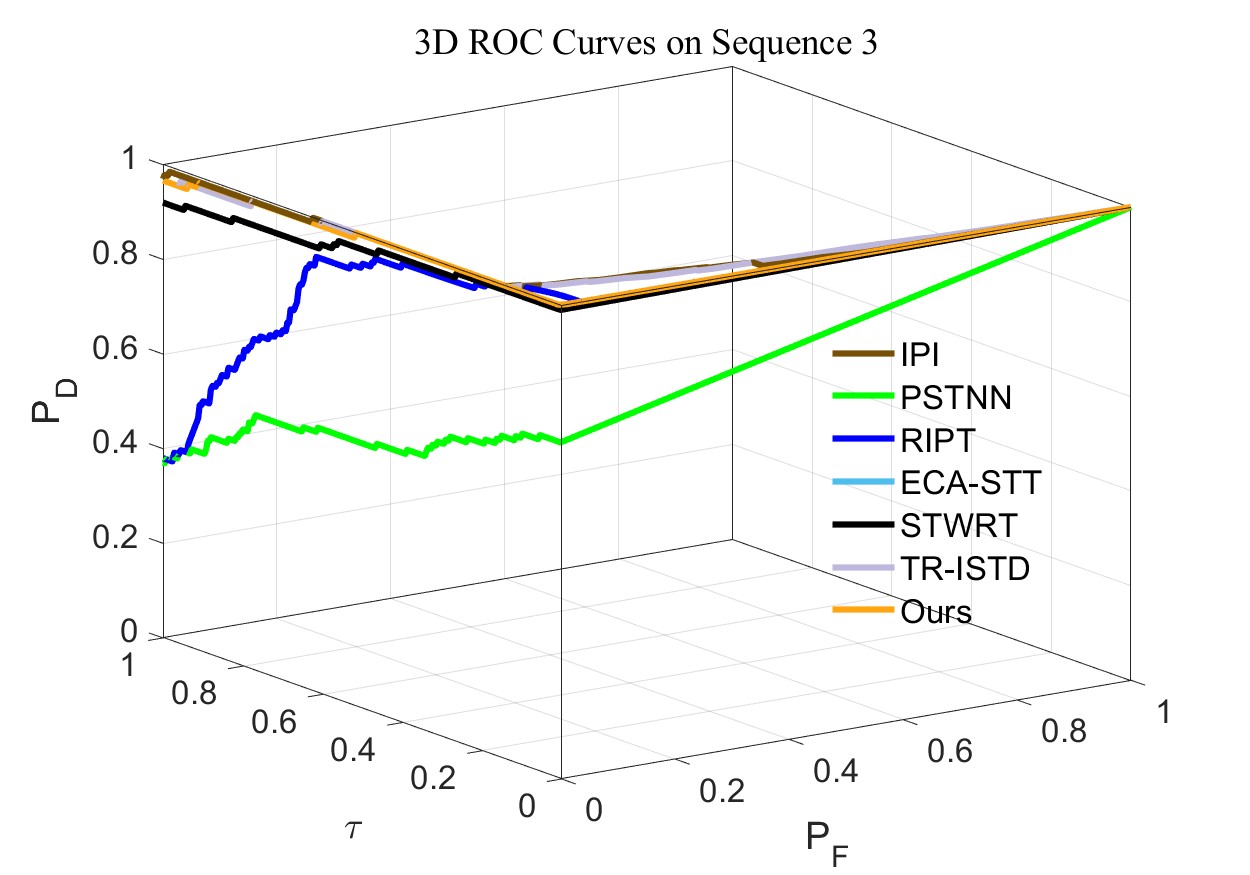}%
\hspace{0.01\linewidth}%
\includegraphics[height=0.15\textheight,width=0.22\linewidth]{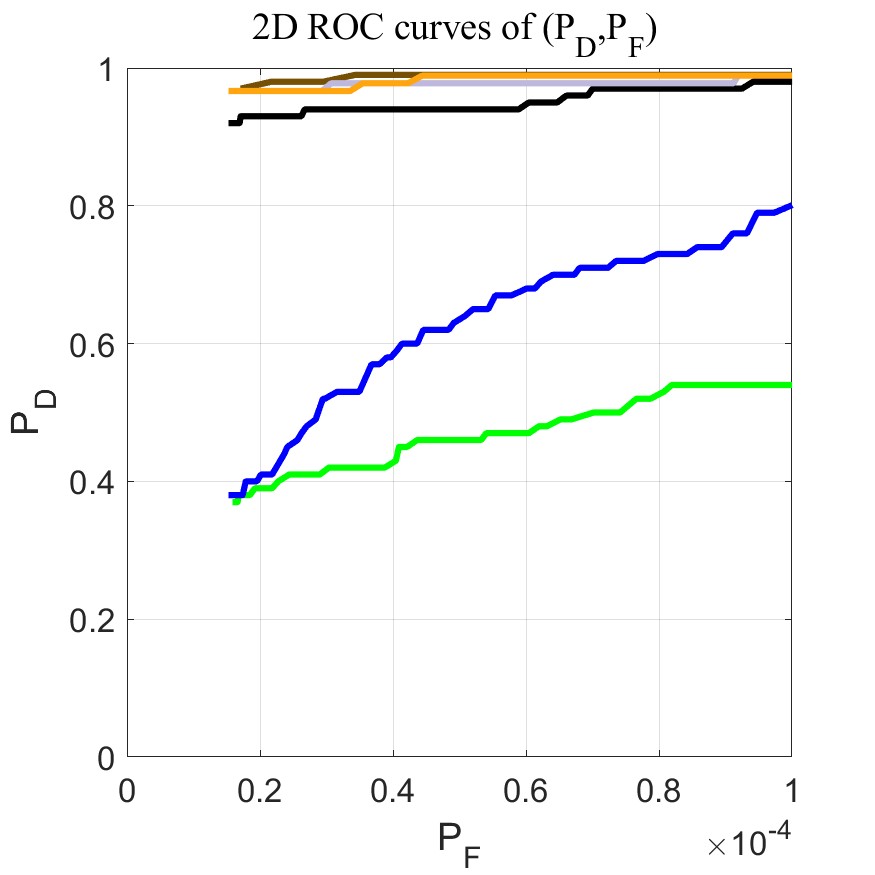}%
\hspace{0.01\linewidth}%
\includegraphics[height=0.15\textheight,width=0.22\linewidth]{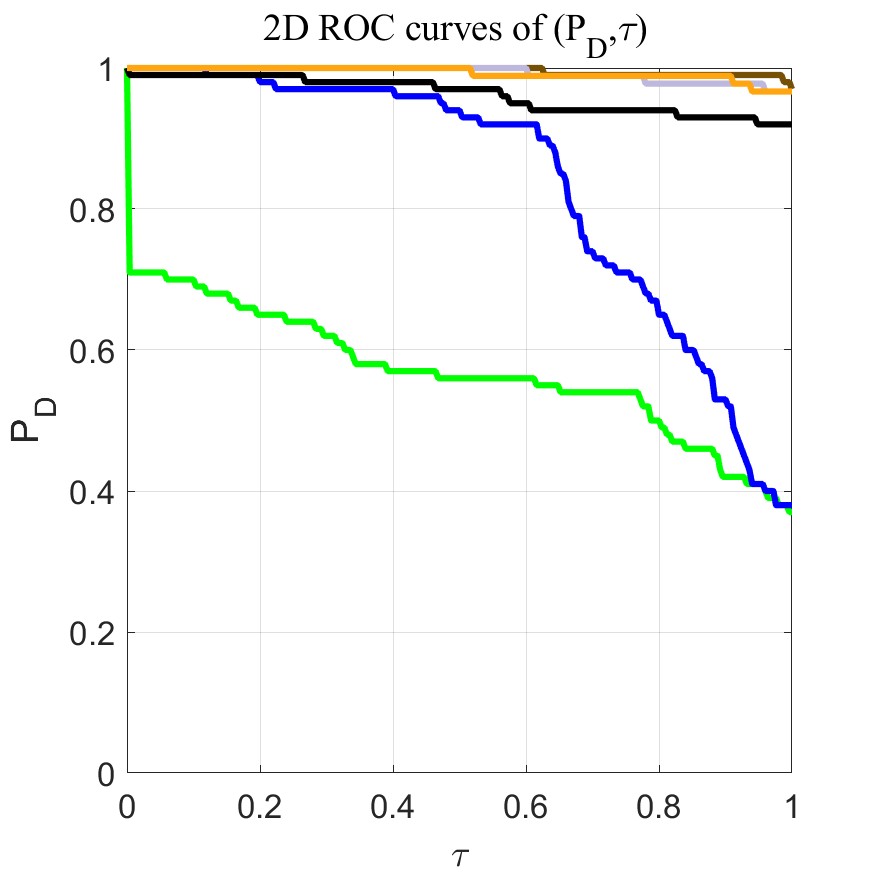}%
\hspace{0.01\linewidth}%
\includegraphics[height=0.15\textheight,width=0.22\linewidth]{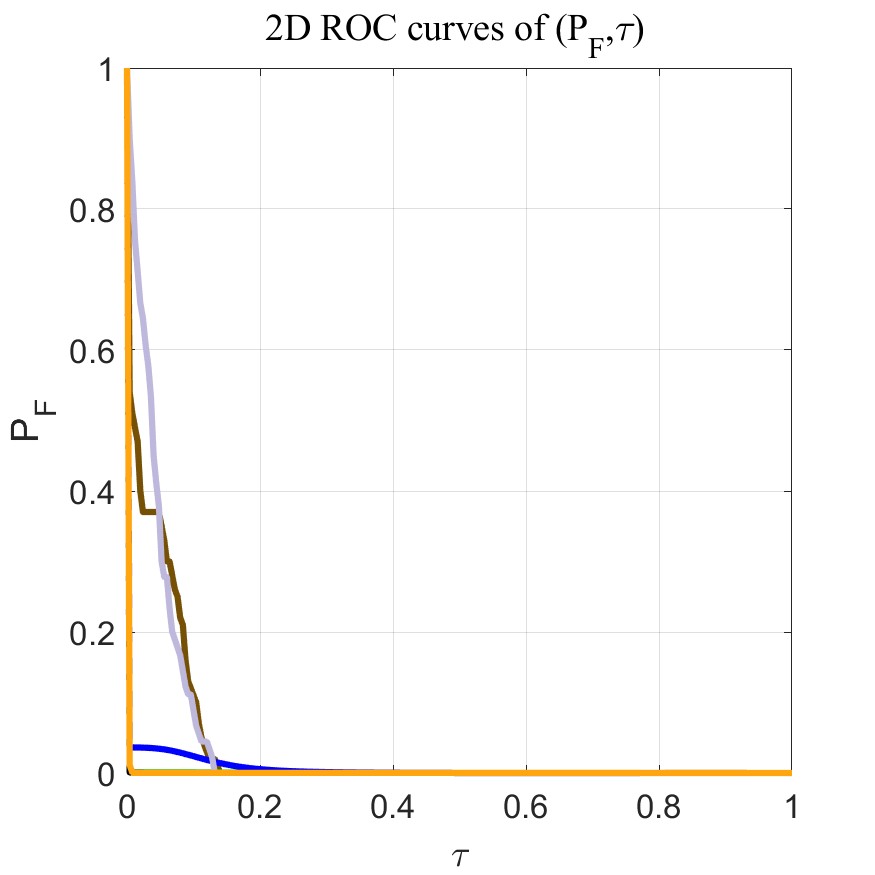}
\caption{Three-dimensional ROC curves of nine competing methods and their resulting 2-D ROC curves in sequence 3.}
\label{fig_row3}
\end{figure*}

\begin{table*}[!t]
\centering
\caption{Quantitative Comparison of Seven Competing Methods Regarding AUC Results Derived from 3-D ROC Curves for Sequence 1–6}
\label{tab:performance_comparison}
\small
\newcolumntype{C}{>{\centering\arraybackslash}m{1.2cm}}
\begin{tabular*}{\linewidth}{@{\extracolsep{\fill}}C C *{7}{c}@{}}
\toprule
\addlinespace[-0.02em]
Methods & Metrics & IPI & PSTNN & RIPT & ECA-STT & STWRT & 4D-TR & BTR-ISTD \\
\addlinespace[-0.3em]
\midrule
\multirow{8}{*}{Seq 1} 
    & $\mathrm{AUC}_{\mathrm{(D,F)}}$ & \textbf{1.0000} & \textbf{1.0000} & \textit{0.9999} & \textbf{1.0000} & 0.9749 & \textbf{1.0000} & \textbf{1.0000} \\
    & $\mathrm{AUC}_{\mathrm{(D,\tau)}}$ & \textbf{1.0000} & \textit{0.9996} & 0.9636 & 0.9993 & 0.9235 & \textbf{1.0000} & \textbf{1.0000} \\
    & $\mathrm{AUC}_{\mathrm{(F,\tau)}}$ & 0.0064 & \textbf{0.0021} & 0.0043 & 0.0072 & \textit{0.0022} & 0.0487 & \textbf{0.0021} \\
    & $\mathrm{AUC}_{\mathrm{(SNPR)}}$ & 156.2500 & \textit{474.2857} & 224.0930 & 138.7917 & 419.7727 & 20.5339 & \textbf{476.1905} \\
     & $\mathrm{AUC}_{\mathrm{(TDBS)}}$ & 0.9936 & \textit{0.9939} & 0.9593 & 0.9921 & 0.9213 & 0.9513 & \textbf{0.9979} \\
      & $\mathrm{AUC}_{\mathrm{(ODP)}}$ & 1.9936 & \textit{1.9939} & 1.9593 & 1.9921 & 1.9213 & 1.9513 & \textbf{1.9979} 
      \\
      
\cmidrule(r){2-9} 
\addlinespace[-0.3em]
     &
    Time(s)&4.4229&\textbf{0.0684}&2.0855&5.6419&0.6958&3.6761&\textit{0.1855}

      \\
\addlinespace[-0.3em]
\midrule
\multirow{8}{*}{Seq 2} 
        & $\mathrm{AUC}_{\mathrm{(D,F)}}$ & \textbf{1.0000} & \textbf{1.0000} & \textbf{1.0000} & \textbf{1.0000} & \textbf{1.0000} & \textbf{1.0000} & \textbf{1.0000} \\
        & $\mathrm{AUC}_{\mathrm{(D,\tau)}}$ & \textit{0.9966} & \textbf{1.0000} & \textbf{1.0000} & \textbf{1.0000} & 0.9789 & \textbf{1.0000} & \textbf{1.0000} \\
    & $\mathrm{AUC}_{\mathrm{(F,\tau)}}$ & 0.0023 & 0.0023 & 0.0027 & 0.0045 & \textit{0.0021} & 0.0194 & \textbf{0.0020} \\
    & $\mathrm{AUC}_{\mathrm{(SNPR)}}$ & 433.3043 & 434.7826 & 370.3704 & 222.2222 & 466.1429 & 51.5464 & \textbf{500.0000} \\
     & $\mathrm{AUC}_{\mathrm{(TDBS)}}$ & 0.9944 & \textit{0.9977} & 0.9973 & 0.9955 & 0.9768 & 0.9806 & \textbf{0.9980} \\
      & $\mathrm{AUC}_{\mathrm{(ODP)}}$ & 1.9944 & \textit{1.9977} & 1.9973 & 1.9955 & 1.9768 & 1.9806 & \textbf{1.9980} 
      \\
\cmidrule(r){2-9} 
\addlinespace[-0.3em]
     &
    Time(s)&13.3380&\textbf{0.0847}&8.5921&6.4335&0.8586&2.5499&\textit{0.14075}
    \\
\addlinespace[-0.3em]
\midrule
\multirow{8}{*}{Seq 3} 
    & $\mathrm{AUC}_{\mathrm{(D,F)}}$ & \textbf{1.0000} & 0.8548 & 0.9997 & \textbf{1.0000} & 0.9950 & \textbf{1.0000} & \textbf{1.0000} \\
    & $\mathrm{AUC}_{\mathrm{(D,\tau)}}$ & \textbf{0.9961} & 0.5687 & 0.8380 & 0.9896 & 0.9624 & 0.9926 & \textit{0.9930} \\
    & $\mathrm{AUC}_{\mathrm{(F,\tau)}}$ & 0.0357 & \textit{0.0023} & 0.0069 & 0.0037 & \textbf{0.0021} & 0.0429 & \textbf{0.0021} \\
    & $\mathrm{AUC}_{\mathrm{(SNPR)}}$ & 27.9020 & 247.2609 & 121.4493 & 267.4595 & \textit{458.2857} & 23.1375 & \textbf{472.8571} \\
     & $\mathrm{AUC}_{\mathrm{(TDBS)}}$ & 0.9604 & 0.5664 & 0.8311 & \textit{0.9859} & 0.9603 & 0.9497 & \textbf{0.9909} \\
      & $\mathrm{AUC}_{\mathrm{(ODP)}}$ & 1.9604 & 1.5664 & 1.8311 & \textit{1.9859} & 1.9603 & 1.9597 & \textbf{1.9909} 
      \\
\cmidrule(r){2-9}  
\addlinespace[-0.3em]
     &
      Time(s)&15.0749&0.0607&0.7618&5.9844&\textit{0.7051}&2.8114&\textbf{0.0225}
      \\
\addlinespace[-0.3em]
\midrule
\multirow{8}{*}{Seq 4} 
    & $\mathrm{AUC}_{\mathrm{(D,F)}}$ & \textbf{1.0000} & \textbf{1.0000} & \textit{0.9900} & \textbf{1.0000} & \textbf{1.0000} & \textbf{1.0000} & \textbf{1.0000} \\
        & $\mathrm{AUC}_{\mathrm{(D,\tau)}}$ & \textbf{1.0000} & \textbf{1.0000} & 0.9207 &\textbf{1.0000} & \textit{0.9898} & \textbf{1.0000} & \textbf{1.0000} \\
    & $\mathrm{AUC}_{\mathrm{(F,\tau)}}$ & 0.0656 & \textit{0.0022} & 0.0025 & 0.0054 & \textbf{0.0021} & 0.0720 & \textbf{0.0021} \\
    & $\mathrm{AUC}_{\mathrm{(SNPR)}}$ & 15.2439 & 454.5455 & 368.2800 & 185.1852 & \textit{471.3333} & 13.8889 & \textbf{476.1905} \\
     & $\mathrm{AUC}_{\mathrm{(TDBS)}}$ & 0.9344 & \textit{0.9978} & 0.9182 & 0.9946  & 0.9877& 0.9280 & \textbf{0.9979} \\
      & $\mathrm{AUC}_{\mathrm{(ODP)}}$ & 1.9344 & \textit{1.9978} & 1.9182 & 1.9946 & 1.9877 & 1.9280 & \textbf{1.9979} 
      \\
\cmidrule(r){2-9} 
\addlinespace[-0.3em]
     &
      Time(s)&4.4263&\textit{0.0984}&3.1024&7.8866&4.9948&3.5737&\textbf{0.0779}
      \\
\addlinespace[-0.3em]
\midrule
\multirow{8}{*}{Seq 5} 
     & $\mathrm{AUC}_{\mathrm{(D,F)}}$ & \textbf{1.0000} & 0.9800 & 0.9849 & \textit{0.9944} & \textbf{1.0000} & \textbf{1.0000} & \textbf{1.0000} \\
    & $\mathrm{AUC}_{\mathrm{(D,\tau)}}$ & \textbf{1.0000} & 0.9440 & 0.9457 & 0.9551 & 0.9413 & \textbf{1.0000} & \textbf{1.0000} \\
    & $\mathrm{AUC}_{\mathrm{(F,\tau)}}$ & 0.1281 & 0.0022 & 0.0025 & 0.0046 & \textit{0.0021} & 0.0063 & \textbf{0.0020} \\
    & $\mathrm{AUC}_{\mathrm{(SNPR)}}$ & 7.8064 & 429.0909 & 378.2800 & 207.6304 & \textit{448.2381} & 158.7302 & \textbf{500.0000} \\
     & $\mathrm{AUC}_{\mathrm{(TDBS)}}$ & 0.8719 & 0.9418 & 0.9432 & 0.9505  & 0.9392& \textit{0.9937 }& \textbf{0.9980} \\
      & $\mathrm{AUC}_{\mathrm{(ODP)}}$ & 1.8719 & 1.9418 & 1.9432 & 1.9505 & 1.9392 & \textit{1.9937} & \textbf{1.9980} 
      \\
\cmidrule(r){2-9}
\addlinespace[-0.3em]
     &
      Time(s)&5.4997&\textit{0.0658}&1.9668&4.9927&0.6954&10.7227&\textbf{0.0314}
      \\
\addlinespace[-0.3em]
\midrule
\multirow{8}{*}{Seq 6} 
    & $\mathrm{AUC}_{\mathrm{(D,F)}}$ & 0.9114 & 0.7615 & \textit{0.9848} & 0.9602 & 0.8888 & 0.9326 & \textbf{0.9667} \\
    & $\mathrm{AUC}_{\mathrm{(D,\tau)}}$ & 0.4222 & 0.2248 & 0.4285 & 0.3544 & 0.2606 & \textit{0.8086} & \textbf{0.9226} \\
    & $\mathrm{AUC}_{\mathrm{(F,\tau)}}$ & 0.2469 & 0.0041 & \textbf{0.0023} & 0.0069 & \textit{0.0028} & 0.1582 & 0.1038\\
    & $\mathrm{AUC}_{\mathrm{(SNPR)}}$ & 1.7100 & 54.8293 & \textbf{186.3043} & 51.3623 & \textit{93.0714} & 5.1112 & \textbf{8.8882} \\
    & $\mathrm{AUC}_{\mathrm{(TDBS)}}$ & 0.1753 & 0.2207 & 0.4262 & 0.3475  & 0.2578& \textit{0.6504 }& \textbf{0.8188} \\
      & $\mathrm{AUC}_{\mathrm{(ODP)}}$ & 1.1753& 1.2207 & 1.4262 & 1.3475 & 1.2578 & \textit{1.6504} & \textbf{1.8188} 
      \\
\cmidrule(r){2-9}
\addlinespace[-0.3em]
     &
      Time(s)&3.6868&\textbf{0.0766}&\textit{0.2071}&5.5222&5.3985&5.5685&0.4578
      \\
\addlinespace[-0.3em]
\bottomrule
\end{tabular*}
\end{table*}
\subsection{Qualitative and Quantitative Evaluation}
Fig. 3 illustrates the visual detection results and  3-D intuitive visualization of Sequences 1–6 processed by all seven evaluated methods. The target region is annotated with a red bounding box and magnified in the bottom-right corner for clarity. As depicted, the proposed method achieves superior background suppression compared to single-frame detection techniques, successfully extracting distinct target signatures. Furthermore, existing multi-frame detection methods exhibit deficiencies in accurately reconstructing edge pixels of dim small targets and incur partial pixel loss. In contrast, the proposed approach effectively mitigates both issues.

For a comprehensive performance evaluation, 3D receiver operating characteristic (3-D ROC) curves are utilized to jointly characterize detection probability, background suppression capability, and false alarm rate.In Figs. 4–9, the 3-D ROC curves and their resulting 2-D ROC curves for sequences 1–6 are presented. Quantitative results derived from the 3-D ROC analysis, including AUC metrics,and the average computational time for each method , are summarized in Table II,with optimal and suboptimal results emphasized in bold and italic fonts, respectively. Both quantitative and qualitative analyses confirm that the proposed BTR-ISTD framework consistently exceeds all competing methods across all evaluation metrics, validating its efficacy and superiority.

As evidenced by the tabulated results, the proposed method attains optimal performance across all metrics in Sequences 5 where targets possess high contrast and backgrounds are relatively uncluttered—while maintaining computational efficiency. Although PSTNN demonstrates reduced computational time on Sequences 1 and 2, it yields inferior detection accuracy and diminished background suppression. In more demanding scenarios, such as Sequences 3-4 and 6 featuring low-contrast dim small targets within complex backgrounds, BTR-ISTD achieves exceptional detection accuracy and operational efficiency, alongside robust background clutter suppression. These advantages originate from the method’s capacity to reduce computational and memory overhead while preserving cross-dimensional local correlations and global dependencies.

\begin{table*}[t]
\centering
\footnotesize
\caption{Performance comparison under different parameters I.}
\begin{threeparttable}

\begin{tabularx}{\textwidth}{ccc *{12}{X}}
\toprule
\multicolumn{3}{c}{\textbf{parameters}} 
& \multicolumn{6}{c}{\textbf{Seq4}}
& \multicolumn{6}{c}{\textbf{Seq3}} \\
\midrule

$N_w$ & $N_t$ & $H$
& $\mathrm{AUC}_{\mathrm{(D,F)}}$ &$\mathrm{AUC}_{\mathrm{(D,\tau)}}$&$\mathrm{AUC}_{\mathrm{(F,\tau)}}$ & $\mathrm{AUC}_{\mathrm{SNPR}}$ & $\mathrm{AUC}_{\mathrm{TDBS}}$&$\mathrm{AUC}_{\mathrm{ODP}}$
& $\mathrm{AUC}_{\mathrm{(D,F)}}$ &$\mathrm{AUC}_{\mathrm{(D,\tau)}}$&$\mathrm{AUC}_{\mathrm{(F,\tau)}}$ & $\mathrm{AUC}_{\mathrm{SNPR}}$ & $\mathrm{AUC}_{\mathrm{TDBS}}$&$\mathrm{AUC}_{\mathrm{ODP}}$ \\
\midrule
40 & 15 & 2.5 & \textbf{1.0000} & \textbf{1.0000} & \textbf{0.0025} & \textbf{400.0000} & \textbf{0.9975} & \textbf{1.9975} & \textbf{1.0000} & 0.9786 & 0.0021 & 466.0000 & 0.9765 & 1.9765 \\
50 & 15 & 2.5 & \textbf{1.0000} & \textbf{1.0000} & 0.0304 & 32.8947 & 0.9696 & 1.9696 & \textbf{1.0000} & \textbf{0.9863} & 0.0036 & 237.9722 & \textbf{0.9827} & \textbf{1.9827 }\\
60 & 15 & 2.5 & \textbf{1.0000} & \textbf{1.0000} & 0.0275 & 36.3636 & 0.9725 & 1.9725 & 0.9723 & 0.9718 & 0.0022 & 441.7273 & 0.9696 & 1.9696 \\
70 & 15 & 2.5 & 0.9344 & 0.9667 & 0.0167 & 57.8862 & 0.9500 & 1.9500 & 0.8301 & 0.9098 & 0.0019 & \textbf{478.8421} & 0.9079 & 1.9079 \\
80 & 15 & 2.5 & 0.9560 & 0.9778 & 0.0253 & 38.6482 & 0.9525 & 1.9525 & 0.7607 & 0.8533 & \textbf{0.0018} & 474.0556 & 0.8515 & 1.8515 \\
\midrule
60 & 5 & 2.5 & 0.6889 & 0.8300 & \textbf{0.0066} & \textbf{125.7576} & 0.8234 & 1.8234 & 0.8988 & 0.9162 & 0.0474 & 19.3291 & 0.8688 & 1.8688 \\
60 & 10 & 2.5 & \textbf{1.0000} & \textbf{1.0000}&0.0100& 100.0000 & \textbf{0.9900} & \textbf{1.9900} & 0.9801 & 0.9746 & 0.0049 & 198.8980 & 0.9697 & 1.9697 \\
60 & 15 & 2.5 & \textbf{1.0000} & \textbf{1.0000} & 0.0275 & 36.3636 & 0.9725 & 1.9725 & 0.9723 & 0.9718 & \textbf{0.0022} & \textbf{441.7273} & 0.9696 & 1.9696 \\
60 & 20 & 2.5 & \textbf{1.0000 }& \textbf{1.0000} & 0.0359 & 27.8552 & 0.9641 & 1.9641 & \textbf{1.0000} & \textbf{0.9946} & 0.0064 & 155.4062 & \textbf{0.9882} & \textbf{1.9882} \\
\midrule
60 & 15 & 1.0 & \textbf{1.0000} & \textbf{1.0000} & 0.0856 & 11.6822 & 0.9144 & 1.9144 & \textbf{1.0000} & \textbf{0.9947} & 0.0262 & 37.9656 & 0.9685 & 1.9685 \\
60 & 15 & 1.5 & \textbf{1.0000} & \textbf{1.0000} & 0.0679 & 14.7275 & 0.9321 & 1.9321 & 0.9779 & 0.9826 & 0.0131 & 75.0076 & 0.9695 & 1.9695 \\
60 & 15 & 2.0 & \textbf{1.0000} & \textbf{1.0000} & 0.0464 & 21.5517 & 0.9536 & 1.9536 & 0.9779 & 0.9806 & 0.0051 & 192.2745 & \textbf{0.9755} & \textbf{1.9755} \\
60 & 15 & 2.5 & \textbf{1.0000} & \textbf{1.0000} & 0.0275 & 36.3636 & 0.9725 & 1.9725 & 0.9723 & 0.9718 & 0.0022 & 441.7273 & 0.9696 & 1.9696 \\
60 & 15 & 3.0 & \textbf{1.0000} & \textbf{1.0000} & 0.0138 & 72.4638 & \textbf{0.9862} & \textbf{1.9862} & 0.9291 & 0.9481 & \textbf{0.0020} & \textbf{474.0500} & 0.9461 & 1.9461 \\
60 & 15 & 3.5 & 0.9344 & 0.9667 & \textbf{0.0063} & \textbf{153.4444 }& 0.9604 & 1.9604 & 0.9131 & 0.9469 & \textbf{0.0020} & 473.4500 & 0.9449 & 1.9449 \\

\bottomrule
\end{tabularx}
\end{threeparttable}
\end{table*}

\begin{table*}[t]
\centering
\footnotesize
\caption{Performance comparison under different parameters II.}
\begin{threeparttable}

\begin{tabularx}{\textwidth}{ccc *{12}{X}}
\toprule
\multicolumn{3}{c}{\textbf{parameters}} 
& \multicolumn{6}{c}{\textbf{Seq4}}
& \multicolumn{6}{c}{\textbf{Seq3}} \\
\midrule

$R_1$ & $R$ & $R_2$
& $\mathrm{AUC}_{\mathrm{(D,F)}}$ &$\mathrm{AUC}_{\mathrm{(D,\tau)}}$&$\mathrm{AUC}_{\mathrm{(F,\tau)}}$ & $\mathrm{AUC}_{\mathrm{SNPR}}$ & $\mathrm{AUC}_{\mathrm{TDBS}}$&$\mathrm{AUC}_{\mathrm{ODP}}$
& $\mathrm{AUC}_{\mathrm{(D,F)}}$ &$\mathrm{AUC}_{\mathrm{(D,\tau)}}$&$\mathrm{AUC}_{\mathrm{(F,\tau)}}$ & $\mathrm{AUC}_{\mathrm{SNPR}}$ & $\mathrm{AUC}_{\mathrm{TDBS}}$&$\mathrm{AUC}_{\mathrm{ODP}}$ \\
\midrule

4 & 3 & 30 & \textbf{1.0000} & \textbf{1.0000} & 0.0699 & 14.3062 & 0.9301 & 1.9301
  & \textbf{1.0000} & 0.9952 & 0.0050 & 199.0400 & 0.9902 & 1.9902 \\

5 & 3 & 30 & \textbf{1.0000} & \textbf{1.0000} & 0.0643 & 15.5512 & 0.9357 & 1.9357
  & \textbf{1.0000} & \textbf{0.9953} & 0.0045 & 221.1778 & \textbf{0.9908} & \textbf{1.9908} \\

6 & 3 & 30 & \textbf{1.0000} & \textbf{1.0000} & \textbf{0.0637} & \textbf{15.6986} & \textbf{0.9363} & \textbf{1.9363}
  & \textbf{1.0000} & 0.9951 & 0.0050 & 199.0200 & 0.9901 & 1.9901 \\

7 & 3 & 30 & \textbf{1.0000} & \textbf{1.0000} & 0.0647 & 15.4560 & 0.9353 & 1.9353
  & \textbf{1.0000} & 0.9951 & 0.0045 & \textbf{221.1333} & 0.9906 & 1.9906 \\

8 & 3 & 30 & \textbf{1.0000} & \textbf{1.0000} & 0.0653 & 15.3139 & 0.9347 & 1.9347
  & \textbf{1.0000} & 0.9951 & \textbf{0.0045} & 221.1333 & 0.9906 & 1.9906 \\

\midrule

6 & 2 & 30 & \textbf{1.0000} & \textbf{1.0000} & 0.0612 & 16.3399 & 0.9388 & 1.9388
  & \textbf{1.0000} & 0.9950 & 0.0050 & 199.0000 & 0.9900 & 1.9900 \\

6 & 3 & 30 & \textbf{1.0000} & \textbf{1.0000} & 0.0637 & 15.6986 & 0.9363 & 1.9363
  & \textbf{1.0000} & \textbf{0.9951} & 0.0050 & 199.0200 & 0.9901 & 1.9901 \\

6 & 4 & 30 & \textbf{1.0000} & \textbf{1.0000} & \textbf{0.0599} & \textbf{16.6945} & \textbf{0.9401} & \textbf{1.9401}
  & \textbf{1.0000} & \textbf{0.9951} & \textbf{0.0047} & \textbf{211.7234} & \textbf{0.9904} & \textbf{1.9904} \\

\midrule

6 & 3 & 20 & \textbf{1.0000} & \textbf{1.0000} & 0.1066 &  9.3809 & 0.8934 & 1.8934
  & \textbf{1.0000} & 0.9937 & 0.0206 &  48.2379 & 0.9731 & 1.9731 \\

6 & 3 & 25 & \textbf{1.0000} & \textbf{1.0000} & 0.0946 & 10.5708 & 0.9054 & 1.9054
  & \textbf{1.0000} & \textbf{0.9955} & 0.0140 &  71.1071 & 0.9815 & 1.9815 \\

6 & 3 & 30 & \textbf{1.0000} & \textbf{1.0000} & 0.0637 & 15.6986 & 0.9363 & 1.9363
  & \textbf{1.0000} & 0.9951 & 0.0050 & 199.0200 & 0.9901 & 1.9901 \\

6 & 3 & 35 & \textbf{1.0000} & \textbf{1.0000} & 0.0422 & 23.6967 & 0.9578 & 1.9578
  & \textbf{1.0000} & 0.9943 & 0.0031 & 320.7419 & \textbf{0.9912} & \textbf{1.9912} \\

6 & 3 & 40 & \textbf{1.0000} & \textbf{1.0000} & \textbf{0.0277} & \textbf{36.1011} & \textbf{0.9723} & \textbf{1.9723}
  & \textbf{1.0000} & 0.9935 & \textbf{0.0025} & \textbf{397.4000} & 0.9910 & 1.9910 \\

\bottomrule
\end{tabularx}
\end{threeparttable}
\end{table*}

\begin{figure*}[!h]
\centering
\includegraphics[height=0.15\textheight,width=0.22\linewidth]{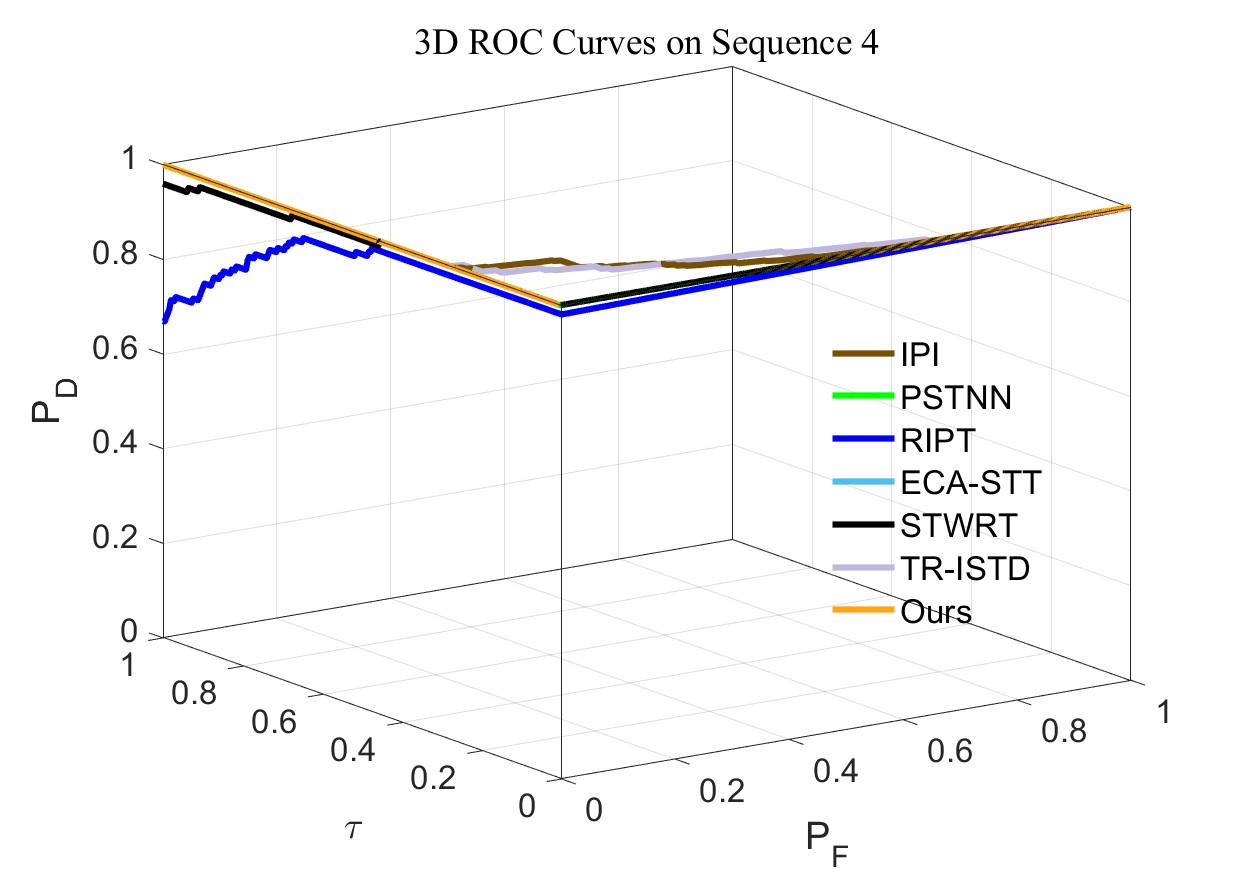}%
\hspace{0.01\linewidth}%
\includegraphics[height=0.15\textheight,width=0.22\linewidth]{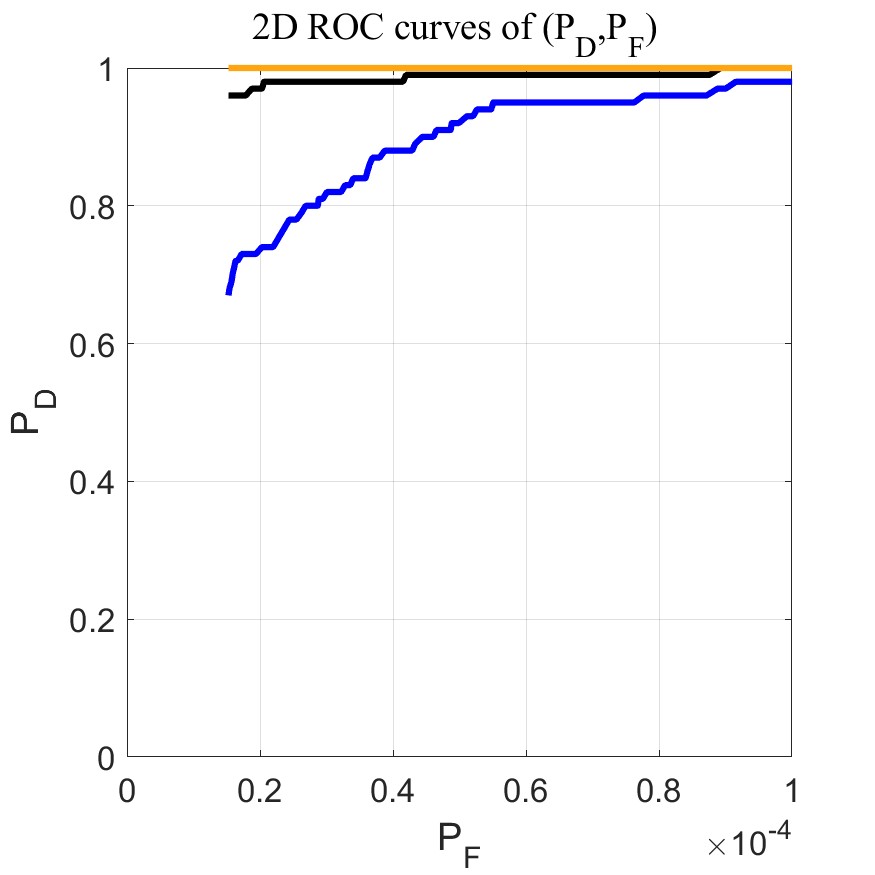}%
\hspace{0.01\linewidth}%
\includegraphics[height=0.15\textheight,width=0.22\linewidth]{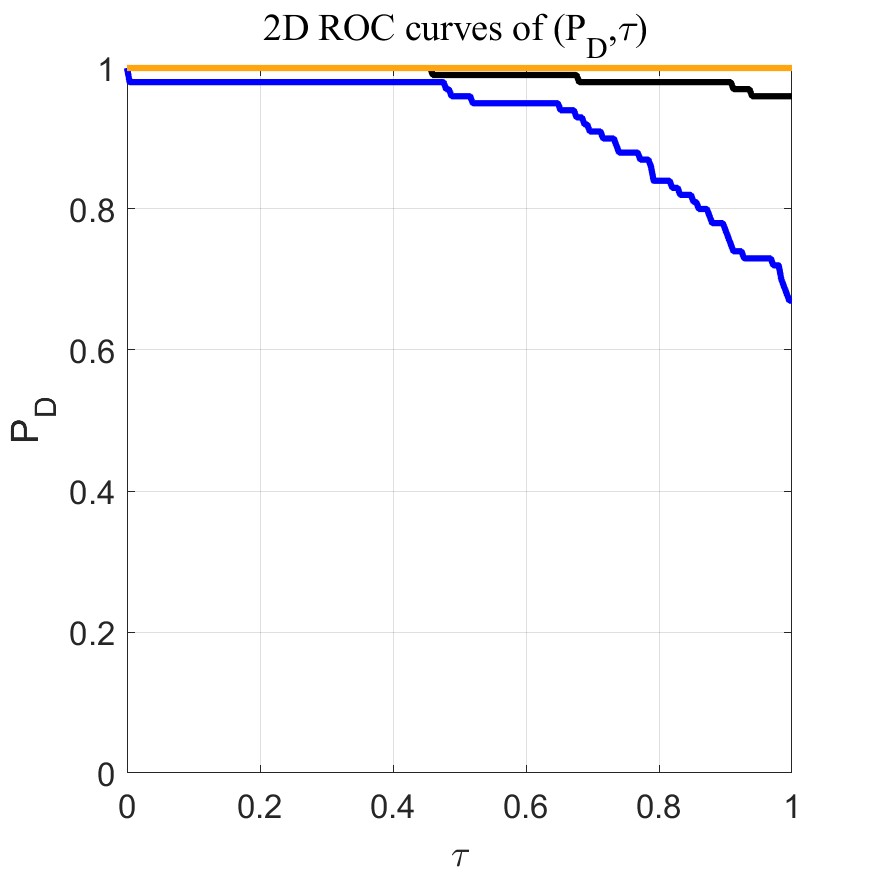}%
\hspace{0.01\linewidth}%
\includegraphics[height=0.15\textheight,width=0.22\linewidth]{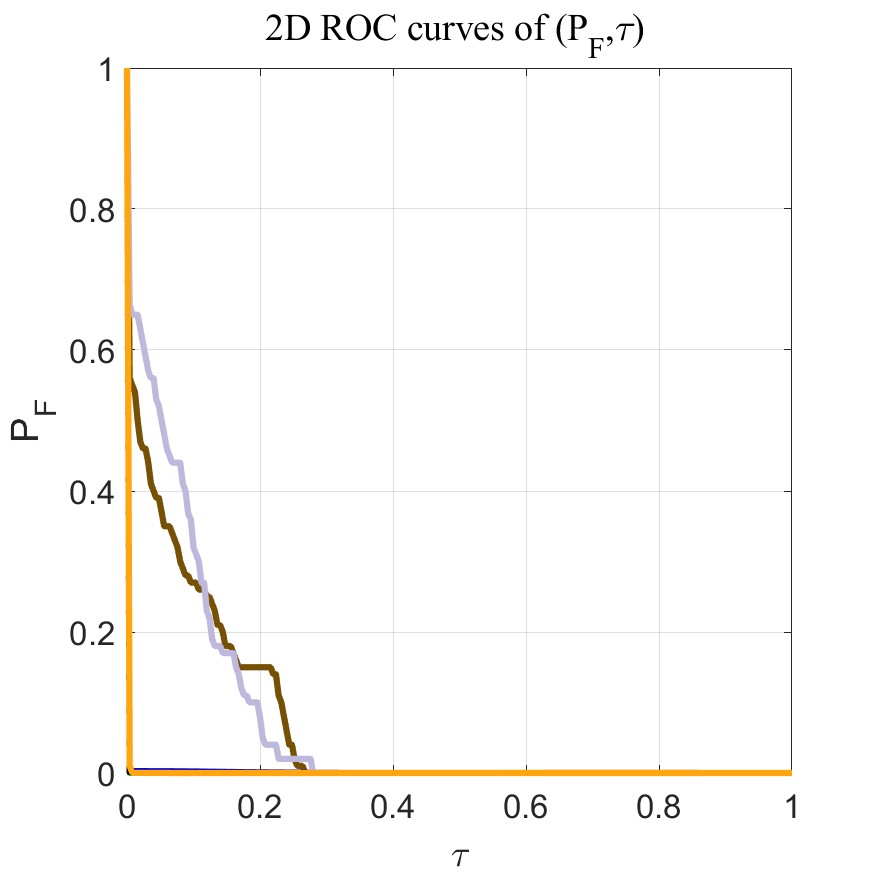}
\caption{Three-dimensional ROC curves of nine competing methods and their resulting 2-D ROC curves in sequence 4.}
\label{fig_row4}
\end{figure*}

\begin{figure*}[!h]
\centering
\includegraphics[height=0.15\textheight,width=0.22\linewidth]{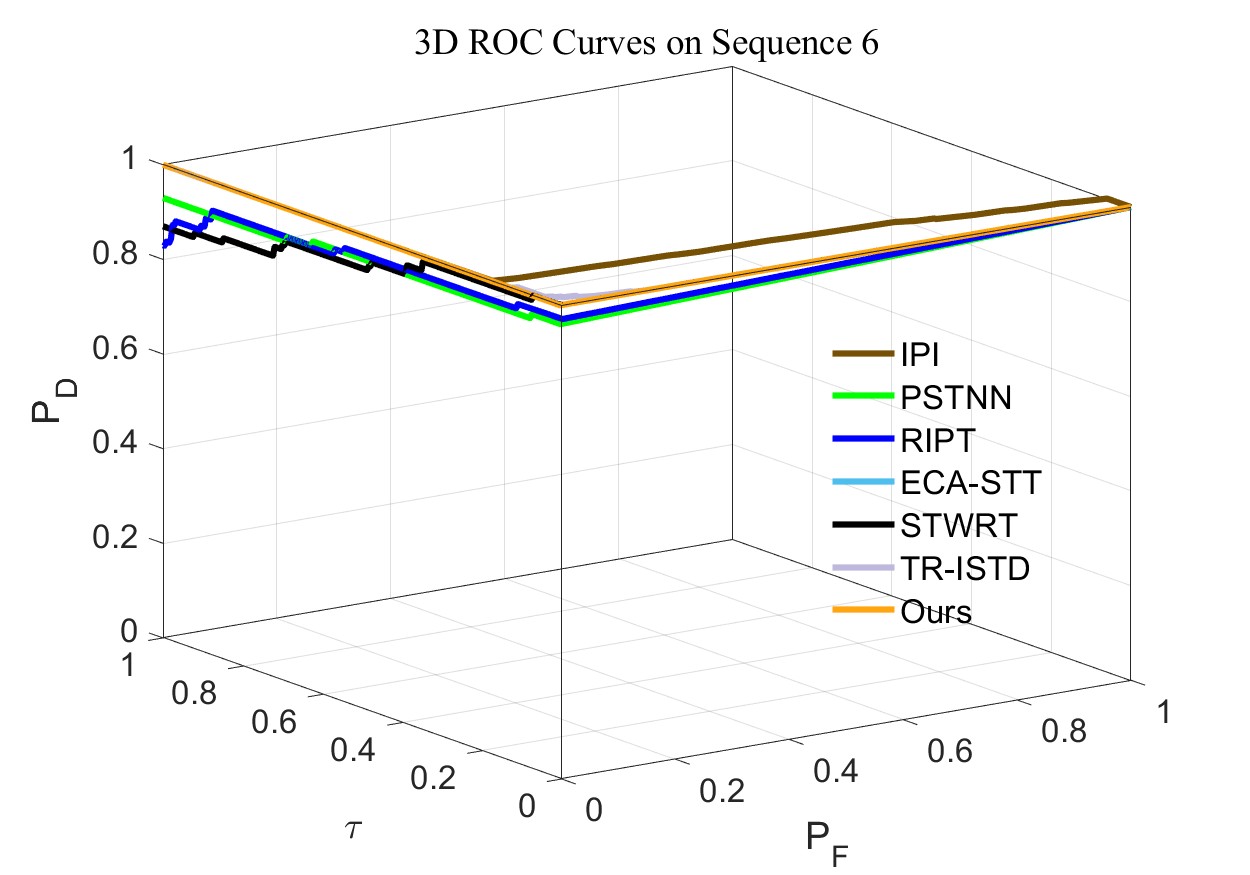}%
\hspace{0.01\linewidth}%
\includegraphics[height=0.15\textheight,width=0.22\linewidth]{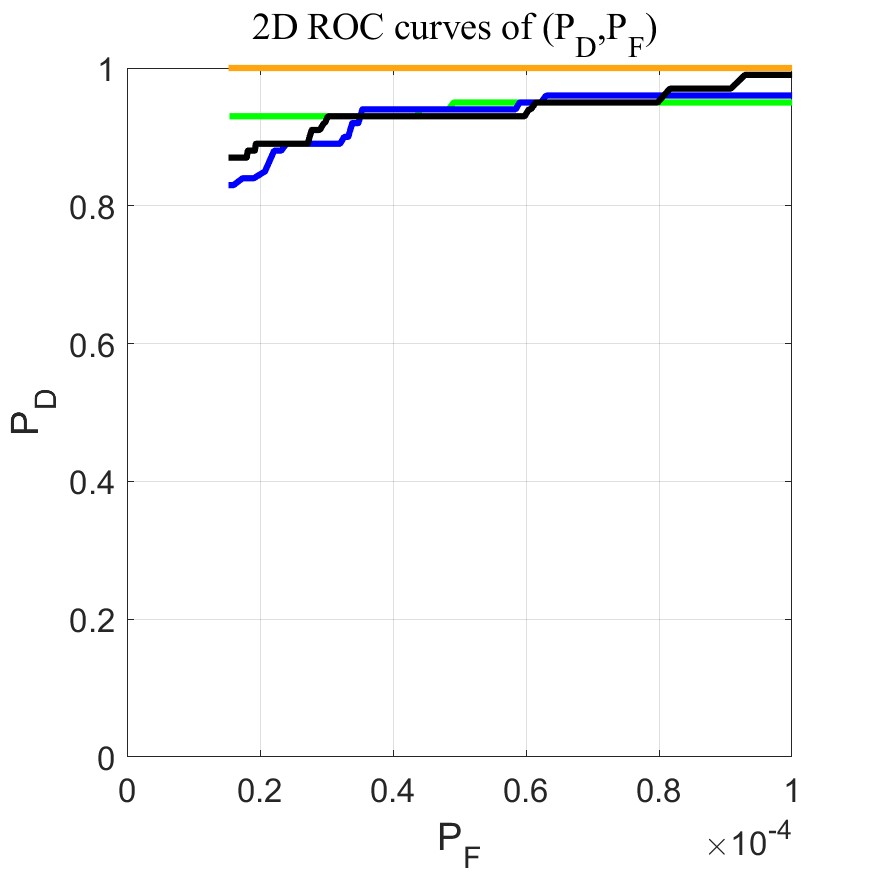}%
\hspace{0.01\linewidth}%
\includegraphics[height=0.15\textheight,width=0.22\linewidth]{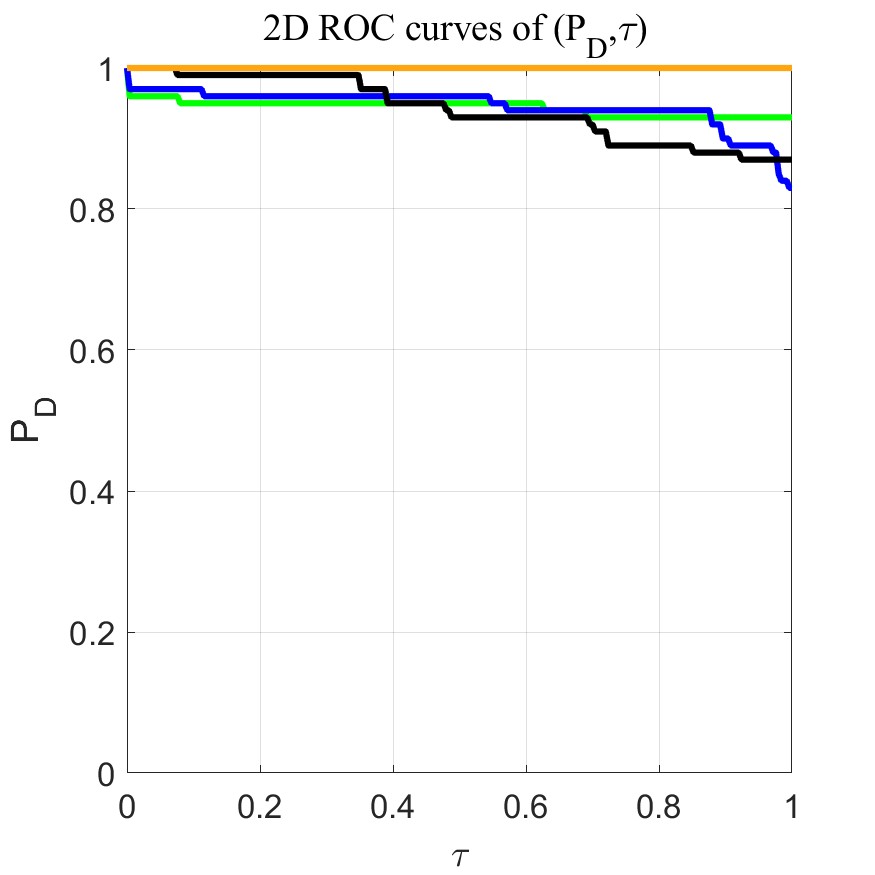}%
\hspace{0.01\linewidth}%
\includegraphics[height=0.15\textheight,width=0.22\linewidth]{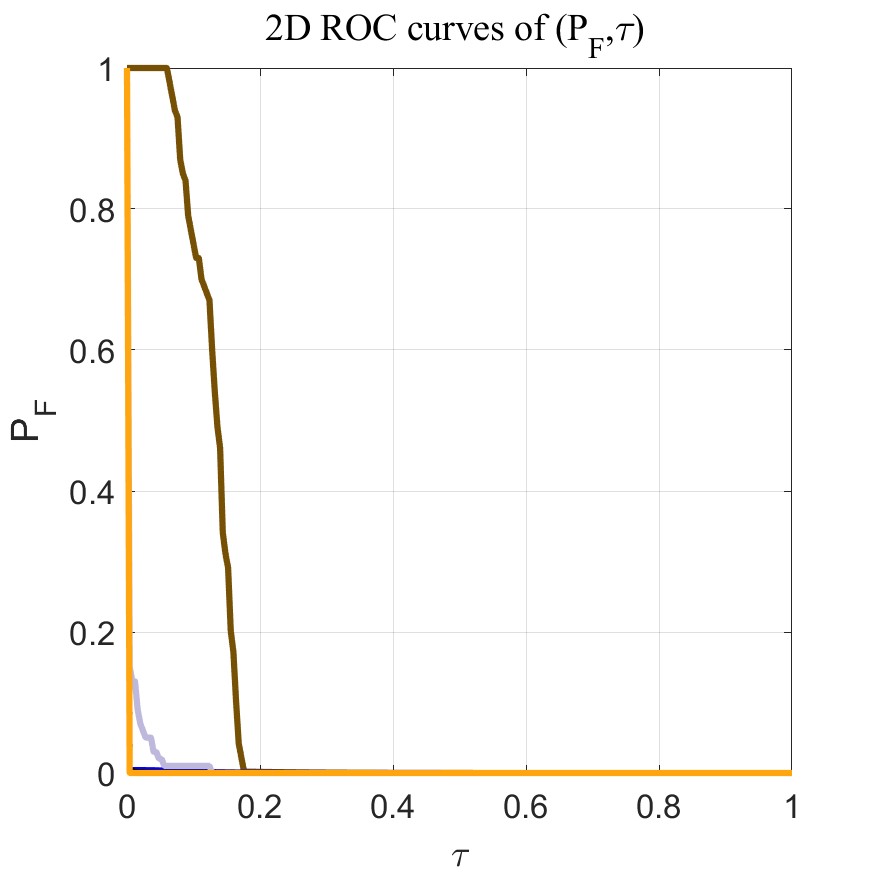}
\caption{Three-dimensional ROC curves of nine competing methods and their resulting 2-D ROC curves in sequence 5.}
\label{fig_row5}
\end{figure*}

\begin{figure*}[!h]
\centering
\includegraphics[height=0.15\textheight,width=0.22\linewidth]{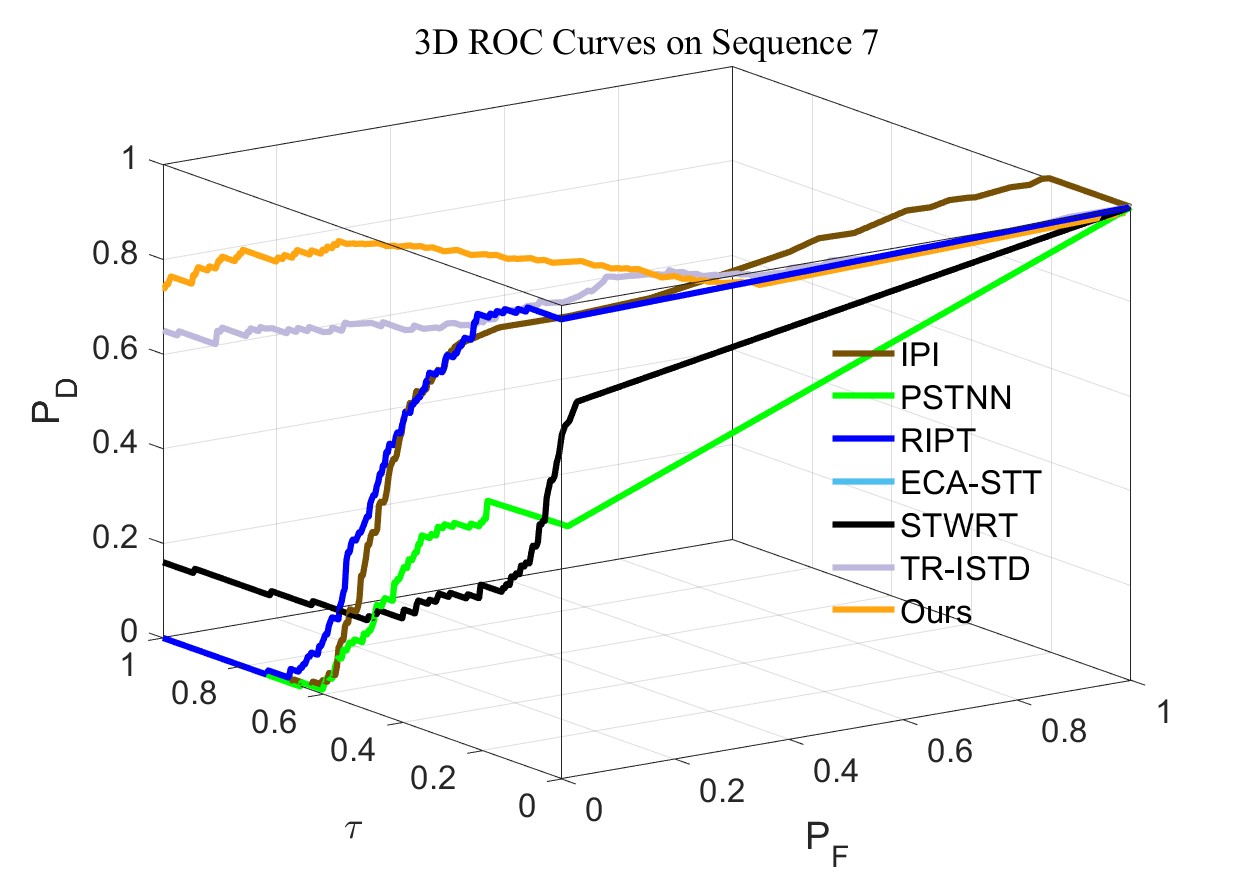}%
\hspace{0.01\linewidth}%
\includegraphics[height=0.15\textheight,width=0.22\linewidth]{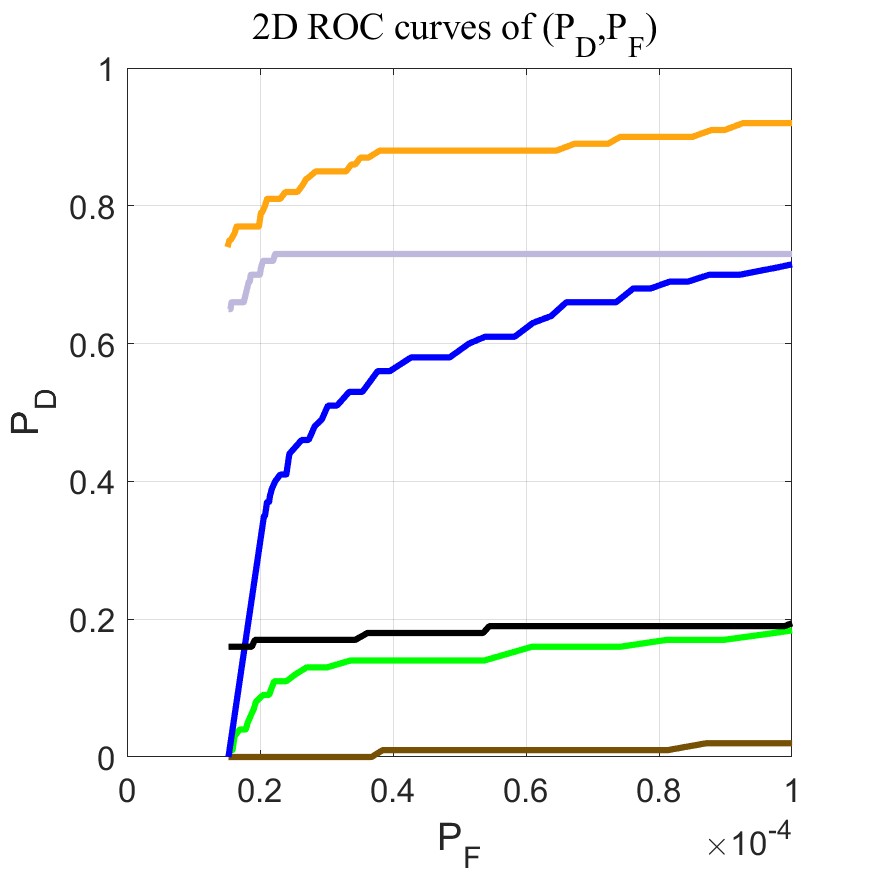}%
\hspace{0.01\linewidth}%
\includegraphics[height=0.15\textheight,width=0.22\linewidth]{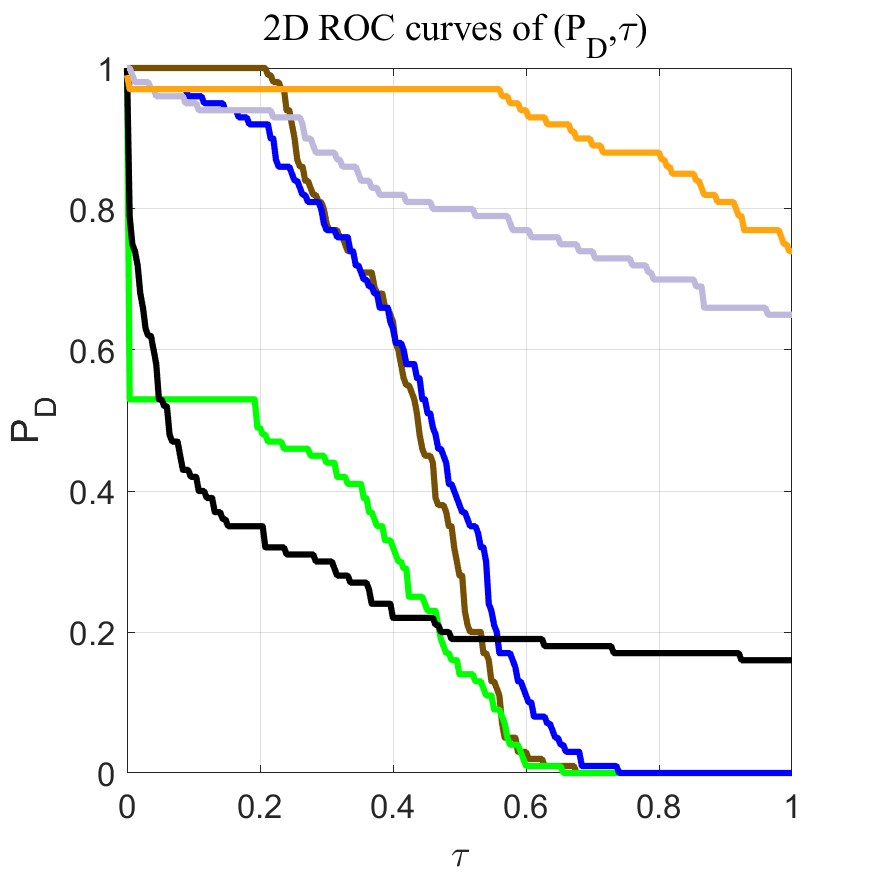}%
\hspace{0.01\linewidth}%
\includegraphics[height=0.15\textheight,width=0.22\linewidth]{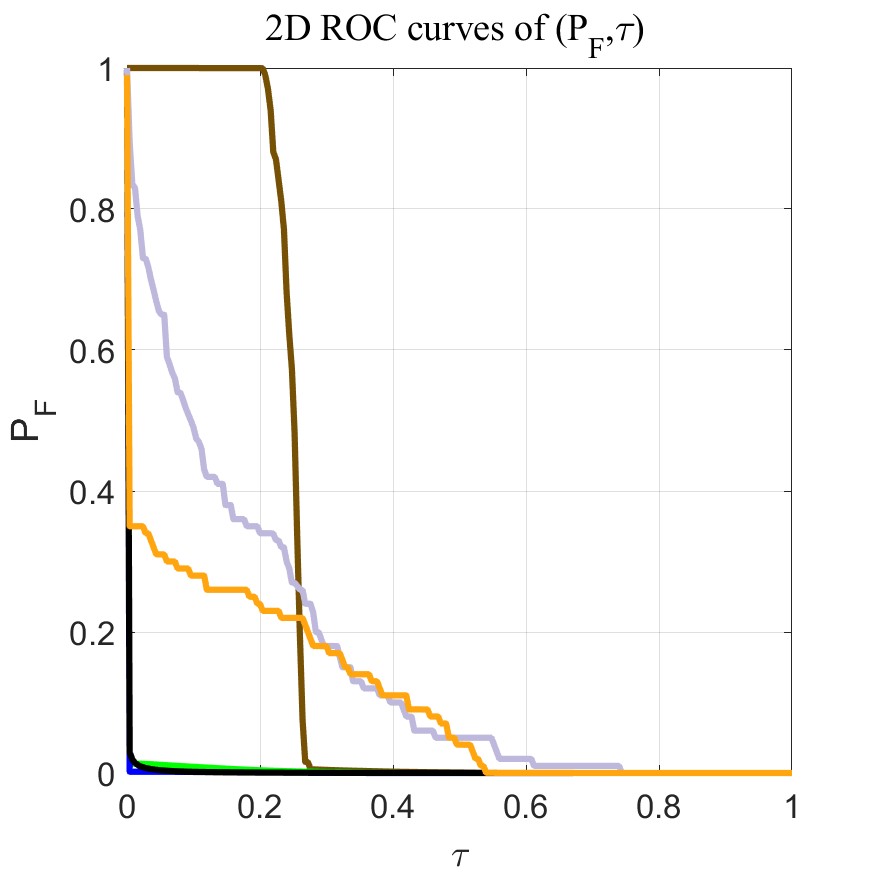}
\caption{Three-dimensional ROC curves of nine competing methods and their resulting 2-D ROC curves in sequence 6.}
\label{fig_row6}
\end{figure*}
\begin{figure*}[!t]
\centering

\subfloat[]{\includegraphics[width=0.15\linewidth]{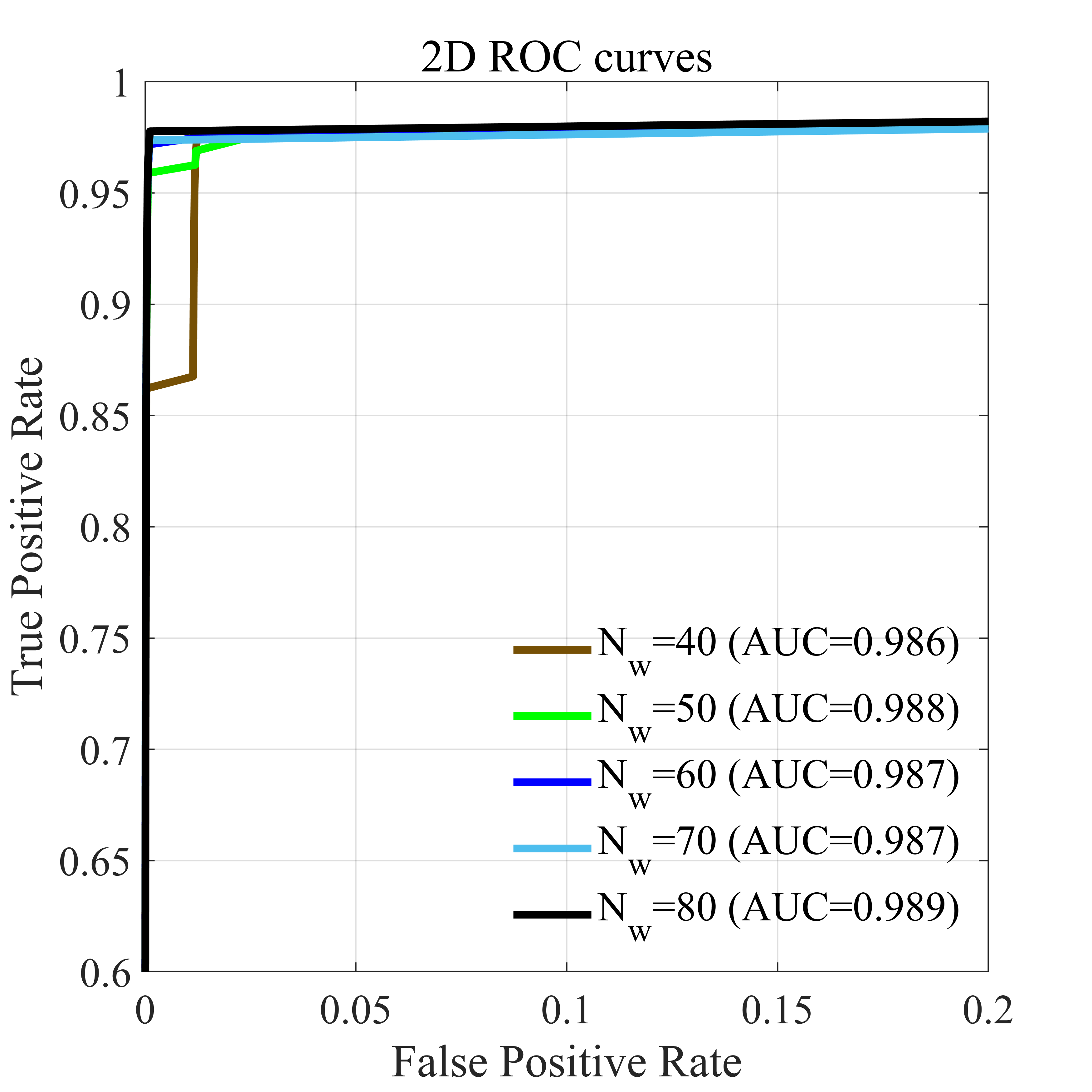}%
\label{fig_first_case}}
\hfill
\subfloat[]{\includegraphics[width=0.15\linewidth]{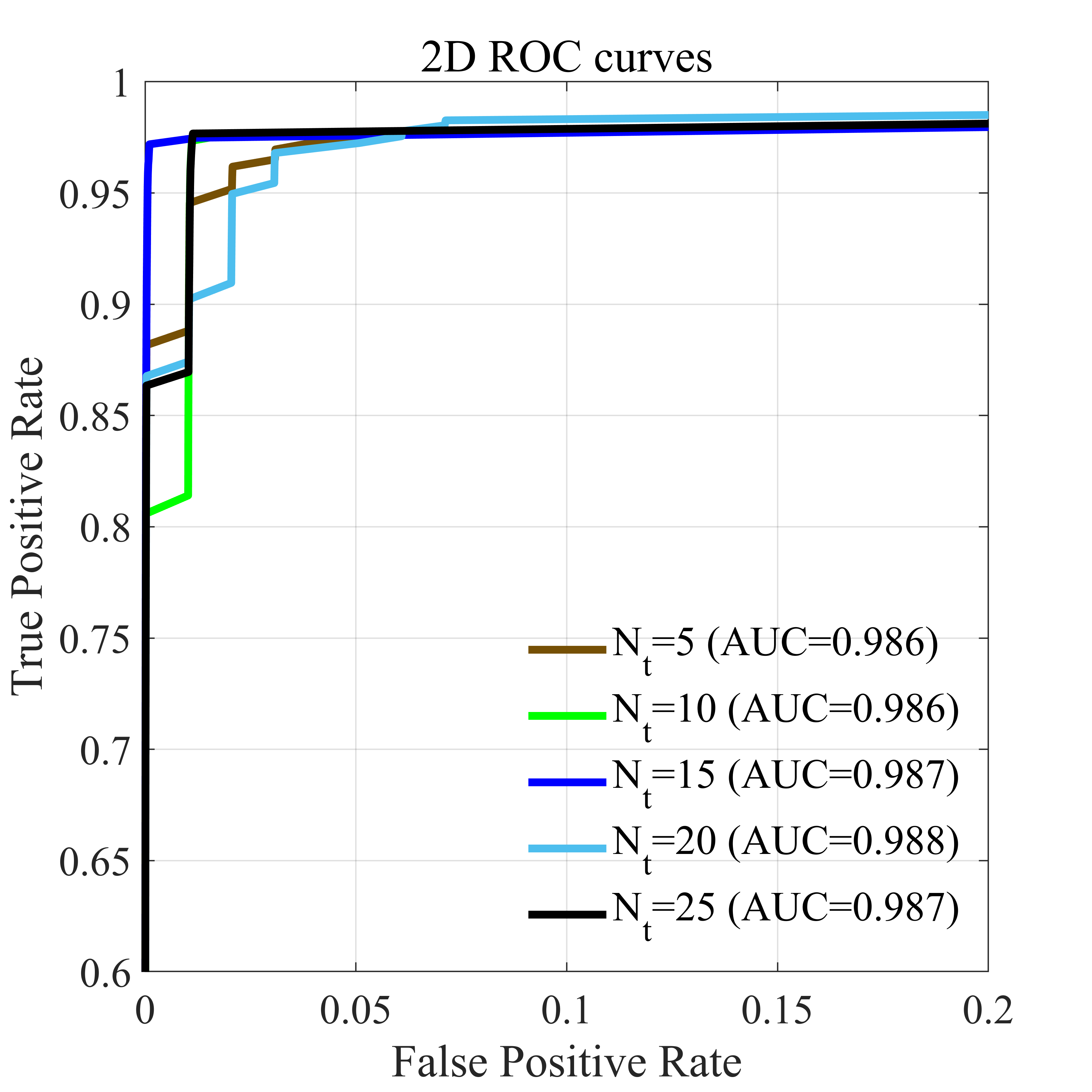}%
\label{fig_second_case}}
\hfill
\subfloat[]{\includegraphics[width=0.15\linewidth]{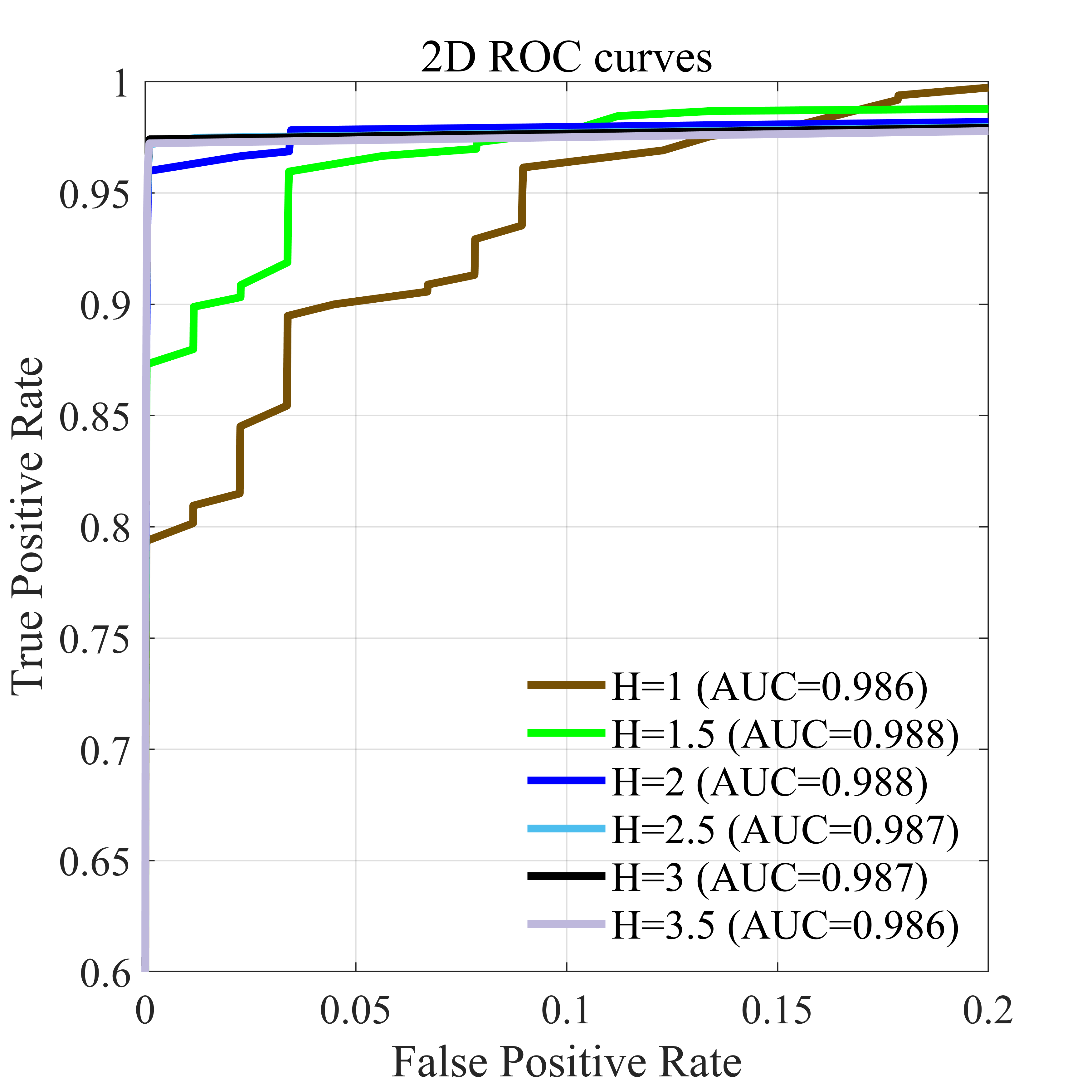}%
\label{fig_third_case}}
\hfill
\subfloat[]{\includegraphics[width=0.15\linewidth]{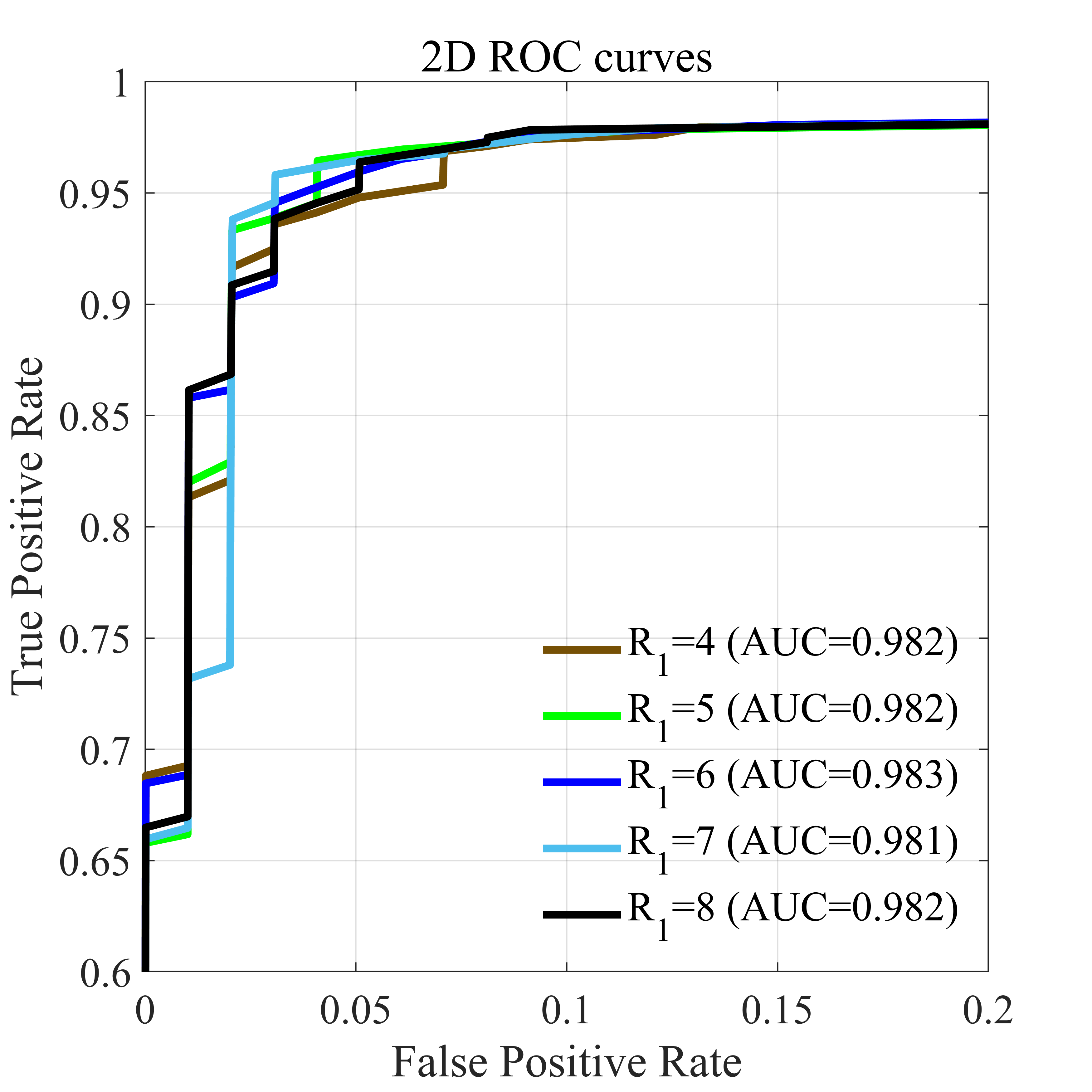}%
\label{fig_fourth_case}}
\hfill
\subfloat[]{\includegraphics[width=0.15\linewidth]{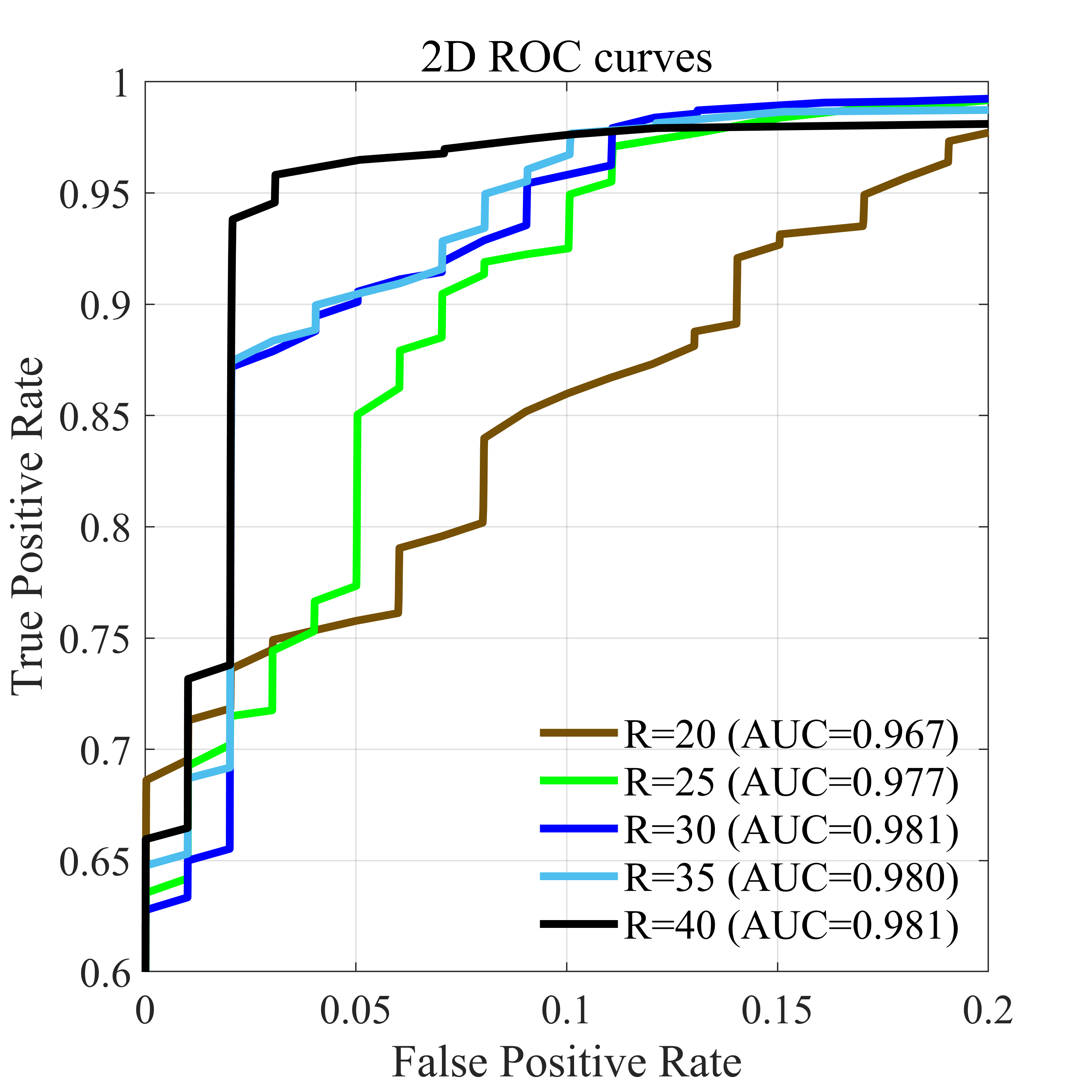}%
\label{fig_fifth_case}}
\hfill
\subfloat[]{\includegraphics[width=0.15\linewidth]{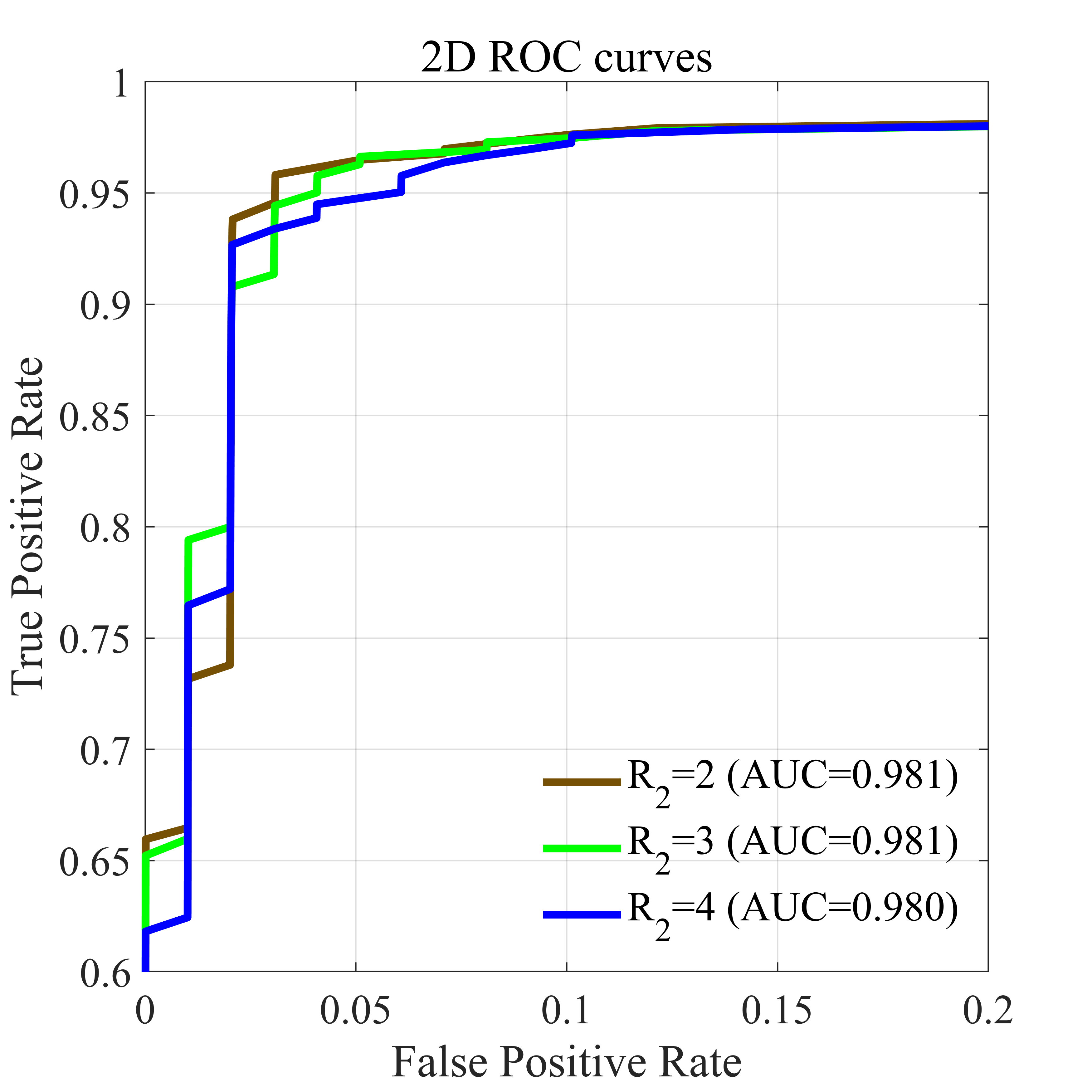}%
\label{fig_sixth_case}}

\caption{ROC curves of six parameters under Seq. 6.}
\label{fig_sim}
\end{figure*}
\subsection{Parameter Analysis}
Based on empirical findings, this study systematically optimized the key hyperparameters of the model. The regularization parameter controlling the sparsity of the target tensor was set to $\lambda_{1}=0.1$; the penalty parameters and corresponding weights were configured as $\beta_{1}=\beta_{2}=\alpha=1$ ; the data fidelity term parameter was set to  $\beta_{3}=2$, and the penalty factor was set to $\rho=0.01$. Other parameters, including the spatial patch size $N_w$, temporal size $N_t$, constant parameter $H$, and the
BTR-related ranks $R_1$, $R$ and $R_2$, significantly influence model performance. Their values were determined through optimization based on ROC metrics on Sequence 3.
\subsubsection{patch size $N_w$ , temporal dimension length $N_t$}
The patch size $N_w$ is a critical parameter that balances algorithmic efficiency and detection performance. Increasing $N_w$ enhances target sparsity but may introduce noise and false detections, while reducing it suppresses noise at the risk of disrupting the low-rank sparse structure between target and background. As shown in Fig. 11(a), detailed numerical results are summarized in Table III, as the patch size increases from 40 to 90 in increments of 10, the ROC curve exhibits a fluctuating upward trend, with the optimal patch size identified as 80. Similarly, the temporal dimension length $N_t$ considerably affects the construction and performance of the four-dimensional tensor: an insufficient number of frames leads to loss of low-rank features and underutilization of temporal information, whereas an excessive number introduces redundancy, compromising spatio-temporal sparsity and degrading performance. Fig. 11(b) indicates that an $N_t$ value of 20 yields the optimal AUC. 
\subsubsection{constant parameter $H$}
The constant parameter $H$, designed for ISTD tasks, exhibits a coupled relationship with $\lambda_{1}$ and possesses a well-defined effective range. Excessively large values may lead to target loss, while insufficiently small values can result in inadequate background suppression. As illustrated in Fig. 11(c)  , detailed numerical results are summarized in Table III, the model achieves optimal performance when $H$ is set to either 1.5 or 2.
\subsubsection{Ranks in BTR decomposition}
Figs. 11(d)–11(f) present the ROC curves and corresponding AUC values obtained by the proposed method on Sequence 3 under different values of $R$, $R_1$, and $R_2$,the specific experimental results are detailed in Table  IV. Among these, $R_1$ and $R_2$ form a dual-rank control mechanism: $R_1$ focuses on the fidelity of low-rank background representation, while $R_2$ enhances the separability of sparse target features. These are synergistically regulated through the interaction rank $R$, providing an interpretable and flexible modeling foundation for infrared small target detection in complex scenarios. The results indicate that the performance remains relatively stable when $R_1\leq6$ and is less sensitive to the choice of $R_2$. In contrast, $R$ has a more substantial impact, with better reconstruction performance achieved when $R_2\geq30$. Therefore, it is recommended to select $R$ from the candidate set $\{30, 35, 40\}$, $R_1$ from $\{4, 5, 6\}$, and $R_2$ from within the range of $\{2, 3\}$.
\section{Conclusion}
In this work, we propose a BTR-ISTD model to address the limitations of existing methods in modeling interdimensional correlations, computational efficiency, and adaptability to complex dynamic backgrounds and long sequences. The proposed approach leverages a bilateral TR structure, which simultaneously captures spatial and temporal–patch correlations while reducing model complexity. To solve the resulting optimization problem, the detection process is formulated within the framework of Proximal Alternating Minimization (PAM) as a low-rank and sparse representation problem. Comparative experimental results demonstrate that the proposed method not only achieves superior target detection and background suppression performance, but also outperforms several state-of-the-art methods in detection speed. However, the model involves a large number of hyperparameters that require manual tuning, making parameter configuration complex and potentially leading to instability in practical applications. Further investigation is needed to optimize and simplify parameter selection.

\typeout{get arXiv to do 4 passes: Label(s) may have changed. Rerun}
\end{document}